\documentstyle[amssymb,aps,multicol,epsf]{revtex}

\begin{document}
\draft
\title{Evolution of networks}

\author{S.N. Dorogovtsev$^{1, 2, \ast}$ and J.F.F. Mendes$^{1,\dagger}$}

\address{
$^{1}$ Departamento de F\'\i sica and Centro de F\'\i sica do Porto, Faculdade de Ci\^encias, 
Universidade do Porto\\
Rua do Campo Alegre 687, 4169-007 Porto, Portugal\\
$^{2}$ A.F. Ioffe Physico-Technical Institute, 194021 St. Petersburg, Russia 
}

\maketitle

\begin{center}
{\em (Submitted to Advances in Physics on 6th March 2001)}
\end{center} 
\vspace{-18pt}$\phantom{i}$

\begin{abstract}
We review the  recent fast progress in statistical physics of evolving networks. 
Interest has focused mainly on the structural properties of 
random 
complex networks in communications, biology, social sciences and economics. 
A number of giant artificial networks of such a kind 
came into existence 
recently. 
This opens a wide field for the study of their topology, evolution, and complex processes 
occurring in them.    
Such networks possess a rich set of scaling properties. 
A number of them are scale-free and  
show striking resilience against random breakdowns. In spite of large sizes of these networks, the distances between most their vertices 
are short --- 
a feature known as the ``small-world'' effect. 
We discuss how growing networks self-organize into scale-free structures  
and the role of the mechanism of preferential linking. We consider the topological and structural properties of evolving networks, 
and 
percolation 
in these networks. We present a number of models demonstrating 
the main features of evolving networks and discuss current approaches for their simulation and analytical study.
Applications of the general results to particular networks in Nature are discussed. 
We demonstrate the generic connections of the network growth processes with the general problems of non-equilibrium physics, econophysics, evolutionary biology, etc.   

\end{abstract}


\vspace{7pt}$\phantom{x}$

\begin{multicols}{2}

\narrowtext



\section*{Contents} 


\begin{small} 

\noindent 
\ref{s-introduction}. \ \ Introduction  
\hfill \pageref{s-introduction}  \\
\ref{s-historical}. \ \ Historical background  
\hfill \pageref{s-historical} \\  
\ref{s-characteristics}. \ \ Structural characteristics of 
evolving networks  
\hfill \pageref{s-characteristics} 

\noindent
\ \ \ \ \ref{ss-degree}. \ \ Degree  
\hfill \pageref{ss-degree}

\noindent
\ \ \ \ \ref{ss-shortest}. \ \ Shortest path 
\hfill \pageref{ss-shortest}

\noindent 
\ \ \ \ \ref{ss-clustering}. \ \ Clustering coefficient 
 \hfill \pageref{ss-clustering}

\noindent 
\ \ \ \ \ref{ss-size}. \ \  Size of the giant component 
 \hfill \pageref{ss-size}

\noindent
\ \ \ \ \ref{ss-many-node}. \ \ Other many-vertex characteristics  
 \hfill \pageref{ss-many-node}
\\
\ref{s-notions}. \ \ Notions of equilibrium and non-equilibrium networks 
 \hfill \pageref{s-notions}
\\
\ref{s-nature}. \ \ Evolving networks in Nature 
 \hfill \pageref{s-nature}

\noindent
\ \ \ \ \ref{ss-citations}. \ \ Networks of citations of scientific papers 
 \hfill \pageref{ss-citations}

\noindent
\ \ \ \ \ref{ss-collaborations}. \ \ Networks of collaborations  
 \hfill \pageref{ss-collaborations}

\noindent
\ \ \ \ \ref{ss-communications}.\,Communications networks, the WWW and Internet 
\hfill \pageref{ss-communications}

\noindent
\ \ \ \ \ \ \ \ref{sss-internet}. \ \ Structure of the Internet 
 \hfill \pageref{sss-internet}

\noindent
\ \ \ \ \ \ \ \ref{sss-www}. \ \ Structure of the WWW  
 \hfill \pageref{sss-www}

\noindent
\ \ \ \ \ref{ss-biological}. \ \ Biological networks 
\hfill \pageref{ss-biological}

\noindent
\ \ \ \ \ \ \ \ref{sss-neural}. \ \ Structure of neural networks 
\hfill \pageref{sss-neural}

\noindent
\ \ \ \ \ \ \ \ref{sss-metabolic}. \ \ Networks of metabolic reactions 
\hfill \pageref{sss-metabolic}

\noindent
\ \ \ \ \ \ \ \ref{sss-protein}. \ \ Protein networks 
\hfill \pageref{sss-protein}

\noindent
\ \ \ \ \ \ \ \ref{sss-ecological}. \ \ Ecological and food webs 
\hfill \pageref{sss-ecological}

\noindent
\ \ \ \ \ \ \ \ref{sss-word_web}. \ \ Word Web of human language 
\hfill \pageref{sss-word_web}
 
\noindent 
\ \ \ \ \ref{ss-circuits}. \ \ Electronic circuits 
\hfill \pageref{ss-circuits}
 
\noindent 
\ \ \ \ \ref{ss-other}. \ \ Other networks 
\hfill \pageref{ss-other}
\\ 
\ref{s-classical}. \ \ Classical random graphs, the Erd\"os-R\'{e}nyi model 
\hfill \pageref{s-classical}
\\
\ref{s-small-world}. \ \ Small-world networks 
\hfill \pageref{s-small-world}

\noindent
\ \ \ \ \ref{ss-watts}. \ \ The Watts-Strogatz model and its variations 
\hfill \pageref{ss-watts}

 
\noindent
\ \ \ \ \ref{ss-smallest}. \ \ The smallest-world network 
\hfill \pageref{ss-smallest}

\noindent
\ \ \ \ \ref{ss-another}. \ \ Other possibilities to obtain large 
clustering coefficient 
\hfill \pageref{ss-another}
\\ 
\ref{s-exponential}. \ \ Growing exponential networks 
\hfill \pageref{s-exponential}
\\
\ref{s-scale-free}. \ \ Scale-free networks 
\hfill \pageref{s-scale-free}

\noindent 
\ \ \ \ \ref{ss-idea}. \ \ Barab\'asi-Albert model and the idea of preferential linking 
\hfill \pageref{ss-idea}

\noindent
\ \ \ \ \ref{ss-masterequation}. \ \ Master equation approach 
\hfill \pageref{ss-masterequation}

\noindent
\ \ \ \ \ref{ss-simplestscale-free}. \ \ A simple model of scale-free networks 
\hfill \pageref{ss-simplestscale-free}

\noindent
\ \ \ \ \ref{ss-relations}. \ \ Scaling relations and cutoff 
\hfill \pageref{ss-relations}

\noindent 
\ \ \ \ \ref{ss-continuous}. \ \ Continuum approach 
\hfill \pageref{ss-continuous}

\noindent 
\ \ \ \ \ref{ss-estimations}. \ \ \ More \,complex \,models \,and \,estimates for \,the WWW 
\hfill \pageref{ss-estimations}

\noindent
\ \ \ \ \ref{ss-types}. Types of preference providing scale-free networks 
\hfill \pageref{ss-types}

\noindent
\ \ \ \ \ref{ss-capture}. \ \ ``Condensation'' of edges 
\hfill \pageref{ss-capture}

\noindent
\ \ \ \ \ref{ss-distributionoflinks}.\,Correlations and distribution of edges over network\hfill\pageref{ss-distributionoflinks}


 
\noindent
\ \ \ \ \ref{ss-accelerating}. \ \ Accelerated growth of networks 
\hfill \pageref{ss-accelerating}

\noindent
\ \ \ \ \ref{ss-decay}. \ \ Decaying networks 
\hfill \pageref{ss-decay}

\noindent
\ \ \ \ \ref{ss-spectrum}. \ \ Eigenvalue spectrum of the adjacency matrix 
\hfill \pageref{ss-spectrum}

\noindent
\ \ \ \ \ref{ss-trees}. \ \ Scale-free trees 
\hfill \pageref{ss-trees}
\\ 
\ \ \ \ \ref{s-non-scale-free}. \ \ Non-scale-free networks with preferential linking 
\hfill \pageref{s-non-scale-free}
\\
\ref{s-percolative}. \ \ Percolation on networks 
\hfill \pageref{s-percolative}
 
\noindent
\ \ \ \ \ref{ss-theory}. \ \ Theory of percolation on undirected equilibrium networks 
\hfill \pageref{ss-theory}
 
\noindent
\ \ \ \ \ref{ss-directed}. \ \ Percolation on directed equilibrium networks 
\hfill \pageref{ss-directed}

\noindent
\ \ \ \ \ref{ss-failure}. \ \ Failures and attacks 
\hfill \pageref{ss-failure}

\noindent 
\ \ \ \ \ref{ss-resilience}. \ \ Resilience against random breakdowns  
\hfill \pageref{ss-resilience}

\noindent 
\ \ \ \ \ref{ss-intentional}. \ \ Intentional damage 
\hfill \pageref{ss-intentional}

\noindent 
\ \ \ \ \ref{ss-spread}. \ \ Disease spread within networks 
\hfill \pageref{ss-spread}

\noindent 
\ \ \ \ \ref{ss-anomalous}. \ \ Anomalous percolation on  
growing networks 
\hfill \pageref{ss-anomalous}
\\ 
\ref{s-soc}. \ \ Growth of networks and self-organized criticality 
\hfill \pageref{s-soc}

\noindent 
\ \ \ \ \ref{ss-sand-pile}. \ \ Linking with sand-pile problems 
\hfill \pageref{ss-sand-pile}

\noindent 
\ \ \ \ \ref{ss-simon}. \ \ Preferential linking and the Simon model 
\hfill \pageref{ss-simon}


\noindent
\ \ \ \ \ref{ss-multiplicative}. \ \ Multiplicative stochastic models and the generalized Lotka-Volterra equation 
\hfill \pageref{ss-multiplicative}
\\ 
\ref{s-concluding}. \ \ Concluding remarks 
\hfill \pageref{s-concluding}
\\ 
Acknowledgements 
\hfill \pageref{s-acknowledgements}
\\
References 
\hfill \pageref{s-references}
\\

\end{small}


\section{Introduction}\label{s-introduction}

The Internet and World Wide Web are perhaps the most impressive 
creatures of our civilization (Baran 1964). 
\cite{b64}. Their influence on 
us is incredible. They are part of our life, of our world. 
Our present and our future are impossible without them. 
Nevertheless, we know much less about them than one may expect. 
We know surprisingly little of their structure and hierarchical 
organization, their global topology, their local properties, 
and various processes occurring 
within them. This knowledge is needed for the most effective 
functioning of the Internet and WWW, for ensuring their safety, 
and for utilizing 
all of their possibilities. Certainly, the understanding of such problems 
is a topic not of computer science and applied mathematics, 
but rather of non-equilibrium statistical physics.

In fact, these wonderful communications nets 
\cite{ajb99,ha99,huppl98,fff99,bkm00,lg98,lg99,cmm99,k99d} 
are only particular examples of a great class of evolving 
networks. Numerous networks, e.g., 
collaboration networks 
\cite{ws98,watbook99,n00,n00f,bjnr01a}, 
public relations nets  
\cite{m67,s91book,wfbook94,g97,ak00a}, 
citations of scientific papers 
\cite{l26,s57,g72,g79book,er90book,ls98,r98}, 
some industrial networks 
\cite{ws98,watbook99,asbs00}, 
transportation networks \cite{mcf96,bmr99}, nets of relations 
between enterprises and agents in financial markets \cite{lm99}, 
telephone call graphs \cite{acl00}, many biological networks 
\cite{k69,bl97,k99e,asbl99,bi99,el00,bs00,ugc00b,jtaob00,itm00,ico01,jmbo01a,w01d}, 
food and ecological webs 
\cite{cbnbook90,wm00,ms00,sm00,cga01,cga01b,wmbdb01}, etc., belong to it. 
The finiteness of these networks sets serious restrictions 
on 
extracting useful experimental data because of strong 
size effects and, often, insufficient statistics.    
The large size of the Internet and WWW and their extensive 
and easily accessible documentation allow reliable and 
informative experimental investigation of their structure and 
properties.     
Unfortunately, 
the statistical theory of neural networks \cite{h82,abook89} 
seems to be rather useless for the understanding of problems 
of the evolution of networks, since this advanced theory does 
not seriously touch on the main question arising for real networks -- 
how networks becomes specifically structured during their growth. 

Quite recently, general features of structural organization of 
such networks were discovered 
\cite{ajb99,ha99,fff99,bkm00,r98,jtaob00,ba99,baj99,s01a}. 
It has become clear that their complex scale-free structure is 
a natural consequence of the principles of their growth. 
Some simple basic ideas 
have been proposed. Self-organization of growing networks and processes 
occurring within them
have been related \cite{ba00a,ceah00a,nsw00,cnsw00,ceah00b} 
to corresponding phenomena (growth phenomena \cite{bsbook95}, 
self-organization \cite{s55,sbook57,kbook93} and self-organized 
criticality \cite{btw87,btw88,bbook97}, percolation 
\cite{e80,sabook91,bhbook94}, localization, etc.) being studied 
by physicists for a long time. 

The goal of our paper is to review the recent rapid progress in understanding the evolution of networks using ideas and methods of statistical physics. The problems that we discuss relate to computer science, mathematics, physics, engineering, biology, economy, and social sciences. Here, we present the point of view of physicists. To restrict ourselves, we do not dwell on Boolean and neural networks.


\section{Historical background}\label{s-historical}

The structure of 
networks has been studied by 
mathematical 
graph theory \cite{bbook85,bbook98,jtrbook00}. Some basic ideas, used 
later by physicists, 
were proposed long ago by the incredibly prolific and outstanding Hungarian mathematician Paul Erd\"{o}s and his collaborator R\'enyi \cite{er59,er60}. Nevertheless, the most intriguing 
type of growing networks, which evolve into 
scale-free structures, 
hasn't been studied by graph theory. 
Most of the results  
of graph theory \cite{fkp89,jklp93} are related to the simplest random graphs 
with Poisson distribution of connections \cite{er59,er60} (classical random graph). 
Moreover, in graph theory, by definition, random graphs are 
graphs with Poisson distribution of connections (we use this 
term in a much more wide sense).
Nevertheless, one should note the very important results 
obtained recently by mathematicians for 
graphs with arbitrary distribution of connections \cite{mr95,mr98}.

The mostly empirical study of specific large random networks such as nets of citations in scientific literature has a long history \cite{l26,s57,g72,g79book}. Unfortunately, their limited sizes 
did not allow to get reliable data and describe their structure until 
recently. 

Fundamental concepts such as  functioning and practical organization of large communications networks were elaborated by the ``father'' of the Internet, Paul Baran, \cite{b64}. 
Actually, many present studies are based on his original ideas and use his terminology. What is the optimal design of communications networks? How may one 
ensure their stability and safety? These and many other vital problems were first studied by P. Baran 
in a practical context. 

By the middle of 90's, the Internet and the WWW had reached 
very large sizes and continued to grow so rapidly that intensively developed search engines failed to cover a great part of the WWW 
\cite{lg98,lg99,cmm99,bp98,lg98c,c99,b00,lg99b,l00a,rm00}. 
A clear knowledge of the structure of the WWW has become vitally 
important  
for its effective operation. 

The first experimental data, mostly for the simplest structural 
characteristics of the communications networks, 
were obtained in 1997-1999 \cite{ajb99,ha99,huppl98,fff99,krrt99,kkrrt99}. 
Distributions of the number of connections in the networks and their surprisingly small average shortest-path lengths were measured.  
A special role of long-tailed, power-law distributions was revealed. After these findings, physicists started intensive study of evolving 
networks in various areas, from communications to biology and public relations.


\section{Structural characteristics of evolving networks}\label{s-characteristics}

Let us start by introducing the objects under discussion. The networks that we consider 
are graphs consisting of vertices (nodes) connected by edges (links). Edges may be directed or undirected (leading to directed and undirected networks, relatively). 
For definition of distances in a network, one sets lengths of all edges to be one. 

Here we do not consider networks with unit loops (edges started and terminated at the same vertex) and multiple edges, i.e., we assume that only one edge may connect two vertices. (One should note that multiple edges are encountered in some collaboration networks \cite{n00f}. Pairs of opposing edges connect some vertices in the WWW, in networks of protein-protein interactions, and in food webs. Also, protein-protein interaction nets and food webs contain unit loops (see below).  
Nets with ``weighted'' edges are discussed in Ref. \cite{yjbt01}.) 

The structure of a network is described by its adjacency matrix, $\hat{B}$, 
whose elements consist 
of zeros and ones.  
An element of the adjacency matrix of a network with undirected edges, $b_{\mu\nu}$, is $1$ if vertices $\mu$ and $\nu$ are connected, and is $0$ otherwise. Therefore, the adjacency matrix of a network with undirected edges is symmetrical. For a network with directed edges, an element of the adjacency matrix, $b_{\mu\nu}$, equals $1$ if there is an edge from the vertex $\mu$ to the vertex $\nu$, and equals $0$ otherwise. 

In the case of a random network, an adjacency matrix describes only a particular member of the entire {\em statistical ensemble} of random graphs. Hence, what one observes is only a particular realization of this statistical ensemble and the adjacency matrix of this graph is only a particular member of the corresponding ensemble of matrices.

The statistics of the adjacency matrix of a random network contains complete information about the structure of the net, and, in principle, one has to study just the adjacency matrix. 
Generally, this is not an easy task,   
so that, instead of this, only a very restricted set of structural characteristics is usually considered. 


\subsection{Degree}\label{ss-degree}

The simplest and the most intensively studied one-vertex characteristic is {\em degree}. 
Degree, $k$, of a vertex is the total number of its connections. (In physical literature, this quantity is often called ``connectivity'' that has a quite different meaning in graph theory. Here, we use the mathematically correct definition.) In-degree, $k_{i}$, is the number of incoming edges of a vertex. Out-degree, $k_{o}$ is the number of its outgoing edges. Hence, 
$k = k_{i} + k_{o}$. 
Degree is actually the number of nearest neighbors of a vertex, $z_1$. 
Total distributions of vertex degrees of an entire network, 
$P(k_{i},k_{o})$ --- the joint in- and out-degree distribution, $P(k)$ --- the degree distribution, $P_{i}(k_{i})$ --- the in-degree distribution, and $P_{o}(k_{o})$ --- the out-degree distribution ---  
are its basic statistical characteristics. Here, 

\begin{eqnarray}
P(k) = & & \sum_{k_{i}}P(k_{i},k-k_{i}) = \sum_{k_{o}}P(k-k_{o},k_{o})
\, ,
\nonumber
\\[5pt]
P_{i}(k_{i}) & = & \sum_{k_{o}}P(k_{i},k_{o})
\, ,
\nonumber
\\[5pt]
P_{o}(k_{o}) & = & \sum_{k_{i}}P(k_{i},k_{o})
\, .  
\label{0-1}
\end{eqnarray} 
For brevity, instead of $P_{i}(k_{i})$ and $P_{o}(k_{o})$ we usually use the notations $P(k_{i})$ and $P(k_{o})$. 
If a network has no connections with the exterior, then the average in- and out-degree are equal:

\begin{equation}
\overline{k}_{i} = \!\!\!\sum_{k_{i},k_{o}}\!\!  k_{i}P(k_{i},k_{o}) = 
\overline{k}_{o} = \!\!\!\sum_{k_{i},k_{o}}\!\!  k_{o}P(k_{i},k_{o})
\, ,  
\label{0-2}
\end{equation} 

Although the degree of a vertex is a local quantity, we shall see that a degree distribution often determines some important global characteristics of random networks. Moreover, if statistical correlations between vertices are absent, $P(k_{i},k_{o})$ totally determines the structure of the network.


\subsection{Shortest path}\label{ss-shortest}

One may define a geodesic distance between two vertices, $\mu$ and $\nu$, of a graph with unit length edges. It is the shortest-path length, 
$\ell_{\mu\nu}$, from the vertex $\mu$ to the vertex $\nu$. If vertices are directed, $\ell_{\mu\nu}$ is not necessary 
equal to $\ell_{\nu\mu}$. 
It is possible to introduce the distribution of the shortest-path lengths between pairs of vertices of a network and the average 
shortest-path length $\overline{\ell}$ of a network. The average here is over all pairs of vertices between which a path exists and 
over all realizations of a network. 

$\overline{\ell}$ is often called the ``diameter'' of a network. It determines the effective ``linear size'' of a network, the average separation of pairs of vertices. For a lattice of dimension $d$ containing $N$ vertices, obviously, $\overline{\ell} \sim N^{1/d}$. In a fully 
connected network, $\overline{\ell} = 1$. One may roughly estimate  
$\overline{\ell}$ of a network in which random vertices are connected. If the average number of nearest neighbors of a vertex is $z_1$, then about $z_1^\ell$ vertices of the network are at a distance 
$\ell$ from the vertex or closer. Hence, $N \sim z_1^{\overline{\ell}}$ and then $\overline{\ell} \sim \ln N /\ln z_1$, i.e., the average shortest-path length value is small even for very large networks. 
This smallness is usually referred to as a {\em small-world effect} \cite{ws98,watbook99,n00g}.  

One can also introduce the maximal shortest-path length over all the pairs of vertices between which a path exists. This characteristic determines the maximal extent of a network.  
(In some papers the maximal shortest path is also referred to as the diameter of the network, so that we 
avoid to use this term.)


\subsection{Clustering coefficient}\label{ss-clustering}

For the description of connections in the environment closest to a vertex, one introduces the so-called {\em clustering coefficient}. For a network with undirected edges, the number of all possible connections of the nearest neighbors of a vertex $\mu$ ($z_1^{(\mu)}$ nearest 
neighbors) equals $z_1^{(\mu)}(z_1^{(\mu)}-1)/2$. Let only $y^{(\mu)}$ of them be present. The clustering coefficient of this vertex, 
$C^{(\mu)} \equiv y^{(\mu)}/[z_1^{(\mu)}(z_1^{(\mu)}-1)/2]$, is the fraction of existing connections between nearest neighbors of the vertex. Averaging $C^{(\mu)}$ over all vertices of a network yields the clustering coefficient of the network, $C$. 
The clustering coefficient is the probability that two nearest neighbors of a vertex are nearest neighbors also of one another.  
The clustering coefficient of the network reflects the ``cliquishness'' of the mean closest neighborhood of a network vertex, that is, the extent to which the nearest neighbors of a vertex are the nearest neighbors of each other \cite{ws98}. One should note that the notion of clustering was introduced in sociology \cite{wfbook94}.

From another point of view, $C$ is the probability that if a triple of vertices of a network is connected together by at least two edges then the third edge is also present. One can check that 
$C/3$ is equal to the number of triples of vertices connected together by three edges divided by the number of all connected triples of vertices.

Instead of $C^{(\mu)}$, it is equally possible to use another related characteristic of clustering, 
$D^{(\mu)} \equiv (z_1^{(\mu)}+y_1^{(\mu)})/[(z_1^{(\mu)}+1)z_1^{(\mu)}/2]$, that is, the fraction of existing connections inside of a set of vertices consisting of the vertex $\mu$ and all its nearest neighbors. 
$D^{(\mu)}$ plays the role of local density of linkage.  
$C^{(\mu)}$ and $D^{(\mu)}$ are connected by the following relations: 

\begin{eqnarray}
D^{(\mu)} & = & C^{(\mu)} + \frac{2}{z_1^{(\mu)}+1}(1-C^{(\mu)})
\, ,
\nonumber
\\[5pt]
C^{(\mu)} & = & D^{(\mu)} - \frac{2}{z_1^{(\mu)}-1}(1-D^{(\mu)})
\, .  
\label{1-1}
\end{eqnarray} 

In a network in which all pairs of vertices are connected (the complete graph) $C=D=1$.   
For tree-like graphs, $C=0$.  
In a 
classical random graph 
$C = M/[N(N-1)/2]= z_1/(N-1)$, 
$D = M/[N(N+1)/2]= z_1/(N+1)$. Here, $N$ is the total number of vertices of the graph, $M$ is the total number of its edges, and $z_1$ is an average number of the nearest neighbors of a vertex in the graph, $M=z_1N/2$. In an ordered lattice, $0 \leq C \leq 1$ depending on its structure. Note that $0 \leq C \leq 1$ but 
$0 < 2/(z_1+1) \leq D \leq 1$.


\subsection{Size of the giant component}\label{ss-size} 

Generally, a network may contain disconnected parts. 
In networks with undirected edges, it is easy to introduce the notion corresponding to the percolating cluster in the case of disordered lattices. 
If the relative size of the largest connected cluster of vertices of a network (the largest connected component) 
approaches a nonzero value when the network is grown to infinite size, the system is above the percolating threshold, and this cluster is called the 
{\em giant connected component} of the network. In this case, the size of the next largest cluster, etc. are small compared to the giant connected component for a large enough network. 
Nevertheless, size effects are usually strong 
(see Sec. \ref{ss-failure}), and for accurate 
measurement of the size of the giant connected component, large networks must be used. 

One may generalize this notion for networks with directed edges. In this case, we have to consider a cluster of vertices from 
each of that one can approach any vertex of this cluster. 
Such a cluster may be called the strongly connected component. 
If the largest strongly connected component
contains a finite fraction of all vertices in the large network limit, it is  
called the {\em giant strongly connected component}. Connected clusters obtained from a directed network by ignoring directions of 
its edges are called weakly connected components, and one can define the {\em giant weakly connected component} of a network.


\subsection{Other many-vertex characteristics}\label{ss-many-node}

One can get a general picture of the distribution of edges between vertices in a network considering the average elements of the adjacency matrix, 
$\overline{b}_{\mu\nu}$ (here, the averaging is over realizations 
of the evolution process, if the network is evolving, or over all configurations, if it is static) although this characteristic is not very informative. 

A local characteristic, degree, $k \equiv k_1 = z_1$ can be easily generalized. 
It is possible to introduce the number of vertices at 
a distance equal $2$ or less from a vertex, $k_2$, the number of second neighbors, $z_2 \equiv k_2 - k_1$, etc.  
Generalization of the clustering coefficient is also straightforward: one has to count all edges between $n$-th nearest neighbors.  

One may consider distributions of these quantities and their average values. 
Often, it is possible to fix a vertex not by 
its label, $\mu$ but only by its in- and out-degrees, therefore, it is reasonable to introduce the probability  
$P(k_{i},k_{o};k_{i}^\prime,k_{o}^\prime)$ 
that a pair of vertices -- the first vertex with the in- and out-degrees $k_{i}$ and $k_{o}$ 
and the second one with the in- and out-degrees $k_{i}^\prime$ and $k_{o}^\prime$ -- are connected by a directed edge going out from the first vertex and coming to the second one \cite{krl00,kr00c}.  

It is easy to introduce a similar quantity for networks with undirected edges, namely the distribution $P(k_1,k_2)$ of the 
degrees of nearest neighbor vertices. This distribution indicates correlations between the degrees of nearest neighbors in a network: 
if $P(k_1,k_2)$, does not factorize,  
these correlations are present \cite{krl00,kr00c}. Unfortunately, it is hard to measure such distributions because of the poor statistics. However, one may easily observe these correlations studying 
a related characteristic -- the dependence $\overline{k}_{nn}(k)$ of the average degree of the nearest neighbors $\overline{k}_{nn}$ on the degree $k$ of a vertex \cite{pvv01a}. 

Similarly, it is difficult to measure a standard joint in- and out- degree distribution $P(k_i,k_o)$. 
However, one may measure the dependences  
$\overline{k}_i(k_o)$ of the average in-degrees $\overline{k}_i$ for vertices of the out-degree $k_o$ 
and $\overline{k}_o(k_i)$ of the average out-degrees $\overline{k}_o$ for vertices of the in-degree $k_i$.

One may also consider the probability, $P_n(k_1,k_n)$, that the number of vertices at a distance $n$ or less from a vertex equals $k_n$, if the degree of the vertex is $k_1$, etc. 
Some other many-node characteristics will be introduced hereafter.


\section{Notions of equilibrium and non-equilibrium networks}\label{s-notions}

From a physical point of view, random networks may be ``equilibrium'' or ``non-equilibrium''. Let us introduce these important notions using simple examples. 

(a) An example of an equilibrium random network: A classical undirected random graph \cite{er59,er60} (see Sec. \ref{s-classical}). 

It is defined by the following rules:

(i) The total number of vertices is fixed. 

(ii) Randomly chosen pairs of vertices are connected via undirected edges. 

Vertices of the classical random graph are statistically independent and equivalent. The construction procedure of such a graph may be thought of as the subsequent addition of new edges between vertices 
chosen at random. When the total number of vertices is fixed, this procedure obviously produces equilibrium configurations. 

(b) The example of a non-equilibrium random network: A simple random graph growing through the simultaneous addition of vertices and edges (see, e.g., Ref. \cite{chk01,dms01f} and Sec. \ref{ss-anomalous}). 

Definition of this graph: 

(i) At each time step, a new vertex is added to the graph. 

(ii) Simultaneously, a pair (or several pairs) of randomly chosen vertices is connected. 

One sees that the system is not in equilibrium. Edges are inhomogeneously distributed over the graph. The oldest vertices are the most connected (in statistical sense), and degrees of new vertices are the smallest. 
If, at some moment, we stop to increase the number of vertices but continue the random addition of edges, then the network will tend to an ``equilibrium state'' but never achieve it. Indeed, edges of the network do 
not disappear, so the inhomogeneity survives. An ``equilibrium state'' can be achieved only if, in addition, we allow old edges to disappear from time to time. 

\begin{figure}
\epsfxsize=55mm
\centerline{\epsffile{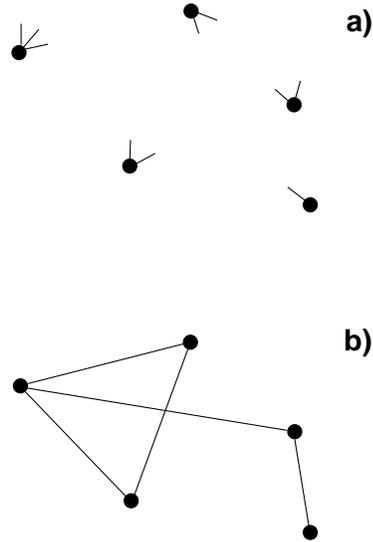}}
\caption{
The construction procedure for an equilibrium random graph with preset arbitrary degree distribution $P(k)$. 
(a) Degrees $\{k_\mu\}$ taken from the distribution are ascribed to the vertices $\{\mu\}$. 
(b) Pairs of random ends sticking out of different vertices are connected. 
}
\label{f0}
\end{figure}

The specific case of equilibrium networks with a Poisson degree distribution was actually the main object of graph theory over more than forty years. Physicists have started the study of non-equilibrium (growing) networks. 
The construction procedure for an equilibrium graph 
with an arbitrary degree distribution $P(k)$ was proposed by Molloy and Reed \cite{mr95,mr98} (note that this procedure cannot be considered as quite rigorous): 

(a) To the vertices $\{\mu\}$ of the graph ascribe degrees $\{k_\mu\}$ taken from the distribution $P(k)$. Now the graph looks like a family of hedgehogs: each vertex has $k_j$ sticking out quills (see Fig. \ref{f0} (a)). 

(b) Connect at random ends of pairs of distinct quills belonging to distinct vertices (see Fig. \ref{f0} (b)).


The generalization of this construction procedure to directed equilibrium graphs with arbitrary joint in- and out-degree distributions $P(k_i,k_o)$ is straightforward.

While speaking about random networks we should keep in mind that a particular network we observe is only one member of a {\em statistical ensemble} of all possible realizations. Hence when we speak about random networks, we actually mean statistical ensembles. The canonical 
ensemble for an undirected network with $N$ vertices has $2^{N(N-1)/2}$ members, i.e. realizations (recall that unit loops and multiple edges are forbidden). 
Each member of the ensemble is a distinct configuration of edges taken with some statistical weight. A rigorous definition 
of a random network must contain a set of statistical weights for all configurations of edges. A grand canonical ensemble of random graphs may be obtained using standard approaches of statistical mechanics. The result, namely the statistical ensemble of equilibrium random networks, is completely determined by the degree distribution. 

The above rather heuristic procedure of Molloy and Reed provides only a particular realization of the equilibrium graph. 
Unfortunately, this procedure is not very convenient for the construction of the entire statistical ensemble, at least, for finite-size networks. 
Surprisingly, the rigorous construction of the statistical ensemble of equilibrium random graphs was made only for classical random graphs (see Sec. \ref{s-classical}), and the problem of strict formal 
construction of the statistical ensemble of equilibrium random graphs with a given degree distribution is still open. (However, see Ref. 
\cite{bck01} for the construction procedure for the statistical ensemble of trees). 

It is possible to construct an equilibrium graph in another way than the Molloy-Reed procedure. Suppose one wants to obtain a large enough 
equilibrium undirected graph with a given set of vertex degrees ${k_\mu}$, where $\mu=1,\ldots,N$. Let us start from an arbitrary 
configuration of 
edges connecting these vertices of degree ${k_\mu}$. We must ``equilibrate'' the graph. 
For this: 

(a) Connect a pair of arbitrary vertices (e.g., $1$ and $2$) by an additional edge. 
Then the degrees of these vertices increase by one ($k_1^\prime=k_1+1$ and 
$k_2^\prime=k_2+1$). 

(b) Choose at random one of edge ends attached to vertex $1$ and rewire it to a randomly chosen vertex $3$. Choose at random one of edge 
ends attached to vertex $2$ and rewire it to a randomly chosen vertex $4$. 
Then $k_1^{\prime\prime}=k_1, k_2^{\prime\prime}=k_2$ and 
$k_3^{\prime\prime}=k_3+1, k_4^{\prime\prime}=k_4+1$.

(c) Repeat (b) until equilibrium is reached. 

Only two vertices of resulting network have degrees greater (by one) than the given degrees ${k_\mu}$. For a large network, this is 
non-essential. If, during our procedure, both the edges under rewiring 
are turned to be rewired to the same vertex, then, at the next step, one may rewire a pair of randomly chosen edges from this vertex.  
Another procedure for the same purpose is described in Ref. \cite{plh01}.

The notion of the statistical ensemble of growing networks may also be introduced in a natural way. This ensemble includes all possible 
paths of the evolution of a network.



\section{Evolving networks in Nature
}\label{s-nature} 

In the present section we discuss some of the most prominent large networks in Nature starting with the most simply organized one.


\subsection{Networks of citations of scientific papers}\label{ss-citations}

\begin{figure}
\epsfxsize=85mm
\epsffile{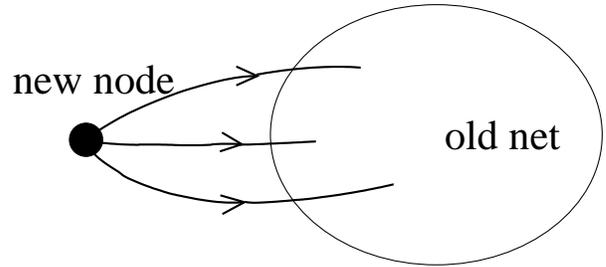}
\caption{
Scheme of the growth of citation networks. 
Each new paper contains references to previously published articles or books. 
It is assumed that old papers are not updated, so new connections between old vertices are impossible.  
}
\label{f1}
\end{figure}

The vertices of these networks are scientific papers, the directed edges are citations. 
The growth process of the citation networks is very simple (see Fig. \ref{f1}). 
Almost each new article contains a nonzero number of references to old ones. This is the only way to create new edges. The appearance of 
new connections between old vertices is impossible 
(one may think that old papers are not updated). The number of references to some paper is the in-degree of the corresponding vertex of the 
network. 


The average number of references in a paper is of the order of $10^1$, so such  networks are sparse. In Ref. \cite{r98}, the data from an ISI 
database for the period 1981 -- June 1997 and citations from Phys. Rev. D {\bf 11-50} (1975-1994) were used to find the distributions of the 
number of citations, i.e., the in-degree distributions. The first network consists of $783\,339$ nodes and $6\,716\,198$ links, the maximum 
number of citations is $\overline{k}_{i}^{(max)}=8\,904$. The second network contains $24\,296$ nodes connected by $351\,872$ links, and its 
maximum in-degree equals $2\,026$. The out-degree is rather small, so the degree distribution coincides with the in-degree one in the range 
of large degree. 

Unfortunately, the sizes of these networks are not sufficiently large to find a conclusive functional form of the distributions. In Ref. 
\cite{r98}, both distributions were fitted by the $k_{i}^{-3}$ dependence. The fitting by the dependence $(k_{i}+const)^{-\gamma}$ was 
proposed in Ref. \cite{ta00}. The exponents were estimated as $\gamma=2.9$ for the ISI net and $\gamma=2.6$ for the Phys. Rev. D citations. 
Furthermore, in Ref. \cite{v01a}, the large in-degree part of the in-degree distribution obtained for the Phys. Rev. D citation graph was 
fitted by a power law with the exponent $\gamma = 1.9 \pm 0.2$. It was found in the same paper that the average number of references per 
paper increases as the citation graphs grow. The out-degree distributions (the distribution of number of references in papers) show 
exponential tails. The factor in the exponential depends on whether or not journals restrict the maximal number of pages in their papers.    

It is possible to estimate roughly the values of the exponent knowing the size $N$ of the network and the cut-off $k_{cut}$ of the 
distribution, $\gamma \approx 1+\ln N/\ln k_{cut}$ (see Sec. \ref{ss-relations}). Using the maximal number of citations as the cut-offs, 
the authors of the papers \cite{krl00,kr00c} got the estimations $\gamma=2.5$ for the ISI net and 
$\gamma=2.3$ for Phys. Rev. D. Moreover, they indicated from similar estimation that these data are also consistent with the 
$k_{i}^{-y}\exp[-const\,k_{i}^{1-y}]$ form of the distribution if one sets $y=0.9$ for the ISI net and $y=0.7$ for Phys. Rev. D.    

In Ref. \cite{ls98}, the very tail of a different distribution was studied. The ranking dependence of the number of citations to the 
$1\,120$ most cited physicists was described by a stretched exponential function. Of course, the statistics of citations collected by 
authors necessarily differ from the statistics of the citations to papers. Also, the form of the tail of the distribution should be quite 
different from its main part.  


In Ref. \cite{jnb01a}, the process of receiving of citations by papers in a growing citation network was empirically studied. $1\,736$ 
papers published in Physical Review Letters in 1988 were considered, and the dynamics of receiving 
$83\,252$ citations was analysed. It was demonstrated that new citations (incoming edges) are distributed among papers (vertices) with 
probability proportional to degree of vertices. This indicates that linear preferential attachment mechanism operates in this citation graph.  



\subsection{Networks of collaborations}\label{ss-collaborations}

\begin{figure}
\epsfxsize=85mm
\epsffile{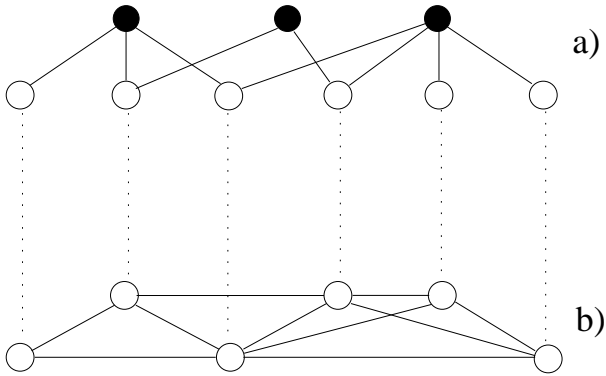}
\caption{
A bipartite collaboration graph (a) and one of its one-mode projections (b) \protect\cite{nsw00}. 
Collaborators are denoted by empty circles, the filled circles depict acts of collaboration.   
}
\label{f2}
\end{figure}

The set of collaborations can be represented by the {\em bipartite graph} containing two distinct types of vertices --- collaborators 
and acts of collaborations (see Fig. \ref{f2},a) \cite{nsw00}. Collaborators connect together through collaboration acts, so in this type of 
a bipartite graph, direct connections between vertices of the same kind are absent. Edges are undirected. 
For instance, in the scientific collaboration bipartite graphs, one kind of vertices corresponds to authors and the other one is scientific 
papers \cite{n00,n00f}. In movie actor graphs, these two kinds of vertices are  
actors and films, respectively \cite{ws98,asbs00,ab00a}. 


Usually, instead of such bipartite graphs, their far less informative one-mode projections are used (for the projection procedure, see 
Fig. \ref{f2},b). In particular, one can directly connect vertices-collaborators without indicating acts of collaboration. Note that the 
clustering coefficients of such one-mode projections are large because each act of collaboration simultaneously creates a number of 
highly connected nearest neighbors. 

Note that, in principle, it is possible to introduce multiple edges if there were several acts of collaboration between the same 
collaborators. Also, one can consider weighted edges accounting for reduction of the ``effect'' of collaboration between a pair of 
collaborators when several participants are simultaneously involved \cite{n00f}. We do not consider these possibilities here. 

Collaboration networks are well documented. For example, in Refs. \cite{ws98,watbook99}, the movie actor one-mode graph consisting 
of $225\,226$ actors is considered. 
The average degree is $\overline{k}=61$, the average shortest path equals $3.65$ that is close to the corresponding value $3.00$ for 
the classical random graph with the same $\overline{k}$. 
The clustering coefficient is large, $C=0.79$ (for the corresponding classical random graph it should be $0.00027$). Note that in 
Ref. \cite{nsw00}, another value, $C=0.199$, for the clustering coefficient of a movie actor graph is given. 

The distribution of the degree of vertices (number of collaborators) in the movie actor network 
($N=212\,250$ and $\overline{k}=28.78$)
was observed to be of a power-law form with the exponent $\gamma=2.3$ 
\cite{ba99}.  
In Ref. \cite{ab00a}, the degree distribution was fitted by the 
$(k+const)^{-\gamma}$ dependence with the exponent $\gamma=3.1$. 
Notice that, in Refs. \cite{ba99} and \cite{ab00a}, TV series were 
excluded from the dataset. The reason for this is that  
each series is considered in the database as a single movie with, sometimes, thousands of actors. In Ref. \cite{asbs00} the full dataset, 
including series, was used, which has yielded exponential form of the degree distribution (for statistical analysis, a cumulative degree 
distribution was used).       

Similar graphs for members of the boards of directors of the Fortune $1\,000$ companies, for authors of several huge electronic archives, 
etc. were also studied \cite{n00,n00f,nsw00}. Distributions of numbers of co-directors, of collaborators that a scientist has, etc. were 
considered in Ref. \cite{nsw00}. Distributions display a rather wide variance of forms, and it is usually hardly possible to observe 
a pure power-law dependence. 

One can find data on structure of large scientific collaboration networks in Refs. \cite{n00,n00f}. The largest one of them, MEDLINE, 
contains $1\,520\,254$ authors with $18.1$ collaborations per author. 
The clustering coefficient equals $0.066$. 
The giant connected component covers $93\%$ of the network. The size of the second largest component equals 
$49$, i.e., is of the order of $\ln N$. The average shortest path is equal to $4.6$ that is close to the corresponding classical random 
graph with the same average degree. The maximal shortest path is several times higher than the average shortest one and equals $24$. 
These data are rather typical for such networks. 

Mathematical (M) ($70\,975$ different authors and $70\,901$ published paper) and neuro-science (NS) ($209\,293$ authors with $3\,534\,724$ 
connections and $210\,750$ papers) journals issued in the period 1991-1998 were scanned in Refs. \cite{bjnr01a,jnb01a}. Degree distributions 
of these collaborating networks were fitted by power laws with exponents $2.4$ (M) and $2.1$ (NS). 
What is important, it was found that the mean 
degrees of these networks were not constant but grew linearly as the numbers of their vertices increased. Hence, the networks became more 
dense. The average shortest-path lengths in these graphs and their clustering coefficients decrease with time. 

New edges were found to be preferentially attached to vertices with the high number of connections. The probability that a new vertex 
is attached to a vertex with a degree $k$ was proportional to $k^y$ with the $y$ exponent equal to $0.8\pm0.1$, so that some deviations from a linear dependence were noticeable. 
However, new edges emerged between the pairs of already existing vertices with the rate proportional to the product of the degrees of vertices in a pair. 

Very similar results were also obtained for the actor collaboration graph consisting of $392\,340$ vertices and $33\,646\,882$ edges \cite{jnb01a}. 

In Ref. \cite{n01b}, the preferential attachment process within collaboration nets of the Medline database (1994-1999: $1\,648\,660$ 
distinct names) and the Los-Alamos E-print Archive (1995-2000: $58\,342$ distinct names) was studied. 
In fact, a relative probability that an edge added at time $t$ connects to a vertex of degree $k$ was measured. 
This probability was observed to be a linear function of $k$ until large enough degrees, so that a linear preferential attachment mechanism 
operates in such networks (compare with Ref. \cite{bjnr01a}). However, the empirical dependence saturated for $k \gtrsim 150$ in the Los-Alamos 
E-print Archive collaboration net or even fell off for $k \gtrsim 600$ in the Medline network.


\subsection{Communications networks, the WWW, and the Internet}\label{ss-communications}

\begin{figure}
\epsfxsize=85mm
\epsffile{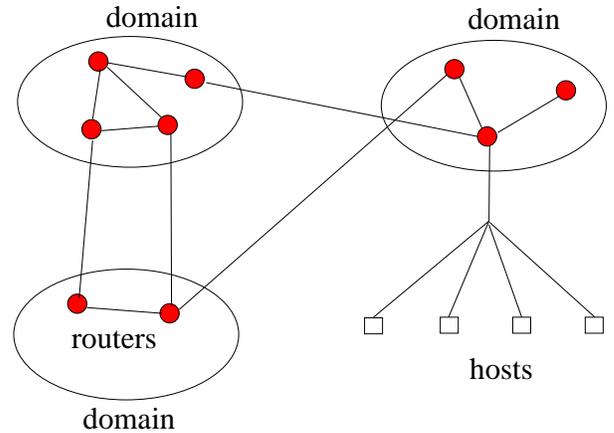}
\caption{
Naive scheme of the structure of the Internet \protect\cite{fff99}.
}
\label{f3}
\end{figure}

Roughly speaking, the Internet is a net of interconnected vertices: hosts (computers of users), servers (computers or programs providing 
a network service that also may be hosts), and routers that arrange traffic across the Internet, see Fig. \ref{f3}. Connections are undirected, 
and traffic (including its direction) changes all the time. Routers are united in domains. In January of 2001, the Internet contained already 
about $100$ millions hosts. However, it is not the hosts that determine the structure of the Internet, but rather, routers and domains. In July of 2000, there were about $150\,000$ routers in the Internet \cite{gt00c}. Latter, the number rose to $228\,265$ 
(data from Ref. \cite{yjb01a}).
Thus, one can consider the topology of the Internet on a router level or inter-domain topology \cite{fff99}. In the latter case, it is actually a small network.


The World Wide Web is the array of its documents plus hyper-links -- mutual references in these documents. 
Although hyper-links are directed, pairs of counter-links, in principle, may produce undirected connections. Web documents are accessible 
{\em through} the Internet (wires and hardware), and this determines the relation between the 
Internet and 
the WWW.


\subsubsection{Structure of the Internet}\label{sss-internet}

On the inter-domain level, the Internet 
is a really small sparse network with the following basic characteristics \cite{fff99}. 
In November of 1997, it consisted of $3\,015$ vertices and $5\,156$ edges, so the average degree was $3.42$, the maximal degree of a vertex 
equaled $590$. In April of 1998, there were $3\,530$ vertices and $6\,432$ edges, the average degree was $3.65$, the highest degree was $745$. 
In December of 1998 there were $4\,389$ vertices and $8\,256$ edges, so the average degree was $3.76$ and the maximal degree equaled $979$. 
The average shortest path is found to be about $4$ as it should be for the corresponding classical random graph, the maximal shortest path is 
about $10$.  


The degree distribution of this network was reported to be of a power-law form, $P(k) \propto k^{-\gamma}$ where $\gamma \approx 2.2$ 
(November of 1997 -- 2.15, April of 1998 -- 2.16, and December of 1998 -- 2.20) \cite{fff99}. In fact, it is hard to achieve this precision for 
a network of such a size.  
One may estimate the value of the exponent using the highest degrees (see Eq. (\ref{11-5}) in Sec. \ref{ss-relations}). Such estimations 
confirm the reported values. For November of 1977, one gets 
$\gamma \approx 1 + \ln 3015/\ln 590 = 2.22$, for April of 1998 -- $\gamma \approx 2.24$, and for December of 1998 -- $\gamma \approx 2.26$.  
One should note that, in paper \cite{fff99}, the dependence 
of a node degree on its rank, $k(r)$, was also studied. A power law (Zipf law) was observed, $k(r) \propto r^{-\zeta}$, but, as one can check, 
the reported values of the $\zeta$ exponent are inconsistent with the corresponding ones of $\gamma$. 

On the router level, according to relatively poor data from 1995 
\cite{fff99,pg98}, 
the Internet consisted of 
3888 vertices and 5012 edges, with the average degree equal to 2.57 and the maximal degree equal to 39. The degree distribution of this network was fitted by a power-law dependence with the exponent, $\gamma \approx 2.5$. Note that the estimation from the maximal degree value gives a quite different value, $\gamma \approx 1 + \ln 3888/\ln 39 = 3.3$, so that the empirical value of the $\gamma$ exponent is not very reliable. 

In 2000, the Internet has already consisted of about $150\,000$ routers connected by $200\,000$ links \cite{gt00c}. The degree distribution 
was found to ``lend some support to the conjecture that a power law governs the degree distribution of real networks'' \cite{gt00c}. If this is true, one can estimate from this degree distribution that its $\gamma$ exponent is about $2.3$.

In Ref. \cite{fff99}, the distribution of the eigenvalues of the adjacency matrix of the Internet graph was studied. The ranking plots for large eigenvalues $\lambda(r)$ was obtained (enumeration is started from the largest eigenvalue). For all three studied inter-domain graphs, 
approximately, $\lambda(r) \propto r^{-0.5}$. From this we get the form of the tail of the eigenvalue spectra, $G(\lambda) \propto \lambda^{-(1+1/0.5)} =  \lambda^{-3.0}$ 
(we used the relation between the exponent of the distribution and the ranking one that is discussed in Sec. \ref{ss-relations}). For the inter-router-95 graph, $\lambda(r) \propto r^{-0.2}$. Note that these dependences were observed for only the $20$ largest eigenvalues. 

More recent data on the structure of the Internet are collected by the National Laboratory for Applied Network research (NLANR). On its Web site 
http://moat.nlanr.net/, one can find extensive Internet routing related information being collected since November 1997. For nearly each day of this period, NLANR has a map of connections of operating ``autonomous systems'' (AS), which 
approximately map to Internet Service Providers. These maps (undirected networks) are closely related to the Internet graph on the inter-domain level. 

For example, on 14.11.1997, there were observed $3042$ AS numbers with $5595$ interconnections, the average degree was $\overline{k}=3.68$; on 09.11.1998, these values were $4\,301$, $8\,589$, and $3.99$, respectively; on 06.12.1999, were $6\,301$, $13\,485$, and $4.28$, but on 08.12.1999, there were only $768$ AS numbers and 
$1\,857$ interconnections (!), so $\overline{k}=4.84$. Hence, fluctuations in time are very strong.  

The statistical analysis of these data was made in Ref. \cite{pvv01a}.
The data were averaged, and for 1997 
the following average values were obtained. The mean degree of the network was equal to $3.47$, the clustering coefficient was $0.18$, and the average shortest-path length was $3.77$. For 1998, the corresponding values were $3.62$, $0.21$, and $3.76$ respectively. For 1999, they were  
$3.82$, $0.24$, and $3.72$ respectively. The average shortest-path lengths are close to the lengths for corresponding classical random graphs 
but the clustering coefficients are very large. Notice that the density of connections increases as the Internet grows. One may say, 
the Internet shows accelerated growth. In Ref. \cite{gkk01e}, the dependence of the total number of interconnections (and the average degree) 
on the number of AS was fitted by 
a power law. Unfortunately the variation ranges of these quantities are too small to reach any reliable conclusion.   

In Ref. \cite{pvv01a}, the following problem was considered. New edges can connect together pairs of new and old, or old and old vertices. 
Were do they emerge, between what particular vertices?  
The mean ratio of the number of new links emerging between new and old vertices and 
the number of new connections between already existing vertices was 
$0.34$, $0.48$, and $0.53$ in 1997, 1998, and 1999,  
respectively. 
Thus the Internet structure is very distinct from citation graphs.  

The degree distributions for each of these three years were found to follow a power law form with the exponent $\gamma \approx 2.2$, which 
is in agreement with Ref. \cite{fff99}. Furthermore, in Ref. \cite{pvv01a}, from the data of 1998, the dependence of the average degree of 
the nearest neighbors of a vertex on its degree, $\overline{k}_{nn}(k)$ was obtained. This slowly decreasing function was approximately 
fitted by a power law with the exponent $0.5$. Such a dependence indicates strong correlations in the distribution of connections over 
the network. Vertices of large degree usually have weakly connected nearest neighbors, and vice versa. 

Notice that the measurement of the average degree of the nearest neighbors of a vertex vs. its degree is an effective way to measure 
correlations between degrees of separate vertices. As explained above, direct measurement of the joint distribution $P(k_1,k_2)$ is 
difficult because of inevitably poor statistics. 

In principle, the behavior observed in Ref. \cite{pvv01a} is typical for citation graphs growing under mechanism of preferential linking (see Sec. \ref{ss-distributionoflinks}). However, as indicated above, most of connections in the Internet emerge between 
already existing sites. If the process of attachment of these edges is preferential, strongly connected sites usually have strongly connected nearest neighbors, unlike what was observed in Ref. \cite{pvv01a} (see Sec. \ref{ss-distributionoflinks}). 
A difficulty is that vertices in the Internet are at least of two distinct kinds. 
In Ref. \cite{pvv01a}, the difference between ``stub'' and ``transit domains'' of the Internet is noticed. 
Stub domains have no connections between them and connect to transit domains, which are, contrastingly, well interconnected.
Therefore, new connections or rewirings are possible not between all vertices. 
This may be reason of the observed correlations. 
A different classification of the Internet sites was used in Ref. \cite{ccmp01}. The vertices of the Internet were separated into two groups, namely ``users'' and ``providers''. Interaction between these two kinds of sites leads to the self-organization of the growing network 
into a scale-free structure.   

The process of the attachment of new edges in these maps of Internet was empirically studied in Ref. \cite{jnb01a}. It was found that 
the probability that a new edge is attached to a vertex is a linear function of the vertex degree. 

A very important feature of the Internet, both on the AS (or the inter-domain) level and on the level of routers, is that its vertices are physically attached to specific places in the world and have their fixed geographic coordinates. The geographic places of vertices and the 
distribution of Euclidean distances are essential for the resulting structure of the Internet. This factor was studied and modeled in a recent paper \cite{yjb01a}. 
It was observed that routers and AS correlate with the population density. All three sets -- population, router, and AS space densities -- form fractal structures in space. The fractal dimensions of these fractals were found to be approximately $1.5$ (the data for North America).   
Maps of AS and the map of $228\,265$ routers were analysed. In particular,   
the average shortest distance between two routers was found to be approximately $9$ \cite{yjb01a}. 

In Ref. \cite{cmp00}, the structure of the Internet was considered using an analogy with river networks. In such an approach, a particular terminal is treated as the outlet of a river basin. The paths from this terminal to all other addresses form the structure of this basin. 
As for usual river networks, the probability, $P(n)$, that a (randomly chosen) point connects $n$ other points uphill, can be introduced. 
In fact, $n$ is the size of the basin connected to some point, and $P(n)$ is the distribution of basin sizes. 
For river networks forming a fractal structure \cite{mbook83}, this distribution is of a power-law form, $P(n) \propto n^{-\tau}$, where 
values of the $\tau$ exponent are slightly lower $3/2$ \cite{hbook89}. For the Internet, it was found that $\tau = 1.9 \pm 0.1$ \cite{cmp00}.


\subsubsection{Structure of the WWW}\label{sss-www}

\begin{figure}
\epsfxsize=85mm
\epsffile{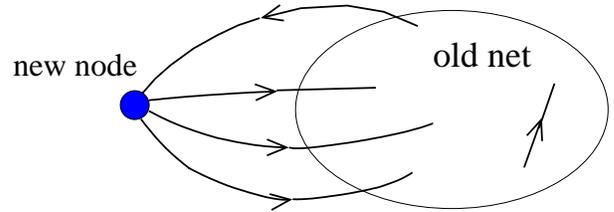}
\caption{
Scheme of the growth of the WWW (compare with Fig. \protect\ref{f1}). 
A new document (page) must have at least one incoming hyper-link to be accessible. Usually it has several references to existing documents of the Web but, in principle, these references may be absent. Old pages can be updated, so new hyper-links can appear between them.  
}
\label{f4}
\end{figure}

Let us first discuss, how the Web grows, that is, how new pages appear in it 
(see Fig. \ref{f4}). 
Here we describe only two simple ways to add a new document. 

(i) Suppose, you want create your own personal home page. First you prepare it, put references to some pages of the Web (usually several references but, in principle, the references may be absent), etc. But this is only the first step. You have to make it accessible in the Web, 
to launch it. You come to your system administrator, he puts a reference to it (usually one reference) in the home page of your institution, and that is more or less all -- your page is in the World Wide Web. 

(ii) There is another way of having new documents appear in the Web. Imagine that you already have your personal home page and want to launch a new document. The process is even simpler than the one described above. You simply insert at least one reference to the document into your page, and that is enough for the document to be included in the World Wide Web. 
We should note also that old documents can be updated, so new hyper-links between them can appear. Thus, the WWW growth is much more complex process than the growth of citation networks.  


The structure of the WWW was studied experimentally in Refs. \cite{ajb99,ha99,huppl98,krrt99,kkrrt99} and the power-law form of various distributions was reported. These studies cover different sub-graphs of the Web and even relate to its different levels. 
The global structure of the entire Web was described in the recent paper \cite{bkm00}. 
In this study, the crawl from Altavista is used. The most important results are the following. 

In May of 1999, from the point of view of Altavista, the Web consisted of  $203\times 10^6$ vertices (URLs, i.e., pages) and $1466\times 10^6$ hyper-links. The average in- and out-degree were $\overline{k}_{i} = \overline{k}_{o} = 7.22$.
In October of 1999 there were already $271\times 10^6$ vertices and $2130\times 10^6$ hyper-links. The average in- and out-degree were $\overline{k}_{i} = \overline{k}_{o} = 7.85$.  
This means that during this period, $68\times 10^6$ pages and $664\times 10^6$ hyper-links were added, that is, $9.8$ extra hyper-links appeared per one additional page. Therefore, the number of hyper-links grows faster than the number of vertices. 

The in- and out-degree distributions are found to be of a power-law form 
with the exponents  
$\gamma_{i}=2.1$ and $\gamma_{o}=2.7$ that confirms earlier data of Albert et al \cite{ajb99} on the nd.edu subset of the WWW ($325\,000$ pages). 
These distributions were also fitted by the dependences 
$(k+c_{i,o})^{-\gamma_{i,o}}$  with some constants $c_{i,o}$ \cite{nsw00}. 
For the in-degree distribution, the fitting provides 
$c_{i}=1.25$ and $\gamma_{i}=2.10$, and for the out-degree distribution,  
$c_{o}=6.94$ and $\gamma_{o}=2.82$. Note that the fit is only for nonzero in-,out-degrees 
$k_{i},k_{o}$. The probabilities $P(k_{i}=0)$ and $P(k_{o}=0)$ were not measured experimentally. The relation between them can be found by employing Eq. (\ref{0-2}). 

\begin{figure}
\epsfxsize=88mm
\epsffile{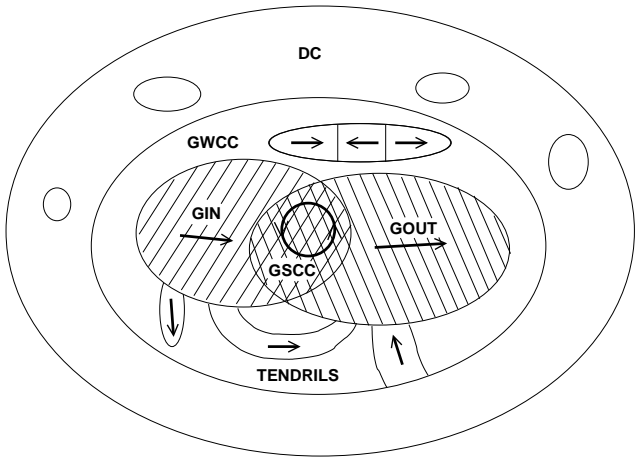}
\caption{
Structure of a directed graph 
when the giant strongly connected component is present \protect\cite{dms01a} (see the text). Also, the structure of the WWW (compare with Fig. 9 of Ref. \protect\cite{bkm00}). 
\ \ \ 
If one ignores the directedness of edges, the network consists of the 
{\em giant weakly connected component} (GWCC) --- actually, the usual percolating cluster --- and disconnected components (DC). 
\ \ \ 
Accounting for the directedness of edges, the GWCC contains the following components: 
\ \ \ 
(a) the {\em giant strongly connected component} (GSCC), that is, the set of vertices reachable from its every vertex by a directed path; 
\ \ \ 
(b) the {\em giant out-component} (GOUT), the set of vertices approachable from the GSCC by a directed path (includes the GSCC); 
\ \ \ 
(c) the {\em giant in-component} (GIN), contains all vertices from which the GSCC is approachable (includes the GSCC); 
\ \ \ 
(d) the {\em tendrils} (TE), the rest of the GSCC, i.e. the vertices which have no access to the GSCC and are not reachable from it. In particular, this part includes something like ``tendrils'' \protect\cite{bkm00} but also there are ``tubes'' and numerous clusters which are only ``weakly'' connected. 
\ \ \ 
Note that our definitions of the GIN and GOUT differ from the definitions of Refs. \protect\cite{bkm00,nsw00}: the GSCC is included into both GIN and GOUT, so the GSCC is the interception of the GIN and GOUT. We shall show in Sec. \protect\ref{ss-directed} that this definition is natural.
}
\label{f5}
\end{figure}

The relative sizes of giant components yield a basic information about the global topology of a directed network, 
and, in particular, about the WWW. 
Let us assume that a large directed graph has both the {\em giant weakly connected component} (GWCC) and the {\em giant strongly connected component} (GSCC) (see Sec. \ref{ss-size}). 
Then its general global structure can be represented in the following form 
(see Fig. \ref{f5}) \cite{bkm00,dms01a}. 





At first, it is possible to extract the GWCC. The rest of the network consists of disconnected clusters -- ``disconnected components''(DC). The GWCC consists of: 

(a) the GSCC -- from each vertex of the GSCC, there exists a directed path to any other its vertex; 

(b) the {\em giant out-component} (GOUT) -- the vertices which are reachable from the GSCC by a directed path, so that GOUT includes GSCC; 

(c) the {\em giant in-component} (GIN) -- 
the vertices from which one can reach the GSCC by a directed path so that GIN includes GSCC; 

 


(d) the {\em tendrils} (T) -- the rest of the GWCC. 
This part consists of the vertices which have no access to the GSCC 
and are not reachable from it. In particular, it includes indeed 
something like ``tendrils'' 
but also there are ``tubes'' 
and numerous clusters which are only weakly connected. 

Notice that, in contrast to Refs. \cite{bkm00,nsw00}, the above defined GIN and GOUT include GSCC. In Sec. \ref{ss-directed} we shall show that this definition is natural. 

One can write 
$$
\mbox{Network} = \mbox{GWCC} + \mbox{DC}
$$ 
and
$$
\mbox{GWCC} = \mbox{GIN} + \mbox{GOUT} - \mbox{GSCC} + \mbox{TE} 
\, .
$$

According to Ref. \cite{bkm00}, in May of 1999, the entire Web, containing
$203\times 10^6$ pages, consisted of 
\\
--- the GWCC, $186\times 10^6$ pages ($91\%$ of the total number of pages), 
and 
\\ 
--- the DC, $17\times 10^6$ pages. 

In turn, the GWCC included: 
\\
--- the GSCC, $56\times 10^6$ pages, 
\\
--- the GIN, $99\times 10^6$ pages, 
\\ 
--- the GOUT, $99\times 10^6$ pages,  
and 
\\ 
--- the TE, $44\times 10^6$ pages. 

Both distributions of the sizes of strongly connected components and of the sizes of weakly connected ones were fitted by power-law dependences with exponents approximately $2.5$. 

The probability that a directed path is present between two random vertices was estimated as $24\%$. 
For pairs of pages of the WWW between which directed paths exist, 
the {\em average shortest-directed-path length} equals $16$. 
For pairs between which at least one undirected path exists, the {\em average shortest-undirected-path length} equals $7$. 

The value of the average shortest-directed-path length estimated from data extracted from the nd.edu subset of the WWW was $19$ \cite{ajb99}. 
This first published value for the ``diameter'' of the Web was obtained in a non-trivial way (it is not so easy to find the shortest path in 
large networks). 
(i) The in-degree and out-degree distributions were measured in the nd.edu domain. 
(ii) A set of small model networks of different sizes $N$ with these in-degree distribution and out-degree distribution was constructed. 
(iii) For each of these networks, the average shortest-path length $\overline{\ell}$ was found. Its size dependence was estimated as 
$\overline{\ell}(N) \approx 0.35+2.06\lg N$. 
(iv) $\overline{\ell}(N)$ was extrapolated to $N=800\,000\,000$, that is, the estimation of the size of the WWW in 1999. 
The result, i.e. $\overline{\ell}(800\,000\,000)\approx19$, is very close to the above cited value $\overline{\ell}(200\,000\,000)=16$ of 
Ref. \cite{bkm00} if one accounts for the difference of sizes. 

 The maximal shortest path between nodes belonging to the GSCC equals $28$. The maximal shortest directed path for nodes of the WWW between 
which a directed path exists is greater than 500 (some estimates indicate that it may be even $1000$).

Although the GSCC of the WWW is rather small, most pages of the WWW belong to the GWCC. Furthermore, even if all links {\em to} pages 
with in-degree larger than $2$ are removed, the GWCC does not disappear. This is clearly demonstrated by the data of Ref. \cite{bkm00}: 

The size of the GWCC of the Web (visible by Altavista in May 1999) 
is $186\times 10^6$ pages.    

If all in-links to pages with 
$k_{i} \geq k_{i}^{(max)} = 1000, 100, 10,5,4,$ and $3$ are removed, 
the size of the retaining GWCC is $177\times 10^6, 167\times 10^6, 
105\times 10^6, 59\times 10^6, 41\times 10^6,$ and $15\times 10^6$ 
pages, respectively. 
 

The Web grows much faster than the possibilities of hardware. 
Even the best search engines index less than one half of all pages of the Web \cite{lg98,lg99,bp98,b00}. Update of files cached by them for 
quick search usually takes many months. 
The only way to improve the situation is indexing of special areas of the WWW, ``cyber-communities'', to provide possibility of an efficient 
specialized search \cite{k99d,lg98c,lg99b,l00a,krrt99,kkrrt99,k98e,gkr98,krrt99b,dh99}. 


\begin{figure}
\epsfxsize=85mm
\epsffile{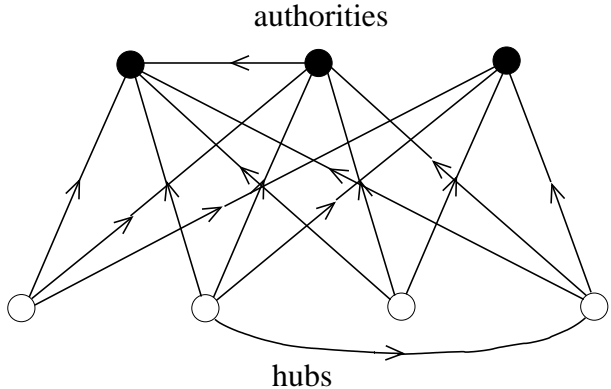}
\caption{
A bipartite directed sub-graph in the Web being used for indexing cyber-communities \protect\cite{krrt99,kkrrt99}.
}
\label{f6}
\end{figure}


Natural objects for such indexing are specific bipartite sub-graphs 
(see Fig. \ref{f6}) \cite{krrt99,kkrrt99}. 
One should note that the directed graphs of this kind have a different structure than the bipartite graphs described in Sec. 
\ref{ss-collaborations}. 
After separation from the other part of a network, they consist of only two kinds of nodes -- ``hubs'' (fans) and ``authorities'' 
(idols). Each hub connects to {\em all} the authorities of this graph.  
Let it be $h$ hubs and $a$ authorities in the bipartite graph. Each of hubs, by definition, must have $a$ links directed to each of 
$a$ authorities. Hence, the number of links between subsets of hubs and authorities equals $ha$. Some extra number of connections may 
be inside of these two subsets. 

The distribution of the number of such bipartite sub-graphs in the Web, $N_b(h,a)$ was studied in Refs. \cite{krrt99,kkrrt99}. For 
a fixed number of hubs, $N_b(h=\mbox{fixed},a)$ resembles a power-law dependence, and for a fixed number authorities, $N_b(h,a=\mbox{fixed})$ 
resembles an exponential one when $h$ is small. We should note that these data are 
poor. 

One can also consider the structure of the Web on another level. In particular, in Ref. \cite{ah00a}, the in-degree distribution for the 
domain level of Web in spring of 1997 was studied, where each vertex (Web site) is a separate domain name, and the value $1.94$ for the 
corresponding exponent was reported. The network consisted of $259\,794$ vertices. 

Measurements of the clustering coefficient of the Web on this level \cite{a99c} have shown that it is much larger than it should be for 
the corresponding classical random graph. 
The data were extracted from the same crawl containing $259\,794$ sites. 

Several other empirical distributions were obtained, which do not relate directly to the 
global structure of the Web but indicate some of its properties.
Huberman and Adamic \cite{ha99} found that the distribution of the number of pages in a Web site also demonstrates a power-law dependence (Web site is a set of linked pages on a Web server). 
From their analysis of sets of $259\,794$ and $525\,882$ Web sites covered by Alexa and Infoseek it follows that the exponent in this power law is about $1.8$. 
Note that the power-law dependence seems not very pronounced in this case.    
A power-law dependence was indicated at the distribution of the number of visits (connections) to the Web sites \cite{a00g}. The value of the corresponding exponent was estimated as $2.0$. The fit is rather poor.

One should stress that usually what experimentalists indicate as a power-law dependence is actually a linear fit for a rather narrow 
range on a log-log plot. It is nearly impossible to obtain some functional form for the degree distribution directly because of strong 
fluctuations. To avoid them, the cumulative distribution 
$P_{cum}(k) = \int_k^\infty dk\,P(k)$ is usually used \cite{asbs00}. Nevertheless, the restricted sizes of the studied networks often 
lead to implausible interpretation (see the discussion of the finite size effects in Secs. \ref{ss-simplestscale-free} and 
\ref{ss-relations}). One has to keep this in mind while working with such experimental data.


\subsection{Biological networks}\label{ss-biological}


\subsubsection{Structure of neural networks}\label{sss-neural}

Let us consider the rich structure of a neural network of a tiny organism, classical {\em C. 
elegans}. $282$ neurons form the network of directed links with average degree $\overline{k}=14$ \cite{ws98,watbook99}. 
The in- and out-degree distributions are exponential. The average shortest-path length measured without account of directness of 
edges is $2.65$, and the clustering coefficient equals $0.26$. Therefore, the network displays the small-world effect, and the 
clustering coefficient is much larger than the characteristic value for the corresponding classical random graph, 
$C=0.26 \gg 14/282 \sim 0.05$.


\subsubsection{Networks of metabolic reactions}\label{sss-metabolic}

The valuable example of a biological network with the extremely rich topological structure is provided by the network of metabolic reactions \cite{k69,bl97,kbook93}. This is a particular case of chemical reactions graphs 
\cite{kbook93,kbook95,kbook00}. At present, such networks are documented for several organisms. Their vertices are substrates -- molecular compounds, and the edges are metabolic reactions connecting substrates. 
According to \cite{jtaob00} (see also \cite{jbto01b}), incoming links for a particular substrate are reactions in which it participates as a product. Outgoing links are reactions in which it is an educt. 

Sizes of such networks in 43 organisms investigated in \cite{jtaob00} are between $200$ and $800$. 
The average shortest-path length is about $3$, $\overline{k}_{i} \sim \overline{k}_{o} \sim 2.5 - 4.0$. Although the networks are very small, the in- and out-degree distributions were interpreted \cite{jtaob00} as scale-free, i.e., of a power-law form with the exponents, $\gamma_{i} \approx \gamma_{o} \approx 2.2$. 

In Ref. \cite{wf00}, one may find another study of the global structure of metabolic reaction networks. 
The networks were treated as undirected. For a network of the {\em Escherichia coli}, consisting of $282$ nodes, the average degree $\overline{k} \sim 7$. The average shortest-path length was found to be equal to $2.9$. The clustering coefficient is $C \approx 0.3$, that is, much larger than for the corresponding classical random network, $7/282 \approx 0.025$. 

The distribution of short cycles in large metabolic networks is considered in Ref. \cite{gswf00}.


\subsubsection{Protein networks}\label{sss-protein} 

A genomic regulatory system can be thought of as an extremely large 
directed network \cite{kbook93}. Vertices in this network are distinct components of the genomic regulatory system, and each directed edge points from the regulating to the regulated component. 


A very important aspect of gene function is protein-protein interactions -- ``the number and identity of proteins with which 
the products of duplicate genes in an organism interact'' (see Ref. \cite{w01d} for a brief introduction in the topic). 
The vertices of the protein-protein interaction network are proteins and the directed edges are, usually, pairwise protein-protein 
interactions. Two vertices may be connected by a pair of opposing edges, and the network also contains unit loops, so that its 
general structure resembles the structure of a food web (see Fig. \ref{f6o}). Recently large maps of protein-protein interaction 
networks were obtained \cite{ugc00b,itm00,ico01} which may be used for structural analysis. 





In Ref. \cite{jmbo01a} (for details see Ref. \cite{jbto01b}), the distribution of connections in the protein-protein interaction 
network of the yeast, {\em S. cerevisiae} was studied using the map from Ref. \cite{ugc00b} (see also Ref. \cite{w01d}). 
The network contains $1\,870$ vertices and $2\,240$ edges. The degree distribution was interpreted as a power-law (scale-free) 
dependence with an exponential cut-off at the point 
$k_c \approx 20$. This value is so small that it is difficult to find the exponent of the degree distribution. The approximate 
value $\gamma \approx 2.5$ was obtained in Ref. \cite{w01d}.

In addition, in Ref. \cite{jmbo01a}, the tolerance of this network against random errors (random deletion of proteins) and its 
fragility against the removal of the most connected vertices were studied. The random errors were found to be rather non-dangerous, 
but single deletion of one of the most connected proteins (having more than $15$ links) was lethal with high probability.


\subsubsection{Ecological and food webs}\label{sss-ecological} 

Food webs of species-rich ecosystems are directed networks, where vertices are distinct species, and directed edges connect pairs --- a specie-eater and its food 
\cite{cbnbook90,wm00,ms00,sm00,cga01,cga01b,lbmv01}. 
In Refs. \cite{ms00,sm00}, structures of three food webs were studied ignoring the directedness of their edges.  

The networks considered in Refs. \cite{ms00,sm00} are {\em very} small. 

(i) The food web of Ythan estuary consists of $N=93$ vertices. 
The average degree is $\overline{k}=8.70$, 
the average clustering coefficient is equal to $C=0.22$, the average shortest-path length is $\overline{\ell} = 2.43$.

(ii) Silwood park web (more precisely speaking, this is a sub-web). 
$N=154$, $\overline{k}=4.75$, $C=0.15$, $\overline{\ell} = 3.40$. 

(iii) The food web of Little Rock lake. $N=182$, $\overline{k}=26.05$, 
$C=0.35$, $\overline{\ell} = 2.22$. 

The clustering coefficients obtained for these networks essentially exceed the corresponding values for the classical random graphs with the same total number of vertices and edges. However, the measured average shortest-path lengths of these webs do not deviate noticeably from the corresponding values for the classical random graphs. 

Furthermore, the degree distributions of the first two webs were fitted by power laws with the exponents $\gamma\approx1.0$ and $\gamma\approx1.1$ for the Ythan estuary web and for the Silwood park web, correspondingly. This allowed authors of Refs. \cite{ms00,sm00} to consider them as scale-free networks 
(however, see Refs. \cite{cga01,cga01b} where the degree distributions in such food webs were interpreted as of an exponential-like form). 
These are the smallest networks for which a power-law distribution was ever reported.
For the third food web, any functional fitting turned to be impossible. 

Additionally, in Ref. \cite{sm00}, the stability of food webs against random or intentional removal of vertices was considered. The results were typical for scale-free networks (see Sec. \ref{ss-failure}). 


\begin{figure}
\epsfxsize=86mm
\epsffile{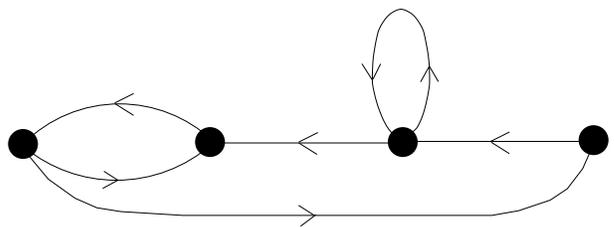}
\caption{
Typical food web. Cannibalism and mutual eating are widespread. 
}
\label{f6o}
\end{figure}


Food webs have a rather specific structure. They are directed, include unit loops, that is, cannibalism, and two opposing edges 
may connect a pair of vertices (mutual eating) \cite{wm00,wmbdb01} (see Fig. \ref{f6o}, compare with the structure of a protein-protein 
interaction network). 
Therefore, the maximal possible number of edges (trophic links) in a food web containing $N$ vertices (trophic species) is equal 
to $2\, N(N-1)/2+N=N^2$. 
Food webs are actually dense: the total number $L$ of edges is high. The values of the ration $L/N^2$ for seven typical food webs 
with $N=25-92$ were found to be in the range between $0.061$ and $0.32$ \cite{wm00}. Authors of Ref. \cite{wmbdb01} observed that 
this leads to an extreme smallness of food webs. Edges were treated as undirected and the average shortest-path lengths were then 
measured to be in the range 
between $1.44$ and $2.55$.  

We should emphasize that it is hard to find well defined and large food webs. This seriously hinders their statistical analysis.


\subsubsection{Word Web of human language}\label{sss-word_web}

Ferrer and Sol\'e (2001) \cite{fs01} constructed a net of fundamental importance, namely the network of distinct words of human language. 
Here we call it Word Web. 
The Word Web is constructed in the following way. The vertices of the web are the distinct words of language, and the undirected edges 
are connections between interacting words. It is not so easy to define the notion of word interaction in a unique way. Nevertheless, 
different reasonable definitions provide very similar structures of the Word Web. For instance, one can connect the nearest neighbors 
in sentences. Without going into details, this means that the edge between two distinct words of language exists if these words are the 
nearest neighbors in at least one sentence in the bank of language. In such a definition, multiple links are absent. One also may connect 
the second nearest neighbors and account for other types of correlations between words \cite{fs01}. In fact, the Word Web displays the 
cooccurrence of the words in sentences of a language. 

\begin{figure}
\epsfxsize=86mm
\epsffile{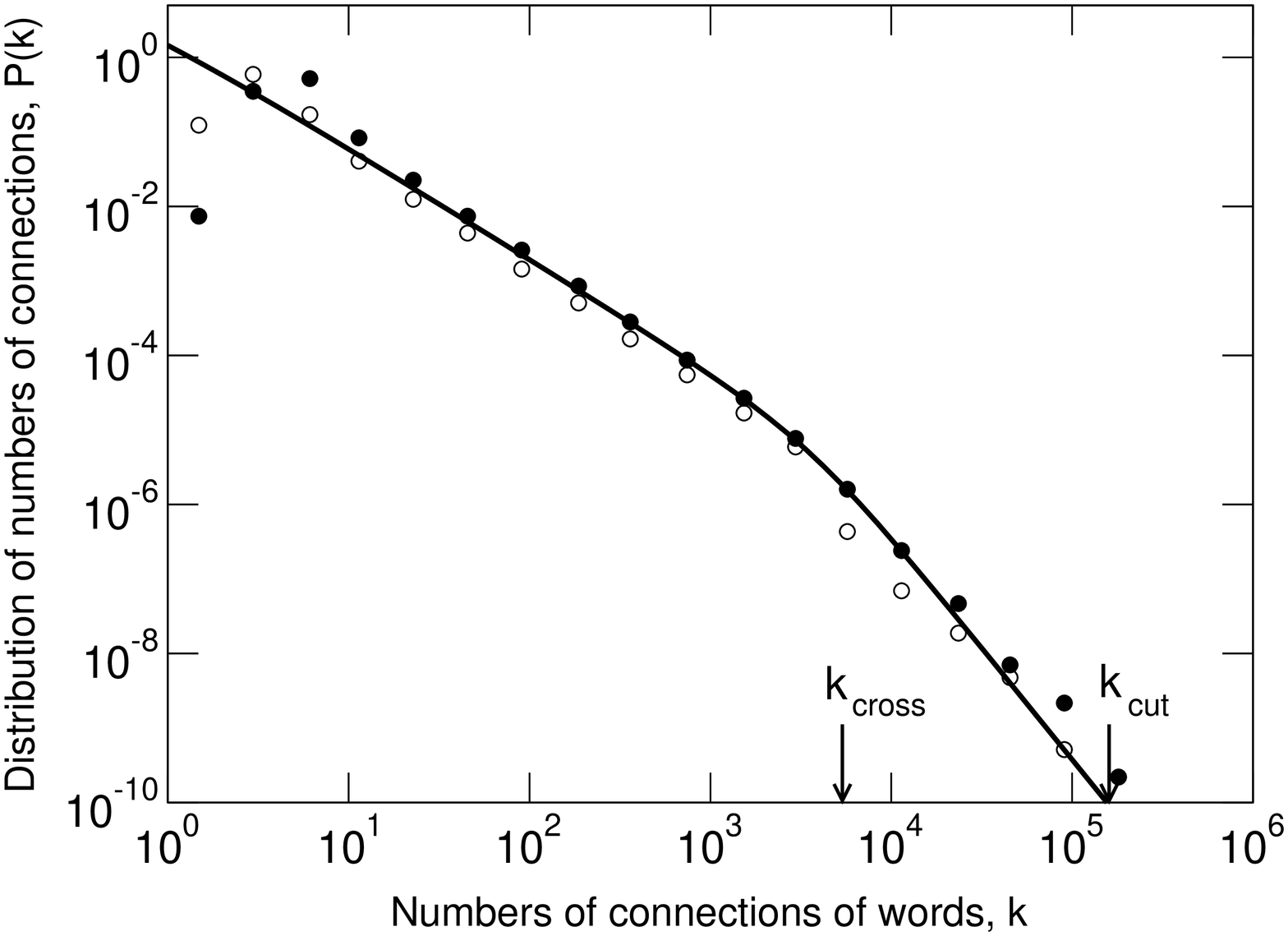}
\caption{
The distribution of the numbers of connections (degrees) of words in the word web in a log-log scale \protect\cite{fs01}. Empty and filled circles show the distributions of the number of connections obtained in Ref. \protect\cite{fs01} for two different methods of the construction of the Word Web. 
The solid line is the result of theory of Ref. \protect\cite{dm01d} (see Sec. \protect\ref{ss-accelerating} where the parameters of the Word Web, namely, the size $t \approx 470\,000$ and the average number of connections of a node, $\overline{k}(t) \approx 72$, were used. 
The arrows indicate the theoretically obtained point of crossover, $k_{cross}$ between the regions with different power laws, 
and the cutoff $k_{cut}$ due to the size effect. For a better comparison, the theoretical curve is displaced upward to exclude 
two experimental points with the smallest $k$ (note that the comparison is impossible in the region of the smallest $k$ where 
the empirical distribution essentially depends on the definition of the Word Web).  
}
\label{f6a}
\end{figure}

Two slightly different methods were used in Ref. \cite{fs01} to construct the Word Web. The two resulting webs obtained after processing 
$3/4$ million words of the British National Corpus (a collection of text samples of both spoken and written modern British English) 
have nearly the same degree distributions (see Fig. \ref{f6a}) and each contains about $470\,000$ vertices. 
The average number of connections of a word (the average degree) is $\overline{k} \approx 72$. As one sees from Fig. \ref{f6a}, 
the degree distribution comprises two distinct regions with quite different power-law dependences. The range of the degree variation 
is really large, so the result looks convincing. The exponent of the power law in the low-degree region is approximately $1.5$, and in the 
high-degree region is close to $3$ (the value $2.7$ was reported in Ref. \cite{fs01})\vspace{3pt}. 


The complex empirical degree distribution of the Word Web was described without fitting using a simple model of the evolution of human language \cite{dm01d} (see Fig. \ref{f6a} and Sec. \ref{ss-accelerating}).


\subsection{Electronic circuits}\label{ss-circuits} 

In Ref. \cite{fjs01}, the structure of large electronic circuits 
was analysed. Electronic circuits were viewed as undirected random graphs. 
Their vertices are electronic components (resistors, diodes, capacitors, etc. in analog circuits and logic gates in digital circuits) and the undirected edges are wires. 
The networks considered in Ref. \cite{fjs01} have sizes $N$ in the range between $20$ and $2\times10^4$ and the average degree between $3$ and $5$. 

For these circuits, the clustering coefficients, the average shortest-path lengths, and the degree distributions were obtained. In all the networks, the values of the average shortest-path length were close to those for the corresponding classical random graphs with the same numbers of vertices and links. 
There was a wide diversity of values of the clustering coefficients. 
However, all the large circuits considered in Ref. \cite{fjs01} ($N > 10^4$) have clustering coefficients that exceed those for the corresponding classical random graphs by more than one order of magnitude. 

The most interesting results were obtained for the degree distributions 
which were found to have power-law tails. 
The degree distributions of the two largest digital circuits were fitted by power laws with the exponent $\gamma \approx 3.0$. Note that the maximal value of the number of connections of a component in these large circuits approaches $10^2$.


\subsection{Other networks}\label{ss-other}

We have listed above only the most representative and well documented 
networks. 
Many kinds of friendship networks may be added \cite{s91book,wfbook94,asbs00}. 
Polymers also form complex networks \cite{sab00,jsb00a,sc01a}. 
Even human sexual contacts were found to form a complex network. 
It was recently
discovered \cite{leasa01} that this marvelous web is scale-free unlike friendship networks \cite{asbs00} which are exponential.  


One can introduce a {\em call graph} generated by long distance telephone calls taken over some time interval \cite{acl00}. Vertices of this network are telephone numbers, and the directed links are completed phone calls (the direction is determined by the initiator of the talk). In Ref. \cite{acl00}, calls made in a typical day were collected, and the network consisting of 
$47\times 10^6$ nodes was constructed (note, however, that this network was probably generated and not obtained from empirical data).  
It was impossible to fit $P(k_{o})$ by any power-law dependence but the fitting of the in-degree distribution $P(k_{i})$ gave $\gamma_{i}\approx 2.1$. 
The size of the giant connected component is of the order of the network size, 
and all others connected components are of the order of the logarithm of this 
size or smaller. The distribution of the sizes of connected components was measured but it was hard to make any conclusion about its functional form. 

Basic data for all networks, in which power-law degree distributions were observed, are summarized in Table \ref{t1} and Fig. \ref{f20}. 
For each such network, the total numbers of vertices and edges, and the degree distribution exponent are presented 
(see discussion of scale-free networks in Sec. \ref{s-scale-free}).  

We finish our incomplete list with a power grid of the Western States Power Grid \cite{ws98,watbook99,asbs00} 
(its vertices are transformers, substations, and generators, and edges are high-voltage transmission lines). 
The number of vertices in this undirected graph is $4\,941$, and the average degree $\overline{k}$ is $2.67$. 
The average shortest-path length equals $18.7$. 
The clustering coefficient of the power grid is much greater than for the corresponding classical random network, 
$C = 0.08 \gg 2.67/4941 \sim 0.0005$ \cite{ws98,watbook99}. The degree distribution of the network is exponential \cite{asbs00}.



\section{Classical random graphs, the Erd\"os-R\'enyi model}\label{s-classical}

The simplest and most studied network with undirected edges was introduced by  
Erd\"os and R\'enyi (ER model) \cite{er59,er60}. In this network: 
  
(i) the total number of vertices, $N$, is fixed; 

(ii) the probability that two arbitrary vertices are connected equals $p$. 

One sees that, on average, the network contains $pN(N-1)/2$ edges. The degree distribution is binomial,

\begin{equation}
P(k) = \left( \begin{array}{c} N-1 \\ k \end{array} \right) 
p^k(1-p)^{N-1-k}
\, ,  
\label{3-1}
\end{equation} 
so the average degree is $\overline{k}=p(N-1)$. For large $N$, the distribution, Eq. (\ref{3-1}) takes the Poisson form, 

\begin{equation}
P(k) = e^{-\overline{k}}\,\overline{k}^{\,k}\!/\,k!
\, .  
\label{3-2}
\end{equation}  
Therefore, the distribution rapidly decreases at large degrees. Such distributions are characteristic for classical random networks. 
Moreover, in the mathematical literature, the term ``random graph'' usually means just the network with a Poisson degree distribution and statistically uncorrelated vertices. 
Here, we prefer to call it ``classical random graph''.

We have already presented the estimate for an average shortest-path length of this network, 
$\overline{\ell} \sim \ln N /\ln [pN]$. 

At small values of $p$, the system consists of small clusters. 
At large $N$ and large enough $p$, the giant connected component appears in the network. 
The percolation threshold is $p_c \cong 1/N$, that is, $\overline{k}_c = 1$. 

In fact, the ER model describes percolation on a lattice of infinite dimension, and the adequate mean-field description is possible.


\section{Small-world networks}\label{s-small-world}

In Sec. \ref{ss-shortest}, we explained that random networks usually show the so-called small-world effect, i.e., their average 
shortest-path length is small. Then, in principle, it is natural to call them small-world networks. 
Watts and Strogatz \cite{ws98} noticed the following important feature of numerous networks in Nature. Although the average 
shortest-path length between their vertices is really small and is of the order of the logarithm of their size, the clustering 
coefficient is much greater that it should be for classical random graphs. 
They proposed a model (the WS model) that demonstrates such a possibility and also called it the small-world network. 
The model belongs to the class of networks displaying a crossover from ordered to random structures and may be treated analytically. 
By definition of Watts and Strogatz, the small-world networks are those with ``small'' average shortest-path lengths and ``large'' 
clustering coefficients. 

This definition seems a bit controversial. 
(i) According to it, numerous random networks with a small clustering coefficient are not small-world networks although they 
display the small-world effect. 
(ii) If one starts from a 1D lattice with interaction only between the nearest neighbors, or from simple square or cubic lattices, 
the initial clustering coefficient is zero and it stays small during the procedure 
proposed by Watts and Strogatz although the network evidently belongs to the same class of nets as the WS model. 
In addition, as we will show, the class of networks proposed by Watts and Strogatz provides only a particular possibility 
to get such a combination of the average shortest-path length and the clustering coefficient (see Sec. \ref{ss-another}). 

Irrespective of the consistency of the definition of the small-world networks \cite{ws98,watbook99} and its relation with real 
networks, the proposed type of networks is very interesting.  
 In fact, the networks introduced by Watts and Strogatz have an important generic feature -- they are constructed from ordered 
lattices by random rewiring of edges or by addition of connections between random vertices. In the present section, we 
consider mainly networks of such kind.



\subsection{The Watts-Strogatz model and its variations}\label{ss-watts}

\begin{figure}
\epsfxsize=89mm
\epsffile{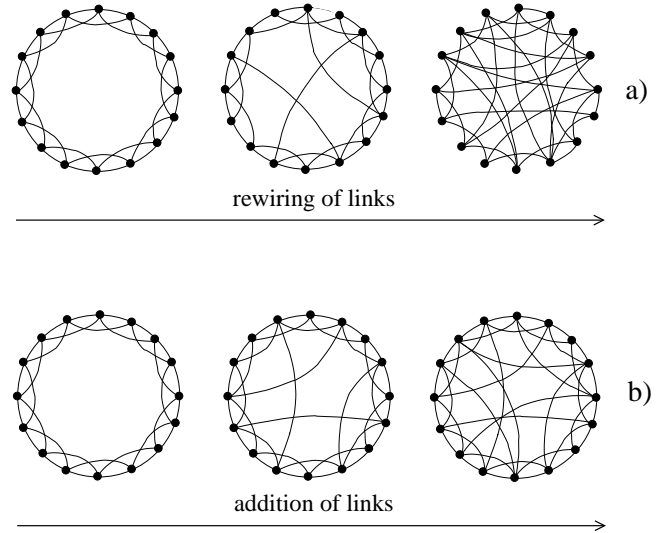}
\caption{
Small-world networks in which the crossover from a regular lattice to a random network is realized. (a) The original Watts-Strogatz model with the rewiring of links \protect\cite{ws98}. (b) The network with the addition of shortcuts \protect\cite{nw99,nw99i}.  
}
\label{f7}
\end{figure}

The original network of Watts and Strogatz is constructed in the following way 
(see Fig. \ref{f7},a). 
Initially, a regular one dimensional lattice with periodical boundary conditions is present. Each of $L$ vertices has $z \geq 4$ 
nearest neighbors ($z=2$ was not appropriate for Watts and Strogatz since, in this case, the clustering coefficient of the original regular 
lattice is zero). Then one takes all the edges of the lattice in turn and with probability $p$ rewires to randomly chosen vertices. 
In such a way, a number of far connections appears. Obviously, when $p$ is small, the situation has to be close to 
the original regular lattice. For large enough $p$, the network is similar to the classical random graph. Note that the 
periodical boundary conditions are not essential. 


\begin{figure}
\epsfxsize=85mm
\epsffile{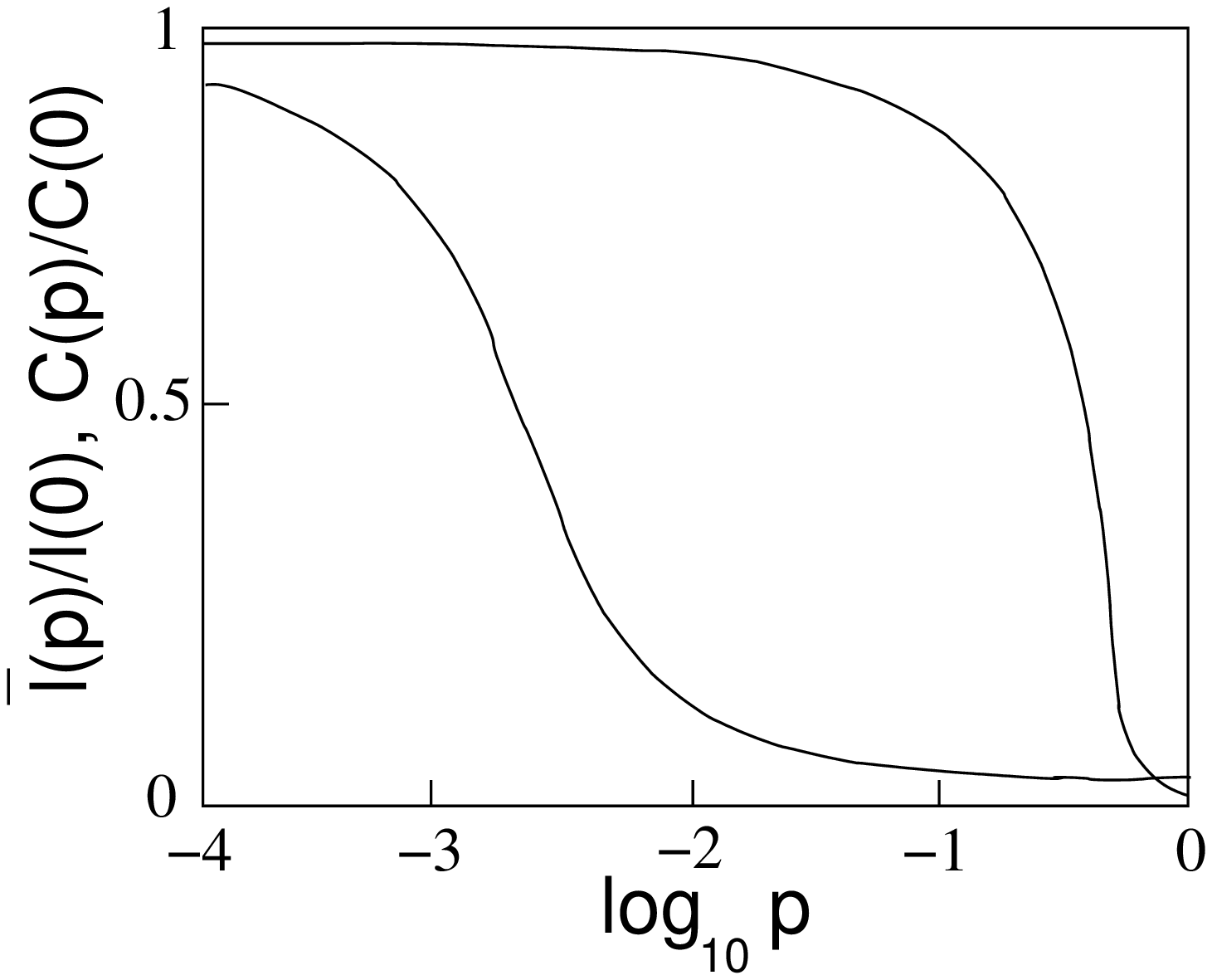}
\caption{
Average shortest-path length $\overline{\ell}$ and clustering coefficient $C$ of the Watts-Strogatz model vs. fraction of the rewired 
links $p$ \protect\cite{ws98}. Both are normalized to their values for the original regular lattice ($p=0$). The network has $1000$ nodes. 
The average number of the nearest neighbors equals $10$. $C$ is practically constant in the range where 
$\overline{\ell}$ sharply diminishes.
}
\label{f8}
\end{figure}

Watts and Strogatz studied the crossover between these two limits. The main interest was in the 
average shortest path, $\overline{\ell}$, and the clustering coefficient (recall that each edge has unit length). The simple but exciting result was the following. 
Even for the small probability of rewiring, when the local properties of the network are still nearly the same as for the original regular lattice and the clustering coefficient does not differ essentially from its initial value, the average shortest-path length is already of the order of the one for classical random graphs (see Fig. \ref{f8}). 


This result seems quite natural. Indeed, the average shortest-path length is very sensitive to the 
short-cuts. One can see, that it is enough to make a few random rewirings to decrease $\overline{\ell}$ by several times. On the 
other hand, several rewired edges cannot crucially change the local properties of the entire network. This means that the global properties of 
the network change strongly already at $pzL \sim 1$, when there is one shortcut in the network, i.e., at $p \sim 1/(Lz)$, when the local 
characteristics are still close to the regular lattice. 

Recall that the simplest local characteristic of nets is degree. Hence, it would be natural to compare, at first, 
the behavior of $\overline{\ell}$ and $\overline{k}$. However, in the originally formulated WS model, $\overline{k}$ is independent on $p$ 
since the total number of edges is conserved during the rewiring. Watts and Strogatz took another characteristic 
for comparison -- the characteristic of the closest environment of a vertex, i.e., the clustering coefficient $C$.  

Using the rewiring procedure, a network with a small average shortest-path length and a large clustering coefficient 
was constructed. Instead of the rewiring of edges, one can add shortcuts to a regular lattice (see Fig. \ref{f7},b) 
\cite{n00g,nw99,nw99i,nmw00}. The main features of the model do not change. One can also start with a regular lattice 
of an arbitrary dimension $d$ where the number of vertices 
$N = L^d$ \cite{k99,k00f}. In this case, the number of edges in the regular lattice is $zL^d/2$. To keep the correspondence 
to the WS model, let us define $p$ in such a way that for $p=1$, $zL^d/2$ random shortcuts are added. Then, the average number 
of shortcuts in the network is $N_s = pzL^d/2$. At small $N_s$, we have two natural lengths in the system, $\overline{\ell}$ 
and $L$, since the lattice spacing is not important in this regime. Their dimensionless ratio can be only a function of $N_s$, 

\begin{equation}
\frac{\overline{\ell}}{L} = f(2N_s) = f(pzL^d)
\, ,  
\label{4-1}
\end{equation}  
where $f(0) \sim 1$ for the original regular lattice and $f(x \gg 1) \sim \ln x/x^{1/d}$. 
From Eq. (\ref{4-1}), one can immediately obtain the following relation, 
$\overline{\ell}(pz)^{1/d} = g(L(pz)^{1/d})$. Here, $\xi = (pz)^{-1/d}$ has the meaning of a length:  
$N_s \xi^d \sim L^d$, it is the average distance between the closest end points of shortcuts measured on the regular lattice. 
In fact, one must study the limit $L \to \infty$, $p \to 0$, as the number of shortcuts $N_s = pzL^d/2$ is fixed.
The last relation for $\overline{\ell}$, in the case $d=1$, was proposed and studied by simulation in Ref. \cite{baam99} and afterwards analytically \cite{b99a,bw00}. 

The WS model and its variations seem exactly solvable. Nevertheless, the only known exact result for the WS model is its degree distribution. 
It was found to be a rapidly decreasing function of a Poisson kind \cite{bw00}. 
The exact form of the shortest-path length distributions has been found only for the simplest model in this class \cite{dm001}, see Sec. \ref{ss-smallest}.  


\begin{figure}
\epsfxsize=85mm
\epsffile{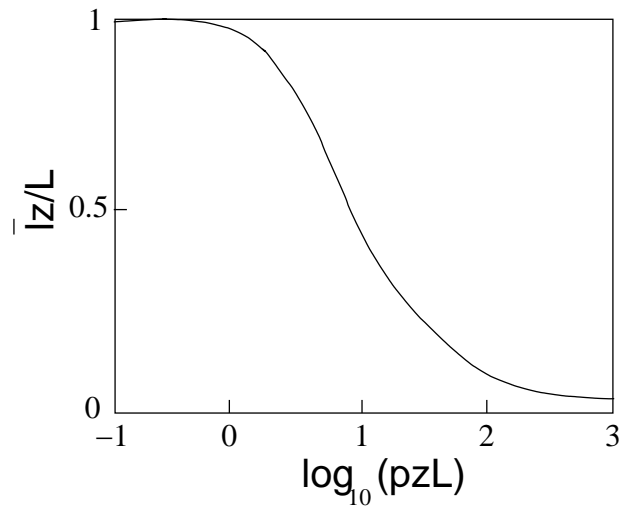}
\caption{
Scaling of the average shortest-path length of ``small-world'' networks \protect\cite{nw99i}. 
The combination $\overline{\ell}z/L$ vs. $pzL$ for the network constructed by the addition of random shortcuts to a one-dimensional lattice of the size $L$ with the coordination number $z$.  
}
\label{f9}
\end{figure}

 
Many efforts were directed to the calculation of the scaling function $f(x)$ describing the crossover between two limiting regimes \cite{nw99,nw99i,nmw00,b99a,bw00,kas00,mm99,m00,mmp00,k00b}. As we have already explained, the average shortest-path length  
rapidly decreases to values characteristic for classical random networks as $p$ grows. Therefore, it is convenient to plot $f(x)$ in log-linear scales 
(see Fig. \ref{f9}). 

\begin{figure}
\epsfxsize=85mm
\epsffile{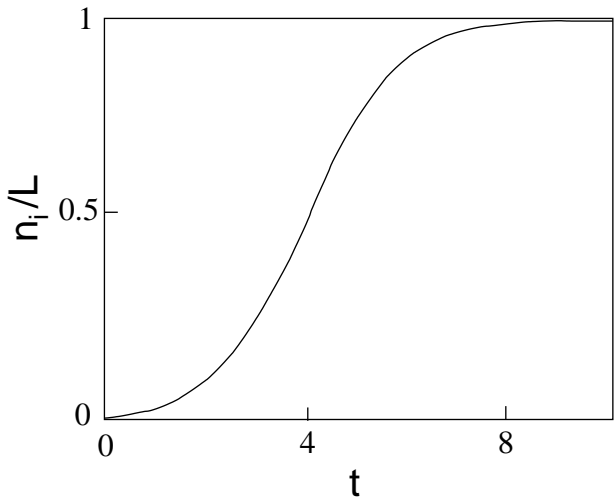}
\caption{
Spreading of diseases in ``small-world'' networks  \protect\cite{nmw00}. 
The average fraction of infected nodes, $n_i/L$, vs. the elapsed time from the instant when the first vertex ``fell ill''. 
}
\label{f10}
\end{figure}

One may study the distribution of diseases on such networks \cite{mn00}. In Fig. \ref{f10}, a portion of ``infected'' nodes, $n_i/L$, in the network is shown vs. time passed after some vertex was infected \cite{nmw00}. At each time step, all the 
nearest neighbors of each infected vertex fall ill. At short times, $n_i/L \propto t^d$ but then, at longer times, it increases exponentially until the saturation at the level $n_i/L=1$. 


It is possible to consider various problems for these networks 
\cite{bw00,pa99,pa00,ml00,kas99,jsb00,jb00,ak00b,bb00a1,br00a,mg00a,lm01a,lhcs00,z01a,zc01a,sj01a}. 
In Refs. \cite{mn00,mn00h}, percolation in them was studied (for infinitely large networks). 
Diffusion in the WS model and other related nets was considered in \cite{m99}. 

It is easy to generalize the procedure of rewiring or addition of edges. In Refs. \cite{k99,k00f}, the following procedure was introduced. New edges between pairs of vertices of a regular $d$-dimensional lattice are added with probability $p(r)$, where $r$ is the Euclidean distance between the pair of vertices. If, e.g., 
$p(r) \propto \exp(-const\ r)$, one gets a disordered $d$-dimensional lattice. 
Much slowly decreasing functions produce the small-world effect and related phenomena. 
In Refs. \cite{k99,k00f}, one may find the study of 
diffusion on a finite size network in the case of a power-law dependence of this probability, $p(r) \propto r^{-\epsilon}$.




\subsection{The smallest-world network}\label{ss-smallest}

\begin{figure}
\epsfxsize=55mm
\centerline{\epsffile{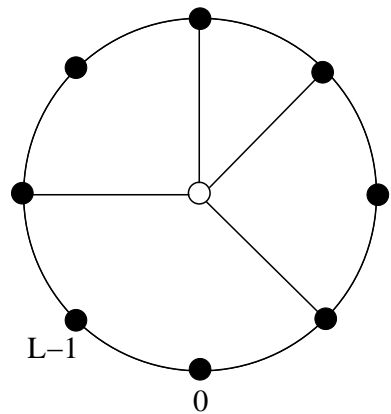}}
\caption{
The ``smallest-world'' network \protect\cite{dm001}. 
$L$ vertices on the circle are connected by unit length edges. 
Each of these vertices is connected to the central one by a half-length edge with probability $p$.
}
\label{f11}
\end{figure}

\begin{figure}
\epsfxsize=85mm
\epsffile{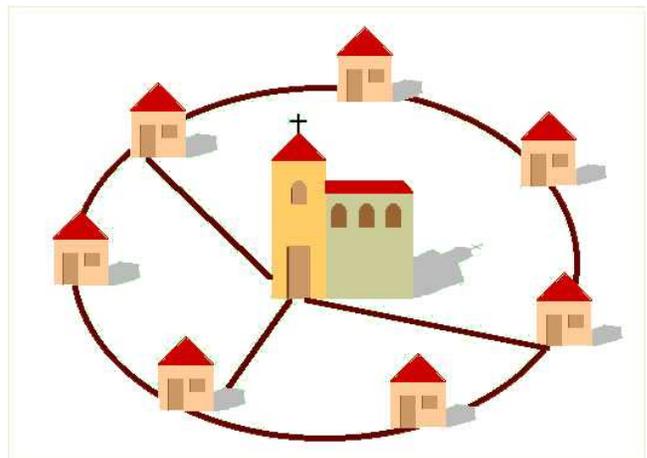}
\caption{
The real ``smallest-world'' network. 
Unsociable inhabitants live in this village. Usually, they contact only with their neighbors but some of them attend the church... 
}
\label{f12}
\end{figure}

Let us demonstrate the phenomena, which we discuss in the present section, using a trivial exactly solvable example, ``the smallest-world network'' (see Fig. \ref{f11})\cite{dm001}. 
We start from $L$ vertices connected in a ring by $L$ links of unit length, that is, the coordination number $z$ equals $2$ and the clustering coefficient is zero. This is not essential for us since we have no intention to discuss its behavior 
(in such a case, instead of the clustering coefficient, one may consider the density of linkage or degree). Then, we add a central vertex and make shortcuts between it and each other vertex with probability $p$. One may assume that lengths of these additional edges equal $1/2$. 
In fact, with probability $p$, we select random vertices and afterwards connect all of them together by edges of unit length. 
For the initial lattice, $\overline{\ell}(p=0)=L/4$, and, for the completely connected one, 
$\overline{\ell}(p=1)=1$. 
One should note that such networks may be rather reasonable in our world where substantial number of connections occurs through 
common centers 
(see Fig. \ref{f12}). 


\begin{figure}
\epsfxsize=90mm
\epsffile{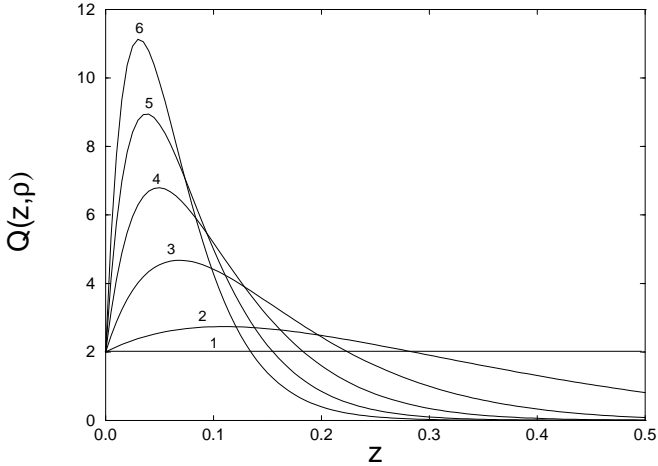}
\caption{
The distribution $Q(z,\rho)= LP(\ell,p)$ of the normalized shortest-path lengths  
$z \equiv \ell/L$  
of the ``smallest-world'' network. Here, $L$ is the size of the network, 
$\rho = pL$. Curves labeled by numbers from $1$ to $6$ correspond to 
$\rho = 0, 2, 5, 8, 11, 14$.
}
\label{f13}
\end{figure}

\begin{figure}
\epsfxsize=85mm
\epsffile{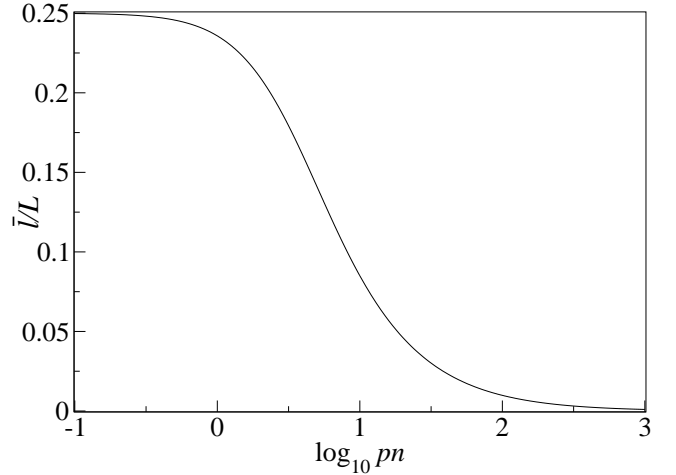}
\caption{
The normalized average shortest-path length $\overline{\ell}/L$ of the ``smallest-world'' network vs. the number $\rho=pL$ of added edges. 
}
\label{f14}
\end{figure}

\begin{figure}
\epsfxsize=85mm
\epsffile{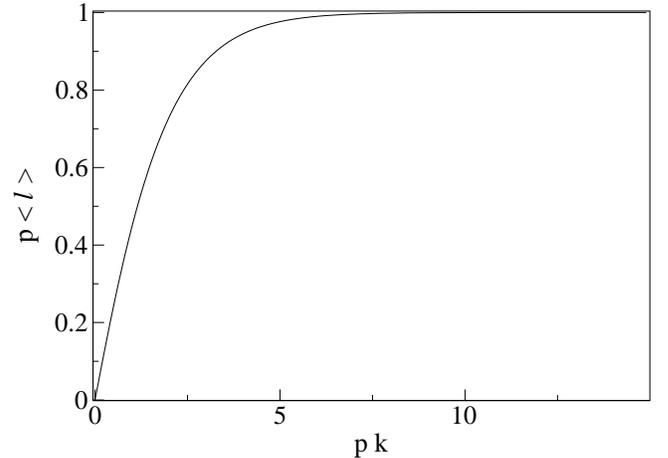}
\caption{
The normalized average shortest-path length $p\langle\ell\rangle$ between two vertices of the ``smallest-world'' 
network separated by the ``Euclidean'' distance $k$ as a function of $pk$.
}
\label{f15}
\end{figure}

One may calculate the distribution $P(\ell)$ of the shortest-path lengths $\ell$ of the network  exactly\cite{dm001}. 
In the scaling limit, $L \to \infty$ and $p \to 0$, while the quantities 
$\rho \equiv pL$ (average number of added edges) and $z \equiv \ell/L$ are fixed, the distribution takes the form, 

\begin{equation}
LP(\ell,p) \equiv Q(z,\rho) = 2[1 + 2\rho z + 2\rho^2 z(1-2z)]e^{-2\rho z}
\, .  
\label{5-1}
\end{equation} 
This distribution is shown in Fig. \ref{f13}. 
The corresponding average shortest-path length between pairs of vertices equals 

\begin{equation}
\frac{\overline{\ell}}{L} \equiv \overline{z} = \frac{1}{2\rho^2}
[2\rho - 3 + (\rho+3)e^{-\rho}]
\, ,  
\label{5-2}
\end{equation} 
that is just the scaling function $f(x)$ discussed in Sec. \ref{ss-watts} (see Fig. \ref{f14}). Hence, $\overline{z}(\rho=0) = 1/4$ and $\overline{z}(\rho \gg 1) \to 1/\rho$, i.e., $\overline{\ell} \to 1/p$.
One may also obtain the average shortest-path length $\langle\ell\rangle(k)$ between two vertices of the network separated by the ``Euclidean'' distance $k$, $k/L \equiv x$. 
In the scaling limit, we have 

\begin{equation}
\frac{\langle\ell\rangle(k,p)}{L} \equiv \langle z\rangle(x) = 
\frac{1}{\rho}
\left[ 1 - (1 + \rho x) e^{-2\rho x} \right]
\, ,  
\label{5-3}
\end{equation} 
(see Fig. \ref{f15}). Obviously, $\langle\ell\rangle(k,p \to 0) \to k$ but saturation is quickly achieved at large $pk$.


\vspace{30mm}$\phantom{x}$

\vspace{30mm}$\phantom{x}$

Eqs. (\ref{5-1})--(\ref{5-3}) actually demonstrate the main features of the crossover phenomenon in the models under discussion although our toy model does not approach the classical random network at large $p$. $\overline{\ell}$ of the model already diminishes sharply in the range of $\rho$ where local properties of the network are nearly the same as of the initial regular structure. 
In Ref. \cite{kr00i}, one can find the generalization of this model -- the probability that a vertex is connected to the center is assumed to be dependent on the state of its closest environment.


\subsection{Other possibilities to obtain large clustering coefficient}\label{ss-another}

The first aim of Watts and Strogatz \cite{ws98} was to construct networks with small average shortest paths and relatively large clustering coefficients 
which can mimic the corresponding behavior of real networks. In their network, the number of vertices is fixed, and only edges are updated (or are added). 
At least most of known networks do not grow like this.  
Let us demonstrate a simple network with a similar combination of these parameters ($\overline{\ell}$ and $C$) but evolving in a different way -- the growth of the network is due to both addition of new vertices and addition of new edges. 

\begin{figure}
\epsfxsize=85mm
\epsffile{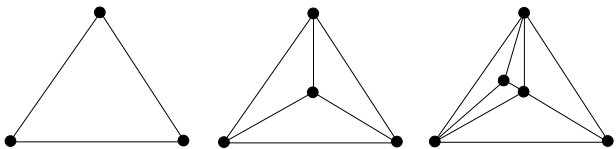}
\caption{
A simple growing network with a large clustering coefficient. 
In the initial configuration, three vertices are present. 
At each time step, a new vertex with three edges is added. 
These edges are attached to randomly chosen triples of nearest neighbor vertices.
}
\label{f16}
\end{figure}

In this model, initially, there are three vertices connected by three undirected edges (see Fig. \ref{f16}). Let at each time step, a new vertex be added. It connects to a randomly chosen triple of nearest neighbor vertices of the network. This procedure provides a network displaying the small-world effect. We will show below that this is a network with preferential linking. Its power-law degree distribution can be calculated exactly \cite{dms003} (see Sec. \ref{ss-simplestscale-free}). 


At the moment, we are interested only in the clustering coefficient. 
Initially, $C=1$ (see Fig. \ref{f16},a).  
Let us estimate its value for the large network. One can see that the number of triangles of edges in the network increases by three each time a vertex is added. Simultaneously, the number of triples of connected vertices increases by the sum of degrees of all three vertices to which the new vertex is connected. This sum may be estimated as $3 \overline{k}$. Here, $\overline{k}=2(3t)/t=6$. Hence, using 
the definition of the clustering coefficient, we get 
$C \approx 3(3t)/(3\overline{k}t) = 3/\overline{k} =1/2$. Therefore, $C$ is much larger than the characteristic value $\overline{k}/t$ for classical random graphs, and this simple network, constructed 
in a quite different way than the WS model, shows both discussed features of many real networks (see also the model with very similar properties in Sec. \ref{ss-simplestscale-free}, Fig. \ref{f18}). The reason for such a large value of the clustering coefficient is the simultaneous connection of a new vertex 
to nearest neighboring old vertices. 
This can partially explain the abundance of networks with large clustering coefficient in Nature. Indeed, the growth process, in which some old nearest neighbors connect together to a new vertex, that is, together ``borne'' it, seems quite natural (see Ref. \cite{n01b}). 

Another possibility to obtain a large clustering coefficient in a growing network is connecting a new vertex to several of its immediate predecessors with high probability (see also models proposed in Refs. \cite{ke01b,ke01c}). 

We should add that the one-mode projections of bipartite random graphs also have large clustering coefficients (see Secs. \ref{ss-collaborations} and \ref{ss-theory}).


\section{Growing exponential networks}\label{s-exponential}

The classical random network considered in Sec. \ref{s-classical} has fixed number of vertices. Let us discuss the simplest random network in which the number of vertices grows \cite{ba99,baj99}. At each increment of time, let a new vertex be added to the network. 
It connects to a randomly chosen (i.e., {\em without any preference}) old vertex 
(see Fig. \ref{f1}). 
Let connections be undirected, although it is inessential here.  
The growth begins 
from the configuration consisting of two connected vertices at time $t=1$, so, at time $t$, the network consists of $t+1$ vertices and $t$ edges. The total degree equals $2t$.  
 One can check that the average shortest-path length in this network is $\overline{\ell} \sim \ln t$ like in classical random graphs. 

It is easy to obtain the degree distribution for such a net. We may label vertices by their birth times, $s=0,1,2,\ldots,t$. Let us introduce the probability, $p(k,s,t)$, that a vertex $s$ has degree $k$ at time $t$.      
The master equation describing the evolution of the degree distribution of individual vertices is

\begin{equation}
p(k,s,t+1) = \frac{1}{t+1} p(k-1,s,t) + \left( 1 - \frac{1}{t+1} \right) p(k,s,t)
\, ,  
\label{6-1}
\end{equation}
$p(k,s=0,1,t=1) = \delta_{k,1}$, $\delta(k,s=t,t\geq 1) = \delta_{k,1}$. 
This accounts for two possibilities for a vertex $s$. (i) With probability $1/(t+1)$, it may get an extra edge from the new vertex and increase its own degree by $1$. (ii) With the complimentary probability $1 - 1/(t+1)$ the vertex $s$ may remain in the former state with the former degree. 
Notice that the second condition above makes Eq. (\ref{6-1}) non-trivial.

The total degree distribution of the entire network is 

\begin{equation}
P(k,t) = \frac{1}{t+1}\sum_{s=0}^{t}p(k,s,t)
\, .  
\label{6-2}
\end{equation} 
Using this definition and applying $\sum_{s=0}^{t}$ to both sides of Eq. (\ref{6-1}), we get 
the following master equation for the total degree distribution, 

\begin{equation}
(t+1)P(k,t+1) - tP(k,t) = P(k-1,t) - P(k,t) + \delta_{k,1}
\, .  
\label{6-3}
\end{equation}
The corresponding stationary equation, i.e., at $t \to \infty$, takes the form

\begin{equation}
2P(k) - P(k-1) = \delta_{k,1}
\,   
\label{6-4}
\end{equation} 
(note that the stationary degree distribution $P(k) \equiv P(k,t \to \infty)$ exists). It has the solution of an exponential form, 

\begin{equation}
P(k) = 2^{-k} 
\, .  
\label{6-4a}
\end{equation} 
Therefore, networks of such a type often are called ``exponential''. This form differs from the Poisson degree distribution of classical random graphs, see Sec. \ref{s-classical}. Nevertheless, both distributions are rapidly decreasing functions, unlike degree distributions of numerous large networks in Nature. 

The average degree of vertex $s$ at time $t$ is 

\begin{equation}
\overline{k}(s,t) = \sum_{k=1}^{\infty}p(k,s,t)
\, .  
\label{6-5}
\end{equation} 
Applying $\sum_{k=1}^{\infty}k$ to both sides of Eq. (\ref{6-1}), we get the equation for this quantity,

\begin{equation}
\overline{k}(s,t+1) = \overline{k}(s,t) + \frac{1}{t+1}
\, .  
\label{6-6}
\end{equation}
The resulting average degree of individual vertices equals 

\begin{equation}
\overline{k}(s,t) = 1 + \sum_{j=1}^{t-s}\frac{1}{s+j} = 1 + \psi(t+1) - \psi(s+1)
\,   
\label{6-7}
\end{equation} 
$[\overline{k}(0,t) = \overline{k}(1,t)]$. Here, $\psi(\ )$ is the $\psi$-function, i.e. the logarithmic derivative of the gamma-function. For $s,t \gg 1$, we obtain the asymptotic form, 

\begin{equation}
\overline{k}(s,t) = 1 - \ln(s/t)
\, ,  
\label{6-8}
\end{equation}  
i.e., the average degree of individual vertices of this network {\em weakly} diverges in the region of the oldest vertex. Hence, the oldest vertex is the ``richest'' (of course, in the statistical sense, i.e., with high probability). 

From Eq. (\ref{6-1}), one can also find the degree
distribution of individual vertices, $p(k,s,t)$, for large $s$ and $t$ and fixed $s/t$,

\begin{equation}
p(k,s,t) = \frac{s}{t} \frac{1}{(k+1)!} \ln^{k+1}\left(\frac{t}{s}\right)
\, .  
\label{6-9}
\end{equation}
One sees that this function decreases rapidly at large values of degree $k$. 

Similar results may be easily obtained for a network in which each new vertex has not one, as previously, but any fixed number of connections with randomly chosen old vertices. In fact, all the results of the present section are typical for growing exponential networks.


\section{Scale-free networks}\label{s-scale-free}

As we saw in Sec. \ref{s-nature}, at least several important large growing networks in Nature are scale-free, i.e., their degree distributions are of a power-law form 
(nevertheless, look at the remark in Sec. \ref{sss-www} concerning the quality of the experimental material). The 
natural question is how they self-organize into scale-free structures while growing. What is the mechanism responsible for such self-organization? 
For explanation of these phenomena, the idea of {\em preferential linking} (preferential attachment of edges to vertices) has been proposed \cite{ba99,baj99}.


\subsection{Barab\'asi-Albert model and the idea of preferential linking}\label{ss-idea}

We have demonstrated in Sec. \ref{s-exponential} that if new connections in a growing network appear 
between vertices chosen without any preference, e.g., between new vertices and randomly chosen old ones, the degree distribution is exponential. Nevertheless, in real networks, linking is very often {\em preferential}. 

For example, when you make a new reference in your own page, the probability that you refer to a popular Web document is certainly higher 
than the probability that this reference is to some poorly known document to that nobody referred before you. Therefore, popular vertices with high number of links are more attractive for new connections than vertices with few links -- {\em popularity is attractive}. 


Let us demonstrate the growth of a network with preferential linking using,  
as the simplest example, the 
Barab\'{a}si-Albert model (the BA model) \cite{ba99}. We return to the model described in Sec. 
\ref{s-exponential} (see Fig. \ref{f1}) and change in it only one aspect. Now a new vertex connects not to a randomly chosen old vertex but to a vertex chosen {\em preferentially}. 

We describe here the simplest situation: The probability that the edge is attached to an old vertex is proportional to the degree of this old vertex, i.e., to the total number of its connections. 
At time $t$, the total number of edges is $t$, and the total degree equals $2t$. Hence, this probability equals $k/(2t)$. One should emphasize that this is only a particular form of a {\em preference function}. However, just the linear type of the preference was indicated in several real networks \cite{bjnr01a,bjnr01a} (see discussion is Secs. \ref{ss-citations}, \ref{ss-collaborations} and \ref{sss-internet}).  
To account for the preferential linking, we must make obvious modifications to the master equation, Eq. (\ref{6-1}). For the BA model, the master equation takes the following form, 

\begin{equation}
p(k,s,t+1) = \frac{k-1}{2t} p(k-1,s,t) + \left( 1 - \frac{k}{2t} \right) p(k,s,t)
\,   
\label{7-1}
\end{equation}     
with the initial condition $p(k,s=0,1, t=1) = \delta_{k,1}$ and the boundary one $p(k,t,t) = \delta_{k,1}$. From Eqs. (\ref{6-2}) and (\ref{7-1}), we get the master equation for the total degree distribution,

\begin{eqnarray}
& & (t+1)P(k,t+1) - tP(k,t) = 
\nonumber 
\\[5pt]
& & 
\frac{1}{2}[(k-1)P(k-1,t) - kP(k,t)] + \delta_{k,1}
\, ,  
\label{7-2}
\end{eqnarray} 
and, in the limit $t \to \infty$, the equation for the stationary distribution, 

\begin{equation}
P(k) + \frac{1}{2}[kP(k) - (k-1)P(k-1)] = \delta_{k,1}
\, .  
\label{7-3}
\end{equation}
In the continuum $k$ limit, this equation is of the form 
$P(k) + (1/2)d[kP(k)]/dk = 0$. The solution of the last equation is $P(k) \propto k^{-3}$. Thus, the preferential linking of the form that we consider provides a scale-free network, 
and the $\gamma$ exponent of its distribution equals $3$ \cite{ba99,baj99,baj00i}. This value is exact, see Ref. \cite{krl00,dms001} and the discussion below. 

We emphasize that the preferential linking mechanism \cite{ba99,baj99} (1999) is the basic idea of the modern theory of evolving networks. 
Notice that preferential attachment may also arise effectively, in an indirect way (e.g., see Sec. \ref{ss-simplestscale-free} and models from Refs. \cite{ke01b,ke01c,v00}).    
The recent empirical data \cite{bjnr01a,jnb01a,n01b} (see Secs. \ref{ss-citations}, \ref{ss-collaborations}, and \ref{ss-communications}) on 
the dynamics of the attachment of new edges in various growing networks 
provide support for this mechanism. 
  


\subsection{Master equation approach}\label{ss-masterequation}

The master equation approach \cite{dms001} is very efficient for problems of the network evolution. Indeed, the linear discrete difference equations that arise (usually of first order) can be easily solved, e.g., 
using $Z$-transform. Let us describe the degree distributions for networks with preferential linking of a more general type than in Sec. \ref{ss-idea}. 


\begin{figure}
\epsfxsize=85mm
\epsffile{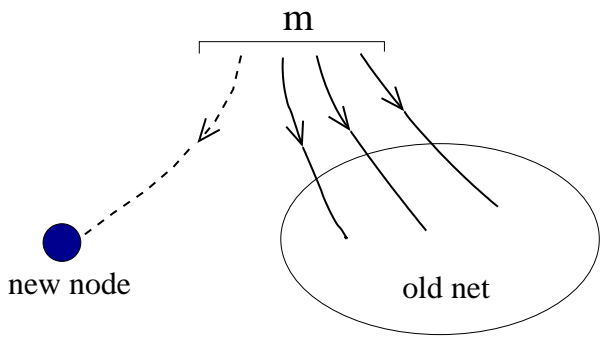}
\caption{
Scheme of the growth of the basic directed network under preferential linking mechanism. 
At each time step a new vertex and $m$ directed edges are added. 
Their source ends may be anywhere. 
The target ends of these edges are attached to vertices of the network according to the rule of preferential linking. 
}
\label{f17}
\end{figure}


Let us consider the following network with {\em directed} edges (see Fig. \ref{f17}). We will discuss here the in-degree distribution, 
so that we use, for brevity, the notations $q(s,t) \equiv k_{i}(s,t)$ and $\gamma$ instead $\gamma_{i}$. 

(i) At each time step, a new vertex is added to the network. 

(ii) Simultaneously, $m$ new directed edges going out of non-specified vertices or 
even from the outside of the network appeared. 

(iii) Target ends of the new edges are distributed among vertices according to the following rule. The probability that a new edge points to some vertex $s$ is proportional to $q(s) + A$. 

The parameter $A \equiv ma$ plays the role of {\em additional attractiveness} of vertices. 
The resulting in-degree distribution does not depend on the place from which new edges go out. If, in particular, each new vertex is the source of all the $m$ new edges (see a citation graph in Fig. \ref{f1}), then $k(s,t) = q(s,t) + m$, and the degree of each vertex is fixed by its in-degree. 
If, in addition, we set $A=m$, i.e., $a=1$, then 
new edges are distributed with probability proportional to $k(s,t)$, and we come to the BA model. 

Let us discuss the general case. The structure of the master equation for the in-degree distribution of individual vertices, $p(q,s,t)$, may be understood from the following. The probability that a new edge comes to a vertex $s$ equals $[q(s,t)+am]/[(1+a)mt]$. Here, $a \equiv A/m$. The probability that a vertex $s$ receives exactly $l$ new edges of the $m$ injected is 
 
\begin{equation}
{\cal P}_s^{(ml)}=
{m \choose l}
\left[\frac{q(s,t)+am}{(1+a)mt}\right]^l \left[ 1 - \frac{q(s,t)+am}{(1+a)mt} \right]^{m-l} 
\, .  
\label{8-1}
\end{equation}  
Hence, the in-degree distribution of an individual vertex of the large network under consideration obeys the following master equation,
\end{multicols}
\widetext
\noindent\rule{20.5pc}{0.1mm}\rule{0.1mm}{1.5mm}\hfill
 
\begin{equation}
p(q,s,t+1) = 
\sum_{l=0}^m{\cal P}_s ^{(ml)}p(q-l,s,t)
=\sum_{l=0}^m {m \choose l}
\left[ \frac{q-l+am}{(1+a) mt}\right] ^l\left[ 1-\frac{q-l+am}{%
(1+a) mt}\right] ^{m-l}p(q-l,s,t) 
\, .  
\label{8-2}
\end{equation}  
Vertices of this simple network are born without incoming edges, so the boundary condition 
for this equation is 
$p(q,t,t) =\delta_{q,0}$, where $\delta_{i,j}$ is the Kronecker symbol. 
The initial condition is fixed by the initial configuration of the network.  
Summing up Eq. (\ref{8-2}) over $s$, at long times, one gets 
the difference-differential equation 


\begin{equation}
(1+a)t \frac{\partial P}{\partial t}(q,t) + (1+a)P(q,t) + 
(q+am)P(q,t) - (q-1+am)P(q-1,t) 
= (1+a) \delta_{q,0} 
\, .
\label{8-4}
\end{equation} 
Excluding from it the term with the derivative, we obtain the equation for 
the stationary in-degree distribution 
$P \left( q\right) = P \left(q,t\to \infty \right)$, that is, for the in-degree distribution of the infinitely large network. (In fact, we have assumed that this stationary distribution exists. In the situation that we consider, this assumption is quite reasonable.) 
\hfill\rule[-1.5mm]{0.1mm}{1.5mm}\rule{20.5pc}{0.1mm}
\begin{multicols}{2}
\narrowtext
One may check by direct substitution that the exact solution of the stationary equation  
is of the form \cite{dms001} 

\begin{equation}
P(q) = (1+a) \frac{\Gamma[1+(m+1)a]}{\Gamma(ma)}\frac{\Gamma(q+ma)}{\Gamma[q+2+(m+1)a]}
\, .  
\label{8-6}
\end{equation} 
Here, $\Gamma(\ )$ is the gamma-function. 
In particular, when $a=1$, that corresponds to the BA model \cite{ba99}, we get 
the expression

\begin{equation}
P(q) = \frac{2m(m+1) }{(q+m)(q+m+1)(q+m+2) }
\, .  
\label{8-7}
\end{equation} 
To get the degree distribution of the BA model, one has only to substitute the degree $k$ instead of $q+m$ into Eq. (\ref{8-7}). 
Hence the continuum approximation introduced in Sec. \ref{ss-idea} indeed produced the proper value $3$ of the exponent of this distribution. 

For $q+ma \gg 1$, the stationary distribution (\ref{8-6}) takes the asymptotic form: 

\begin{equation}
P(q) \propto
(q+ma)^{-(2+a)}
\,.  
\label{8-8}
\end{equation}
Therefore, the scaling exponent $\gamma $ of the distribution depends on the additional attractiveness in the following way: 

\begin{equation}
\gamma = 2+a = 2+A/m
\, .  
\label{8-9}
\end{equation} 
Since $A > 0$, $\gamma$ varies between $2$ and $\infty$. 
This range of the $\gamma$ exponent values is natural for networks with constant average degree. In such a case, the first moment of the degree distribution must be finite, so that $\gamma>2$ (see discussion in Sec. \ref{ss-accelerating}).  

For this network, one may also find the in-degree distribution of individual vertices. At long times, the equation for it follows from Eq. (\ref{8-2}), 
\end{multicols}
\widetext
\noindent\rule{20.5pc}{0.1mm}\rule{0.1mm}{1.5mm}\hfill
 
\begin{equation}
p(q,s,t+1) =\left[ 1-\frac{q+am}{(1+a)t}\right]
p(q,s,t) +\frac{q-1+am}{(1+a)t}p(
q-1,s,t) + {\cal O}\left( \frac p{t^2}\right)   
\, .
\label{8-10}
\end{equation}
Assuming that the scale of time variation 
is much larger than $1$, at long times (large sizes of the network) 
we can replace the finite $t$-difference with a derivative:

\begin{equation}
(1+a) t\frac{\partial p}{\partial t}(q,s,t) = (q-1+am) p(q-1,s,t) - (q+am) p(q,s,t) 
\,.  
\label{8-11}
\end{equation}
\hfill\rule[-1.5mm]{0.1mm}{1.5mm}\rule{20.5pc}{0.1mm}
\begin{multicols}{2}
\narrowtext
The solution of Eq. (\ref{8-11}), 
i.e., the in-degree distribution of individual vertices, is

\begin{equation}
p(q,s,t) = \frac{\Gamma(am+q) }{\Gamma (am) q!}
\left( \frac st\right)^{am/(1+a)}
\left[ 1-\left( \frac st\right) ^{1/(1+a)}\right] ^q
\,.  
\label{8-12}
\end{equation}
Hence, this distribution has an exponential tail. 
One may also get the expression for the average in-degree of a given vertex:

\begin{equation}
\overline{q}(s,t) 
= \sum_{q=0}^\infty q \, p(q,s,) = a\,m\left[ \left( \frac st\right)^{
-1/(1+a)}-1\right] 
\, .  
\label{8-13}
\end{equation} 
Unlike a weak logarithmic divergence of average degree for oldest vertices of the exponential network (see Eq. (\ref{6-8})), here, 
at fixed time $t$, the average in-degree of an old vertex $s \ll t$ diverges as $s^{-\beta}$, where the exponent $\beta = 1/(1+a)$. 
One sees that for the BA model, $\beta=1/2$. 
The average degree of the oldest vertices is the highest, so the rule ``the oldest is the richest'' is certainly fulfilled here. The singularity is strong, so the effect is pronounced. 
From Eqs. (\ref{8-9}) and (\ref{8-13}), we obtain the following relation between the exponents of the network \cite{dm002}: 

\begin{equation}
\beta (\gamma - 1) = 1
\, ,  
\label{8-14}
\end{equation} 
We will show in Sec. \ref{ss-relations} that the relation, Eq. (\ref{8-14}), is universal for scale-free networks  
and can be obtained from the general considerations (nevertheless, see discussion of a particular case of violation of this relation in Sec. \ref{ss-decay}). 

In the scaling limit, when $q, s, t \to \infty$, $s \ll t$, and the scaling variable $\xi \equiv q(s/t)^{\beta}$ is fixed, the in-degree distribution, Eq. (\ref{8-12}), takes the form

\begin{equation}
p(q,s,t) = \left(\frac{s}{t}\right)^{\beta} f\left[ q \left(\frac{s}{t}\right)^{\beta} \right]
\, ,  
\label{8-15}
\end{equation} 
where the scaling function is 

\begin{equation}
f(\xi) = 
\frac{1}{\Gamma(am)}\, 
\xi^{am-1}\exp(-\xi)
\label{8-16}
\, .
\end{equation}  
Note that a particular form of the scaling function is model-dependent.


\subsection{A simple model of scale-free networks}\label{ss-simplestscale-free}

The results of Secs. \ref{ss-idea} and \ref{ss-masterequation} were obtained for large networks. 
Let us discuss a simple scale-free growing net for which exact answers may be obtained for an arbitrary size, without passing to the limit of large networks \cite{dms003}.


\begin{figure}
\epsfxsize=85mm
\epsffile{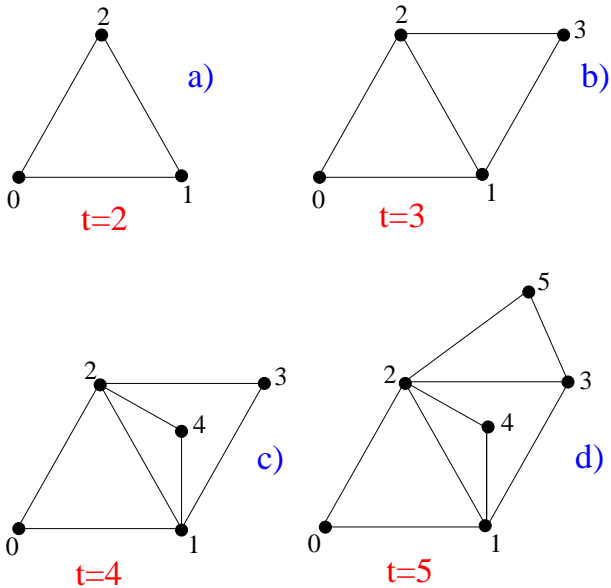}
\caption{
Illustration of a simple model of a scale-free growing network \protect\cite{dms003}. 
In the initial configuration, $t=2$, three vertices are present, $s=0,1,2$ $(a)$. 
At each increment of time, a new vertex with two edges is added. These edges are attached to the ends of a randomly chosen edge of the network.
}
\label{f18}
\end{figure}


We introduce the growing network with {\em undirected} edges (see Fig. \ref{f18}). Initially ($t=2$), three vertices are present, $s=0,1,2$, each with degree $2$. 

(i) At each increment of time, a new vertex is added.

(ii) It is connected to both ends of a randomly chosen edge by two undirected edges.  

The preferential linking arises in this simple model not because of some special rule 
including a function of vertex degree 
as in Refs. \cite{ba99,dms001}  
but quite naturally. Indeed, in the model that we consider here, the probability that a vertex has a randomly chosen edge attached to it is equal to the 
ratio of the degree $k$ of the vertex and the total number of edges, $2t-1$. Therefore, the evolution of the network is described by the following master equation for the degree distribution of individual vertices,

\begin{equation}
p(k,s,t+1) = \frac{k-1}{2t-1} p(k-1,s,t) + 
\frac{2t-1-k}{2t-1}
p(k,s,t)
, 
\label{9-1}
\end{equation}
with the initial condition, $p(k,s=\{0,1,2\},t=2)=\delta_{k,2}$. Also, $p(k,t,t)=\delta_{k,2}$.
This master equation and all the following ones in this subsection are exact for all $t \geq 2$. Eq. (\ref{9-1}) has a form similar to that  
of the BA model, Eq. (\ref{7-1}). Therefore, the scaling exponents of these models have to coincide.  

From Eq. (\ref{9-1}), there follows a number of exact relations for this model.  
In particular, from Eq. (\ref{9-1}), one may find the equation for the average degree of an individual vertex, $\overline{k}(s,t)\equiv\sum_{k=2}^{t-s+2} k p(k,s,t)$: 

\begin{equation}
\overline{k}(s,t+1) = \frac{2t}{2t-1}\overline{k}(s,t) \ \ \ \ , \ \ \ \ \overline{k}(t,t)=2
\,  
\label{9-2}
\end{equation}
with the following solution:

\begin{equation}
\overline{k}(s,t) = 2^{t-s+1} \frac{(t-1)!}{(s-1)!} \frac{(2s-3)!!}{(2t-3)!!} \,
\stackrel{s,t \gg 1}{\cong} \,
2\sqrt{\frac{t}{s}}
\, .
\label{9-3}
\end{equation}
Here, $s \geq 2$ and $\overline{k}(0,t)=\overline{k}(1,t)=\overline{k}(2,t)$. 
Hence, the scaling exponent $\beta$, 
defined through the relation  
$\overline{k}(s,t) \propto (s/t)^{-\beta}$, equals $1/2$ as for the BA model. 


The scaling form of $p(k,s,t)$ for $k,s,t \gg 1$ and $k\sqrt{s/t}$ fixed is 

\begin{equation}
p(k,s,t) = \sqrt{\frac{s}{t}} \left( k\sqrt{\frac{s}{t}} \right) 
\exp\left(-k\sqrt{\frac{s}{t}} \right)
\, . 
\label{9-6}
\end{equation}
(compare with Eqs. (\ref{8-15}) and (\ref{8-16})).

\end{multicols}
\widetext
\noindent\rule{20.5pc}{0.1mm}\rule{0.1mm}{1.5mm}\hfill

The matter of interest is the total degree distribution, 
$P(k,t)\equiv \sum_{s=0}^{t}p(k,s,t)/(t+1)$. The equation for it follows from Eq. (\ref{9-1}),

\begin{equation}
P(k,t) = \frac{t}{t+1}
\left[
\frac{k-1}{2t-3} P(k-1,t-1) + \left(1 - \frac{k}{2t-3}  \right) P(k,t-1)
\right]
+ \frac{1}{t+1} \delta_{k,2}
\, 
\label{9-7}
\end{equation}
with the initial condition $P(k,2)=\delta_{k,2}$. 
\hfill\rule[-1.5mm]{0.1mm}{1.5mm}\rule{20.5pc}{0.1mm}
\begin{multicols}{2}
\narrowtext

In the limit of the large network size, $t \to \infty$, $P(k,t)$ approaches a stationary degree distribution $P(k)$ which  
is very similar to the degree distribution of the BA model,

\begin{equation}
P(k) = \frac{12}{k(k+1)(k+2)}
\, , 
\label{9-9}
\end{equation}
that is, $\gamma=3$.

\begin{figure}
\epsfxsize=75mm
\epsffile{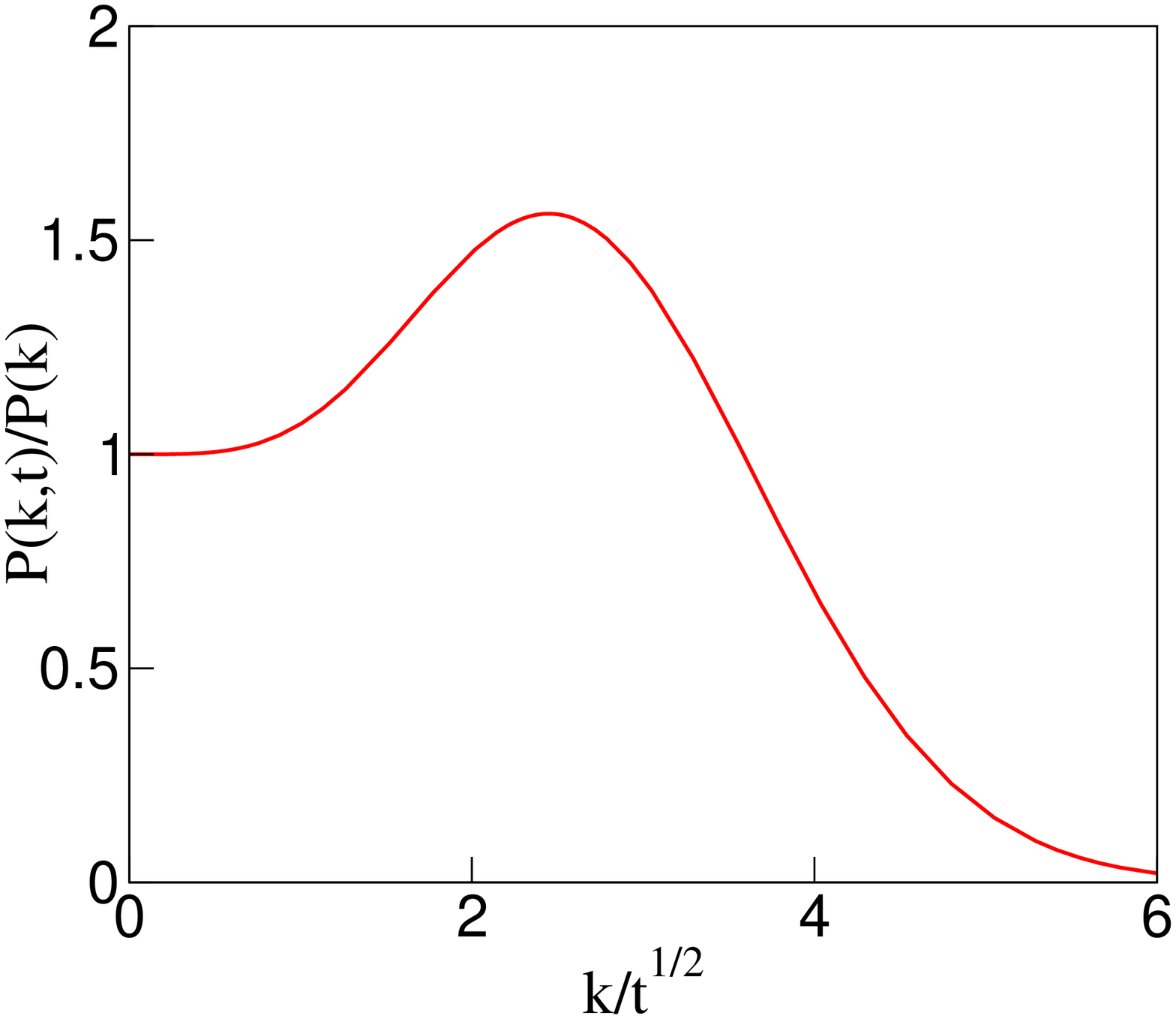}
\def\figcapt{
Deviation of the degree distribution of the finite-size network 
from the stationary one, 
$P(k,t)/P(k,t \to \infty)$, vs. $k/\sqrt t$.  
The form of the hump depends on the initial configuration of the network.
}
\caption{\protect\figcapt}
\label{f19}
\end{figure}

How does the degree distribution approach this stationary limit? 
We do not write down the cumbersome exact solution of Eq. (\ref{9-7}) \cite{dms003} but only write its scaling form for large $k$ and long time $t$ with $k/\sqrt{t}$ fixed: 

\begin{equation}
P(k,t) = P(k) 
\left[ 1 + \frac{1}{4}\frac{k^2}{t} + \frac{1}{8}\left(\frac{k^2}{t}\right)^2  \right] \exp\left\{ -\frac{1}{4}\frac{k^2}{t} \right\}
\, .
\label{9-10}
\end{equation}
The factor $P(k,t)/P(k) \equiv g(k/\sqrt t)$ depends only on the combination $k/\sqrt t$. 
Therefore, the peculiarities of the distribution induced by the size effects never disappear but only move with increasing time in the direction of large degree.
The function $g(k/\sqrt t)$ is 
shown in Fig. \ref{f19}.
Thus, the power-law dependence of the degree distribution of the finite size network 
is observable only in a rather narrow region, $1 \ll k \ll \sqrt t$. The cut-off 
at $k_{cut} \sim \sqrt t = t^{1/(\gamma-1)}$ and the hump impede observation of scale-free behavior.  



\vspace{35mm}$\phantom{x}$ 

One can check that the form of the hump in Fig. \ref{f19} depends on the initial conditions. 
In our case, the evolution starts from the configuration shown in  Fig. \ref{f19},a.
If the growth starts from another configuration, the form of the hump will be different. 
Note that this trace of the initial conditions is visible at any size of the network. 
Similar humps (or peaks) at the cut-off position were also observed recently \cite{zm00} in the non-stationary distributions of the Simon model \cite{s55,sbook57} (see Sec. \ref{ss-simon}).  

In Sec. \ref{sss-www} we have mentioned a kind of bipartite sub-graphs (bipartite cliques) which are used for indexing of cyber-communities in a large directed graph --- the WWW. In such a bipartite sub-graph, all $ha$ directed edges connecting $h$ 
hubs to $a$ authorities are present (see Fig. \ref{f6}). One may easily check that in a large equilibrium random graph the total number of these bipartite sub-graphs is negligible \cite{krrs00}. This is not the case for the growing networks. 
In the model under discussion, the statistics of the bipartite sub-graphs is very simple, so that we use this model as an illustrating example. 

Let us slightly modify the model to get a directed network. For this, let new edges be directed from new vertices to old ones. The possible number of authorities in the bipartite subgraphs of our graph is fixed: $a=2$. Each pair of nearest neighbor vertices plays the role of authorities of a bipartite sub-graph based on them. 
At each time step, a new hub (a new vertex) is added to a randomly chosen bipartite clique and two new cliques (two new edges) emerge. 
The total number $N_b(h,a=2,t)\equiv N_b(h,t)$ of bipartite sub-graphs with $h$ hubs in the network in time $t$ satisfy the following simple equation 

\begin{equation}
N_b(h,t+1) = 2\delta_{h,0} + N_b(h,t) + \frac{1}{t} N_b(h-1,t) - \frac{1}{t} N_b(h,t)
\, .
\label{9-11}
\end{equation}  
The first term on the right-hand part of Eq. (\ref{9-11}) is a contribution from two new edges, the third and fourth terms are due to addition of a new hub, that is, a new vertex, to the network. 

The probability that a randomly chosen vertex belongs to the bipartite sub-graphs with $h$ hubs is $G(h,t)=N_b(h,t)/t$. From Eq. (\ref{9-11}) we have 

\begin{equation}
(t+1)G(h,t+1) - tG(h,t) = 2\delta_{h,0} - G(h,t) + G(h-1,t)
\, .
\label{9-12}
\end{equation} 
Its stationary solution $G(h) \equiv G(h, t\to\infty)$ is $G(h) = 2^{-h}$. Hence, the total number of the bipartite sub-graphs with $h$ hubs in the large network is large (proportional to $t$) and decreases exponentially as $h$ grows:  

\begin{equation}
N_b(h,t)= 2^{-h}t
\, .
\label{9-13}
\end{equation}  
This result agrees with an estimate made in Ref. \cite{krrs00} for citation graphs growing under mechanism of preferential linking and correlates with the measurements \cite{krrt99,kkrrt99} of the distribution of the bipartite subgraphs in the WWW (see Sec. \ref{sss-www}). 
    
Note that the above model is very close to the network growing under mechanism of linking to triples of the nearest neighbor nodes (see Sec. \ref{ss-another}). The degree distributions of these networks are very similar ($\gamma=3$). Both these scale-free networks have large clustering coefficients.   


\begin{figure}
\epsfxsize=70mm
\centerline{\epsffile{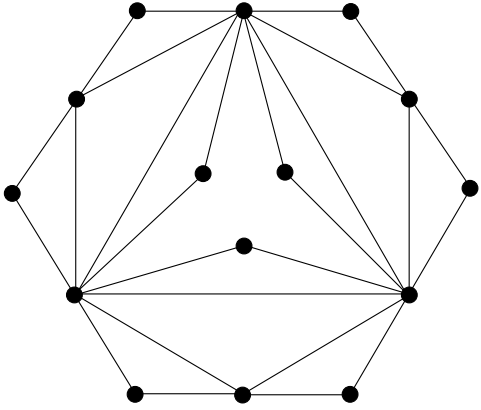}}
\caption{
A simple deterministic growing graph. At time $t=0$, 
the graph is a triangle. 
At each time step every edge of the graph generates 
a new vertex which connects to both ends of the edge.
}
\label{f19a}
\end{figure}
 

One may slightly modify the model under consideration and obtain a deterministic growing graph which has a discrete spectrum of degrees. ``Scale-free'' networks of this kind were recently proposed in Ref. \cite{br01a}. At each time step, let every edge of the graph generate a new vertex which connects to both ends of the edge (see Fig. \ref{f19a}). The growth starts from a triangle ($t=0$). 
Then the total number of vertices at time $t$ is $N_t=3(3^t+1)/2$ and the total number of edges is $L_t=3^{t+1}$, so that the average degree 
$\overline{k}_t=4/(1+3^{-t})$ approaches $4$ in the large-graph limit. 
The ``perimeter'' of the graph (see Fig. \ref{f19a}) is $P_t = 3\times 2^t$, hence 
$N_t \sim L_t \sim P_t^{\ln3/\ln2}$ when $t$ is large. 
The clustering coefficient of the graph is large: $C \to 4/5$ as $t \to \infty$. 

The spectrum of degrees of the graph is discrete: 
at time $t$, the number $n(k,t)$ of vertices of degree 
$k=2,2^2,2^3, \ldots, 2^{t-1},2^t,2^{t+1}$ is equal to $3^t, 3^{t-1},3^{t-2},\ldots,3^2,3,3$, respectively.  
Other values of degree are absent in the spectrum. 
Clearly, for the large network, $n(k,t)$ decreases as a power law of $k$, so the network may be called ``scale-free''. It is easy to introduce the exponent $\gamma$ for this  discrete situation where degree points are inhomogeneously spread over the $k$ axis. 
For this one may calculate the corresponding cumulative distribution 
$P_ {cum}(k_j) \propto k_j^{-\ln3/\ln2} \propto k_j^{1-\gamma}$. Here $k_j$ are points of the discrete degree spectrum. 
Then we obtain  

\begin{equation}
\gamma=1+\frac{\ln3}{\ln2}>2
\, .
\label{9-14}
\end{equation}  
Compare this expression with the exponent (the fractal dimension) in the relation between the ``mass'' and the perimeter of the graph. Also, notice that the maximal degree of a vertex is $k_{cut} = 2^{t+1} \sim N_t^{\ln2/\ln3} = N_t^{1/(\gamma-1)}$.  
Other deterministic versions of the same simple model produce various discrete distributions (exponential and others).


\subsection{Scaling relations and cutoff}\label{ss-relations}

In Secs. \ref{ss-masterequation} and \ref{ss-simplestscale-free} we found that a number of quantities of particular scale-free networks may be written in a scaling form, and the scaling exponents involved are connected by a simple relation. Can these forms and relations be applied to all scale-free networks? 

Let us proceed with general considerations. In this subsection, it is not essential, whether we consider degree, in-degree, or out-degree. Hence we use one general notation, $k$. When one speaks about scaling properties, a continuum treatment is sufficient, so that we can use the following expressions
 
\begin{equation}
P(k,t) = \frac{1}{t}\int_{t_0}^{t}ds\,p(k,s,t)
\,   
\label{11-1}
\end{equation} 
and 

\begin{equation}
\overline{k}(s,t) = \int_0^\infty dk\, k p(k,s,t)
\, .  
\label{11-2}
\end{equation} 
In addition, we will need the normalization condition for $p(k,s,t)$, 

\begin{equation}
\int_0^\infty dk\,p(k,s,t) = 1
\, .  
\label{11-3}
\end{equation} 

If the stationary distribution exists, than from Eq. (\ref{11-1}), it follows that $p(k,s,t)$ has to be of the form  
$p(k,s,t)=\rho(k,s/t)$. From the normalization condition, Eq. (\ref{11-3}), we get 
$\int_0^\infty dk\,\rho(k,x) = 1$, so $\rho(k,x) = g(x) f(kg(x))$, where $g(x)$ and $f(x)$ are arbitrary functions. 

Let us assume that the stationary distribution $P(k)$ and the average degree $\overline{k}(s,t)$ exhibit scaling behavior, that is,  
$P(k) \propto k^{-\gamma}$ for large $k$ and $\overline{k}(s,t) \propto s^{-\beta}$ 
for $1 \ll s \ll t$. Then, from Eq. (\ref{11-2}), one sees that 
$\int_0^\infty dk\,k\rho(k,x) \propto x^{-\beta}$. Substituting $\rho(k,x)$ into this relation, 
one obtains $g(x) \propto x^\beta$. Of course, without loss of generality, one may set $g(x)=x^\beta$, so that we obtain the following scaling form of the degree distribution of individual vertices,

\begin{equation}
p(k,s,t)=(s/t)^\beta f(k(s/t)^\beta) 
\, .  
\label{11-4}
\end{equation}
Finally, assuming the scaling behavior of $P(k)$, i.e., $\int_0^\infty dx\, \rho(k,x) \propto k^{-\gamma}$, 
and using Eq. (\ref{11-4}), we obtain $\gamma=1+1/\beta$, i.e., relation (\ref{8-14}) between the exponents is universal for scale-free networks. 
Here we used the rapid convergence of $\rho(k,x)$ at large $x$ (see Eqs. (\ref{8-16}) and (\ref{9-6})).
One should note that in this derivation we did not use any approximation.

The relation between the $\gamma$ and $\beta$ exponents looks precisely the same as that for the $\gamma$ exponent of the degree distribution and the corresponding exponent of Zipf's law, $\nu$. One can easily understand the reason of this coincidence.  
Recall that, in Zipf's law, the following dependence is considered: $k = f(r)$. Here $r$ is the rank of a vertex of the degree $k$, i.e., $r \propto \int_k^\infty dk\,P(k) \equiv P_{cum}(k)$. 
If Zipf's law is valid, $k \propto r^{-\nu}$, then 
$r \propto k^{-1/\nu} \propto k^{-\gamma+1}$, and  
we get $\gamma=1+1/\nu$. Therefore, 
the $\beta$ exponent equals the exponent of the Zipf's law, $\beta=\nu$. 

Now we can discuss the size-effects in growing scale-free networks. Accounting for the rapid decrease of the function $f(z)$ in Eq. (\ref{8-15}), one sees that the power-law dependence of the total degree distribution has a cut-off at the characteristic value, 

\begin{equation}
k_{cut} \sim t^{\beta} = t^{1/(\gamma-1)}
\, .
\label{11-5}
\end{equation} 
In fact, $k_{cut}$ is the generic scale of all ``scale-free'' networks. 
It also follows from the condition 
$t\int_{k_{cut}}^\infty dk P(k) \sim 1$, i.e., $t\int_{k_{cut}}^\infty dk k^{-\gamma} \sim 1$. 
This means that only one vertex in a network has degree above the cutoff. 
A more precise estimate is $k/k_0 \sim t^{1/(\gamma-1)}$, where $k_0$ is the lower boundary of the power-law region of the degree distribution.  
Eq. (\ref{11-5}) can be used to estimate the $\gamma$ exponent if the maximal degree in a network is known from empirical data \cite{krl00,kr00c}. We have already applied Eq. (\ref{11-5}) in Sec. \ref{s-nature} to check the quality of reported values of some real networks. 

We have shown (see Sec. \ref{ss-simplestscale-free}) that a trace of initial conditions at $k \sim k_{cut}$ may be visible in a degree distribution measured for any network size \cite{dms003}. 
The cutoff (and the trace of initial conditions) sets strong restrictions for observations of power-law distributions since there are few really large networks in Nature. 

In fact, measurement of degree distributions is always hindered by  
strong fluctuations at large $k$. 
The reason of such fluctuations is the 
poor statistics in this region. One can easily estimate 
the characteristic value of degree, $k_f$, above which the fluctuations are strong. 
 If $P(k) \sim k^{-\gamma}$, 
$t k_f^{-\gamma} \sim 1$. Therefore, $k_f \sim t^{1/\gamma}$. One may improve the situation using the cumulative distributions, $P_{cum}(k) 
\equiv \int_k^\infty dk P(k)$, instead of $P(k)$. 
Also, in simulations, one may make a lot of runs to improve the statistics. 
Nevertheless, one can not exceed the cut-off, $k_{cut}$, that we discuss. 
This cut-off is the real barrier for the observation of the power-law dependence. 
(One should note that accounting for the aging of nodes, break of links, or disappearance of nodes suppresses the effect of the initial conditions and removes the hump \cite{dm002,dm00e,dm003}.)

No scale-free networks with large values of $\gamma$ were observed. The reason for this is clear.   
Indeed, the power-law dependence of the degree distribution can be observed only if it exists for at least $2$ or $3$ decades of degree. For this, the networks have to be large: their size should be, at least,  
$t > 10^{2.5(\gamma-1)}$. Then, if $\gamma$ is large, one practically has no chances to find the scale-free behavior. 

\begin{figure}
\epsfxsize=85mm
\epsffile{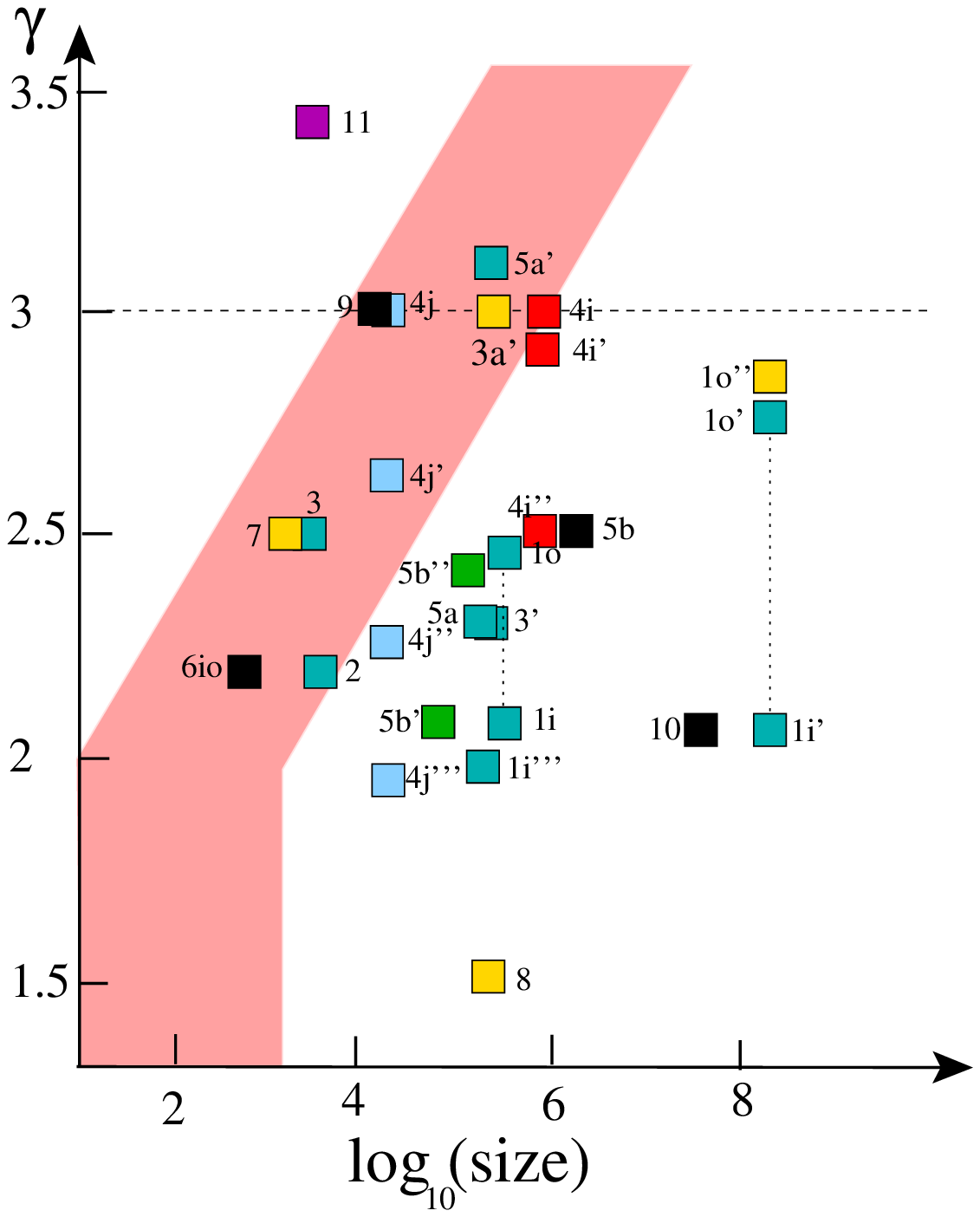}
\def\figcapt{
Log-linear plot of the $\gamma$ exponents of all the networks reported as having power-law (in-, out-) 
degree distributions (i.e., scale-free networks) vs. their sizes. 
The line $\gamma \sim 1 + \log_{10} t /2.5$ is the estimate of the finite-size boundary for the observation of the power-law 
degree distributions for $\gamma>2$. Here $2.5$ is the range of degrees (orders) which we believe is necessary to observe a power law. 
The dashed line, $\gamma=3$, is the resilience boundary (see Sec. \protect\ref{ss-resilience}). 
This boundary is 
important for  
networks which must be stable to random breakdowns. 
The points are plotted using the data from Tab. \protect\ref{t1}. 
Points for $\gamma_o$ and $\gamma_i$ from the same set of data are connected. 
The precision of the 
right points is about $\pm 0.1$ (?) and is much worse for points in the grey region. 
There exists a chance that some of these nets are actually not in the class of scale-free networks. 
\ \ \ 
The points: 
\ \ 
$1i$ and $1o$ are obtained from in- and out-degree distributions of the complete map of the nd.edu domain of the WWW \protect\cite{ajb99};  
\ \ 
$1i^\prime$ and $1o^\prime$ are from in- and out-degree distributions of the pages of the WWW scanned by Altavista in October of 1999 
\protect\cite{bkm00,krrt99};  
\ \ 
$1o^{\prime\prime}$ is the $\gamma_{o}$ value from another fitting of the same data \protect\cite{nsw00}; 
\ \ 
$1i^{\prime\prime\prime}$ is $\gamma_{i}$ for domain level of the WWW in spring 1997 \protect\cite{ah00a}); 
\ \ 
$2$ is $\gamma$ for the inter-domain level of the Internet in December 1998 \protect\cite{fff99}; 
\ \ 
$2^\prime$ is $\gamma$ for the network of operating AS in one of days in December 1999 \protect\cite{pvv01a};
\ \
$3$ is $\gamma$ for the router level of the Internet in 1995 \protect\cite{fff99};  
\ \
$3^\prime$ is $\gamma$ for the router level of the Internet in 2000 \protect\cite{gt00c}; 
\ \
$4i$ is $\gamma_i$ for citations of the ISI database 1981 -- June 1997 \protect\cite{r98}; 
\ \
$4i^\prime$ is the result of the different fitting of the same data \protect\cite{ta00};
\ \ 
$4i^{\prime\prime}$ is another estimate obtained from the same data \protect\cite{krl00,kr00c};
\ \
$4j$ is $\gamma_i$ for citations of the Phys. Rev. D {\bf 11-50} (1975-1994) \protect\cite{r98}; 
\ \ 
$4j^\prime$ is the different fitting of the same data \protect\cite{ta00};
\ \ 
$4j^{\prime\prime}$ is another estimate from the same data \protect\cite{krl00,kr00c};
\ \ 
$4j^{\prime\prime\prime}$ is $\gamma_i$ for citations of the Phys. Rev. D (1982-June 1997) \protect\cite{v01a}; 
\ \
$5a$ is the $\gamma$ exponent for the collaboration network of movie actors \protect\cite{ba99};
\ \ 
$5a^\prime$ is the result of another fitting for the same data \protect\cite{ab00a};
\ \ 
$5b$ is $\gamma$ for the collaboration network of MEDLINE \protect\cite{n00};
\ \ 
$5b^\prime$ is $\gamma$ for the collaboration net collected from mathematical journals \protect\cite{bjnr01a};
\ \ 
$5b^{\prime\prime}$ is $\gamma$ for the collaboration net collected from neuro-science journals \protect\cite{bjnr01a}; 
\ \ 
$6io$ is $\gamma_i=\gamma_o$ for networks of metabolic reactions \protect\cite{jtaob00};
\ \ 
$7$ is $\gamma$ of the network of protein-protein interactions (yeast proteome) if it is treated as undirected \protect\cite{jmbo01a,w01d};
\ \ 
$8$ is $\gamma$ of the degree distribution of the word web in the range below the crossover point \protect\cite{fs01};
\ \ 
$9$ is $\gamma$ of large digital electronic circuits \protect\cite{fjs01};
\ \ 
$10$ is $\gamma_i$ of the telephone call graph \protect\cite{acl00} (the out-degree distribution of this graph cannot be fitted by a power-law dependence);
\ \ 
$11$ is $\gamma$ of vertices in the web of human sexual contacts \protect\cite{leasa01}.
}
\caption{\protect\figcapt}
\label{f20}
\end{figure}

In Fig. \ref{f20}, in the log-linear scale, we present  
the values of the $\gamma$ exponents of all the networks reported as having power-law degree distributions vs. their sizes (see also Tab. \ref{t1}). 
One sees that almost all the plotted points are inside of the region restricted by the lines: $\gamma=2$, $\log_{10} t \sim 2.5(\gamma-1)$, and by the logarithm of 
the size of the largest scale-free network -- the World-Wide Web -- $\log_{10} t \sim 9$. 

In a similar way, we obtain the following general form of $P(k,t)$ for scale-free networks in the scaling regime:

\begin{equation}
P(k,t) =  k^{-\gamma} F(kt^{-\beta}) = k^{-\gamma} F(kt^{-1/(\gamma-1)})
\, .
\label{11-7}
\end{equation} 
Here $F(x)$ is a scaling function. We have obtained this form for an exactly solvable network in Sec. \ref{ss-simplestscale-free}. 

\newpage


\end{multicols}
\widetext
\noindent\rule{20.5pc}{0.1mm}\rule{0.1mm}{1.5mm}\hfill


\begin{table}
\begin{tabular}{l|c|c|c|c}
                           &                  &                  &           &        \\
network or subgraph                   & $\!\!\!\!$number of vertices \  & $\!\!\!\!$number of edges \  & $\gamma$  & Refs.   \\
                           &                  &                  &           &        \\
\cline{1-5}
                           &                  &                  &           &        \\
 complete map of the nd.edu domain of the Web & $325,729$  & $1,469,680$ & $\gamma_{i}=2.1$  & \protect\cite{ajb99}  \\
                           &                  &                  & $\gamma_{o}=2.45$ &        \\[5pt]
 pages of World Wide Web scanned by Altavista  & $2.711\ 10^8$  & $2.130\ 10^9$ 
& $\gamma_{i}=2.1$  & \protect\cite{bkm00,krrt99}  \\
in October of 1999    &                  &                  & $\gamma_{o}=2.7$ &        \\[3pt]
``------------'' (another fitting of the same data)  &   &  
& $\gamma_{i}=2.10$  & \protect\cite{nsw00}  \\
                           &                  &                  & $\gamma_{o}=2.82$ &        \\[5pt]
 domain level of the WWW in spring 1997 & $2.60\ 10^5$  & --- & $\gamma_{i}=1.94$  & \protect\cite{ah00a}   \\[5pt]
 inter-domain level of the Internet in December 1998  & $4389$  & $8256$ & $2.2$  & \protect\cite{fff99}  \\[5pt] 
net of operating ``autonomous systems'' in Internet$\,^1$  & $6374$ & $13641$ & $2.2$ &  \protect\cite{pvv01a}  \\[5pt]      
router level of the Internet in 1995  & $3888$  & $5012$ & $2.5$  & \protect\cite{fff99}  \\[5pt] 
router level of the Internet in 2000$\,^2$ & $\sim 150,000$  & $\sim 200,000$ & $\sim2.3$  & \protect\cite{gt00c}  \\[5pt]
citations of the ISI database 1981 -- June 1997 & $783,339$  & $6,716,198$ & $\gamma_{i}=3.0$  & \protect\cite{r98}  \\[3pt] 
``------------'' (another fitting of the same data) &   &   & $\gamma_{i}=2.9$  & \protect\cite{ta00}  \\[3pt] 
``------------'' (another estimate from the same data) &   &   & $\gamma_{i}=2.5$  & \protect\cite{krl00,kr00c}  \\[5pt] 
citations of the Phys. Rev. D {\bf 11-50} (1975-1994) & $24,296$  & $351,872$ & $\gamma_{i}=3.0$  & \protect\cite{r98}  \\[3pt] 
``------------'' (another fitting of the same data) &   &   & $\gamma_{i}=2.6$  & \protect\cite{ta00}  \\[3pt]
``------------'' (another estimate from the same data) &   &   & $\gamma_{i}=2.3$  & \protect\cite{krl00,kr00c}  \\[3pt] 
citations of the Phys. Rev. D (1982-June 1997) & ---  & --- & $\gamma_{i}=1.9$  & \protect\cite{v01a}  \\[5pt]
collaboration network of movie actors & $212,250$  & $61,085,555$ & $2.3$  & \protect\cite{ba99}  \\[3pt] 
``------------'' (another fitting of the same data) &   &  & $3.1$  & \protect\cite{ab00a}  \\[5pt] 
collaboration network of MEDLINE & $1,388,989$  & $1.028\ 10^7$ & $2.5$  & \protect\cite{n00}  \\[5pt] 
collaboration net collected from mathematical journals & $70,975$  & $0.132\times 10^6$ & $2.1$  & \protect\cite{bjnr01a}  \\[5pt]
collaboration net collected from neuro-science journals  & $209,293$  & $1.214\times 10^6$ & $2.4$  & \protect\cite{bjnr01a}  \\[5pt]
networks of metabolic reactions & $\sim 500-800$  & $\sim 1500-3000$ & $\gamma_{i}=2.2$  & \protect\cite{jtaob00}  \\[3pt]
                                &  &  & $\gamma_{o}=2.2$  &     \\[5pt] 
net of protein-protein interactions (yeast proteome)$\,^3$ & $1870$ & $2240$ & $\sim 2.5$ &  \protect\cite{jmbo01a,w01d} \\[5pt] 
word web$\,^4$                        & $470,000$ & $17,000,000$ & $1.5$ & \protect\cite{fs01}  \\[5pt] 
digital electronic circuits & $2\times 10^4$ & $4\times 10^4$ & $3.0$ & \protect\cite{fjs01}  \\[5pt]
telephone call graph$\,^5$ & $47\times 10^6$  & $8\times 10^7$ & $\gamma_{i}=2.1$  & \protect\cite{acl00}  \\[5pt]
web of human sexual contacts$\,^6$ & $2810$  & --- & $3.4$  & \protect\cite{leasa01}  \\[5pt]
food webs$\,^7$ & $93 - 154$  & $405 - 366$ & $\sim 1$  & \protect\cite{ms00,sm00}  \\[3pt]
\end{tabular}
\caption{
Sizes and values of the $\gamma$ exponent of the networks or subgraphs reported as having power-law (in-, out-) degree distributions. 
For each network (or class of networks) data are presented in more or less historical order, so that the recent exciting progress is visible. 
Errors are not shown (see the caption of Fig. \protect\ref{f20}). They depend on the size of a network and on the value of $\gamma$. We recommend our readers to look at the remark at the end of Sec. \protect\ref{sss-www} before using these values.
\ \ \ 
$^1$The data for the network of operating AS was obtained for one of days in December 1999. 
\ \ 
$^2$The value of the $\gamma$ exponent was estimated from the degree distribution plot in Ref. \protect\cite{gt00c}. 
\ \
$^3$The network of protein-protein interaction is treated as undirected.  
\ \ 
$^4$The value of the $\gamma$ exponent for the word web is given for the range of degrees below the crossover point (see Fig. \protect\ref{f6a}). 
\ \ 
$^5$The out-degree distribution of the telephone call graph cannot be fitted by a power-law dependence (notice the remark in Sec. \protect\ref{ss-other}). 
\ \ 
$^6$In fact, the data was collected from a small set of vertices of the web of human sexual contacts. These vertices almost surely have no connections between them.  
\ \
$^7$These food webs are truly small. In Refs. \protect\cite{cga01,cga01b} degree distributions of such food webs were interpreted as exponential-like.
}
\label{t1}
\end{table}

\hfill\rule[-1.5mm]{0.1mm}{1.5mm}\rule{20.5pc}{0.1mm}
\begin{multicols}{2}
\narrowtext


$\phantom{x}$
\newpage

\end{multicols}
\widetext
\noindent\rule{20.5pc}{0.1mm}\rule{0.1mm}{1.5mm}\hfill


\hfill\rule[-1.5mm]{0.1mm}{1.5mm}\rule{20.5pc}{0.1mm}
\begin{multicols}{2}
\narrowtext

$\phantom{x}$
\newpage  


\subsection{Continuum approach}\label{ss-continuous}

As we have already seen in Sec. \ref{ss-idea}, the continuum approximation produces the exact value of $\gamma$ for the BA model. 
The first results for the exponents \cite{ba99} were obtained just using this approximation (in Refs. \cite{ba99,baj99} it was called ``mean field''). Such an approach gives the exact values of the exponents for numerous models of growing scale-free networks and allows us to describe easily main features of the network growth \cite{dm002,dm00e}. 

Let us briefly describe this simple technique. Passing to the continuum limits of $k$ and $t$ in any of written above master equations for the degree distributions of individual vertices (e.g., in Eq. (\ref{6-1}) for the exponential network or Eq. (\ref{7-1}) for the BA model) we get the linear partial differential equations of the first order which have the following solution

\begin{equation}
p(k,s,t) = \delta(k - \overline{k}(s,t))
\, .
\label{13-1}
\end{equation} 
Of course, the form of this solution is rather far from the solutions of the corresponding exact master equations. Nevertheless, this $\delta$-function ansatz works effectively both for exponential and scale-free networks \cite{dm002,dm00e}. 

One may even not use master equations but proceed in the following way. 
In the simplest example, the BA model with one vertex and one edge added at each time step, the ansatz (\ref{13-1}) immediately leads to the equation for the average degree of vertices:

\begin{equation}
\frac{\partial \overline{k}(s,t)}{\partial t} = \frac{\overline{k}(s,t)}{\int_0^t du\, \overline{k}(u,t)}
\, .  
\label{13-2}
\end{equation}
Equation (\ref{13-2}) also follows from the continuum limit of the master equation for $p(k,s,t)$ of this model, Eq. (\ref{7-1}). 
It has a simple meaning -- 
new edges are distributed among vertices proportionally to their degrees as it is fixed by the rule of preferential linking. 
The initial condition is $\overline{k}(0,0)=0$, 
and the boundary one, $\overline{k}(t,t)=1$.
One sees that Eq. (\ref{13-2}) is consistent. Indeed, applying  
$\int_0^t ds$ to Eq. (\ref{13-2}) we obtain 

\begin{equation}
\frac{\partial}{\partial t} \int_0^t ds\,\overline{k}(s,t) = 
\int_0^t ds\,\frac{\partial}{\partial t} \overline{k}(s,t) + \overline{k}(t,t) 
= 1 + 1
\, , 
\label{13-3}
\end{equation} 
from which the proper relation follows, 

\begin{equation}
\int_0^t ds\,\overline{k}(s,t) = 2t
\, , 
\label{13-4}
\end{equation}
that is, the total degree in this case equals double the number of edges. 
Therefore, Eq. (\ref{13-3}) takes the form  

\begin{equation}
\frac{\partial \overline{k}(s,t)}{\partial t} = 
\frac{1}{2} \frac{\overline{k}(s,t)}{t}
\, .  
\label{13-5}
\end{equation}
Its general solution is 

\begin{equation}
\overline{k}(s,t) = C(s) t^{1/2}
\, ,  
\label{13-6}
\end{equation}
where $C(s)$ is arbitrary function of $s$. 
Accounting for the boundary condition, $\overline{k}(t,t)=1$, one has 

\begin{equation}
\overline{k}(s,t) = \left(\frac{s}{t}\right)^{-1/2}
\, .  
\label{13-7}
\end{equation}
Hence, the scaling exponent $\beta$ equals $1/2$, as we have seen before. 

In the continuum approach, the expression for the total degree distribution is of the form 

\begin{eqnarray}
P(k,t) & = & \frac{1}{t}\int_0^t ds\, \delta(k-\overline{k}(s,t)) = 
\nonumber
\\[5pt] 
& &  -\frac{1}{t}\left( \frac{\partial \overline{k}(s,t)}{\partial s} \right)^{-1}\!\!\!\!\![s=s(k,t)]
\, ,  
\label{13-8}
\end{eqnarray}
where $s(k,t)$ is a solution of the equation, $k = \overline{k}(s,t)$.
Using Eq. (\ref{13-8}), one may immediately reproduce the scaling relation between the exponents, so $\gamma=1+1/\beta$. Therefore, in the present case, $\gamma=3$.


\subsection{More complex models and estimates for the WWW}\label{ss-estimations}

One may consider more complex growing networks \cite{dm00e,dm003}. 
We will demonstrate that scale-free nets may be obtained even without ``pure'' preferential linking. 
It is convenient to consider incoming edges here, so we 
use the following notation for in-degree, $q \equiv k_{i}$. 


\begin{figure}
\epsfxsize=85mm
\epsffile{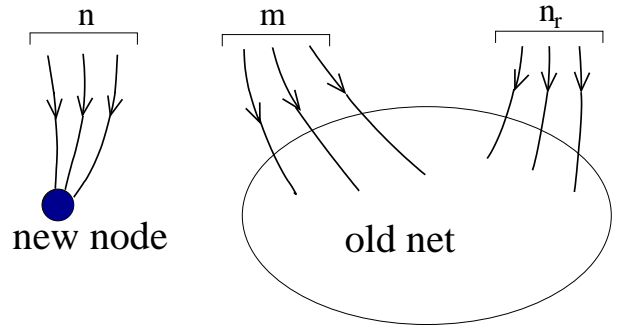}
\caption{
Scheme of the growth of the network with a mixture of the preferential and random linking 
(compare with the schematic Fig. \protect\ref{f4} for the WWW growth). 
At each time step, a new vertex with $n$ incoming edges is added. 
Simultaneously, the target ends of 
$m$ new edges are distributed among vertices according to a rule of preferential linking, and, in addition, the target ends of 
$n_r$ new edges are attached to randomly chosen vertices. The source ends of each edge may be anywhere.
}
\label{f21}
\end{figure}


Let us describe the model (see Fig. \ref{f21}): 

(i) At each time step, a new vertex is added. 

(ii) It has $n$ incoming edges which go out from arbitrary vertices or even from some external source. 

(iii) Simultaneously, $m$ extra edges are distributed with preference. This means that 
they go out from non-specified vertices or from an external source but a target end 
of each of them is attached to a vertex chosen preferentially -- probability to choose some particular vertex is proportional to 
$q+A$. $A$ is a constant which we call {\em additional attractiveness} (see Sec. \ref{ss-masterequation}). We shall see that its reasonable 
values are $A>-n-n_r$.  

(iv) In addition, at each time step, the target ends of $n_r$ edges are distributed among vertices randomly, without any preference. Again, these edges may go out from anywhere. 

In the continuum approach, one can assume that $m$ and $n$ are not necessarily integer numbers but are 
positive. 
Note that here we do not include into consideration the source ends of edges, since we are studying only in-degree distributions. 

The equation for the average in-degree of vertices in this network has the form, 

\begin{equation}
\frac{\partial \overline{q}(s,t)}{\partial t} = 
\frac{n_r}{t} + 
m \frac{\overline{q}(s,t)+A}{\int_0^t du [\overline{q}(u,t) + A]}
\,   
\label{15-1}
\end{equation}
with the initial condition, $\overline{q}(0,0)=0$, and the boundary one, $\overline{q}(t,t)=n$. The first term on the right-hand side accounts for linking without preference, the second one -- for the preferential linking. 
In this case, 
$\int_0^t ds\,\overline{q}(s,t) = (n_r+m+n)t$. 
It follows from Eq. (\ref{15-1}) that $\beta = m/(m+n_r+n+A)$, so $0 < \beta < 1$, and 

\begin{equation}
\gamma_i = 2 + \frac{n_r+n+A}{m}
\, .  
\label{15-2} 
\end{equation} 
Thus, the additional fraction of randomly distributed edges does not suppress the power-law dependence of the degree distributions but only increases $\gamma_i$ which is in the range between $2$ and infinity.  

This model allows one to obtain some estimates for the exponents of in- and out-degree distributions of the WWW \cite{be00,dms002}. Let us discuss, first, the in-degree distribution. We have already explained how new pages appear in the Web (see Sec. \ref{sss-www}). 
The introduced model, at least, resembles this process.
The problem is that we do not know the values of the quantities on the left-hand side of Eq. (\ref{15-2}). 

The constant $A$ may take {\em any} values between $-(n_r+n)$ and infinity, the number of the randomly distributed edges, $n_r$, in principle, may be not small (there exist many individuals making their references practically at random), and $n$ is not fixed. From the experimental data \cite{bkm00} (see Sec. \ref{sss-www}) we know more or less the sum $m+n+n_r \sim 10 \gg 1$ (between $7$ and $10$, more precisely), and that is all.   

The only thing we can do, is to fix the scales of the quantities. The natural characteristic values for $n_r+n+A$ in Eq. (\ref{15-2}) are (a) $0$, (b) $1$, (c) $m \gg 1$, and (d) infinity. In the first case, 
all new edges are attached to the oldest vertex since only this one is attractive for linking, and $\gamma_i \to 2$. In the last case, there is no preferential linking, and the network is not scale-free, $\gamma_i \to \infty$. Let us consider the truly important cases (b) and (c). 

(b) \ Let us assume that the process of the appearance of each document in the Web is as simple as the procedure of the creation of your personal home page described in Sec. \ref{sss-www}. 
If only one reference to the new document $(n=1)$ appears, and if one forgets about the terms $n_r$ and $A$ in Eq. (\ref{15-2}), 
than, for the $\gamma_{i}$ exponent of the in-degree distribution, we immediately get the estimate $\gamma_{i}-2 \sim 1/m \sim 10^{-1}$. 
This estimate indeed coincides with the experimental value $\gamma_{i}-2=0.1$ \cite{bkm00} (see Sec. \ref{sss-www}). 
Therefore, the estimation looks good. 
Nevertheless, we should repeat, that this estimate follows only from the fixation of the scales of the involved quantities, 
and many real processes are not accounted for in it. 

(c) \ Above we discussed the distribution of incoming links. Eq. (\ref{15-2}) may also be applied for the distribution of links which go out from documents of the Web, since the model of the previous section can easily be reformulated for outgoing edges of vertices.  
In this case all the quantities in Eq. (\ref{15-2}) take other values which are again unknown. However, we can estimate them.
As we explained in Sec. \ref{sss-www}, there are usually several citations ($n$) in each new WWW document.  
In addition, one may think that the number of the links distributing without any preference, $n_r$, is not small now. Indeed, even beginners proceed by linking of their pages. 
Hence, $n+n_r \sim m$ --- we have no other available scale, --- and $\gamma_{o}-2 \sim m/m \sim 1$. 
We can compare this estimate with the experimental value, $\gamma_{o}-2=0.7$ \cite{ajb99,bkm00}. 

Unfortunately, numerous channels of linking make similar contributions to the values of the exponents of the degree-distributions, so quite ``honest'' estimates are impossible. Let us introduce the ``general'' model of a growing directed network. 
In this model we account for the main channels of linking which yield contributions of the same order to $\gamma_{i}$ and $\gamma_{o}$. 
This will demonstrate the complexity of the problem. 


\begin{figure}
\epsfxsize=90mm
\epsffile{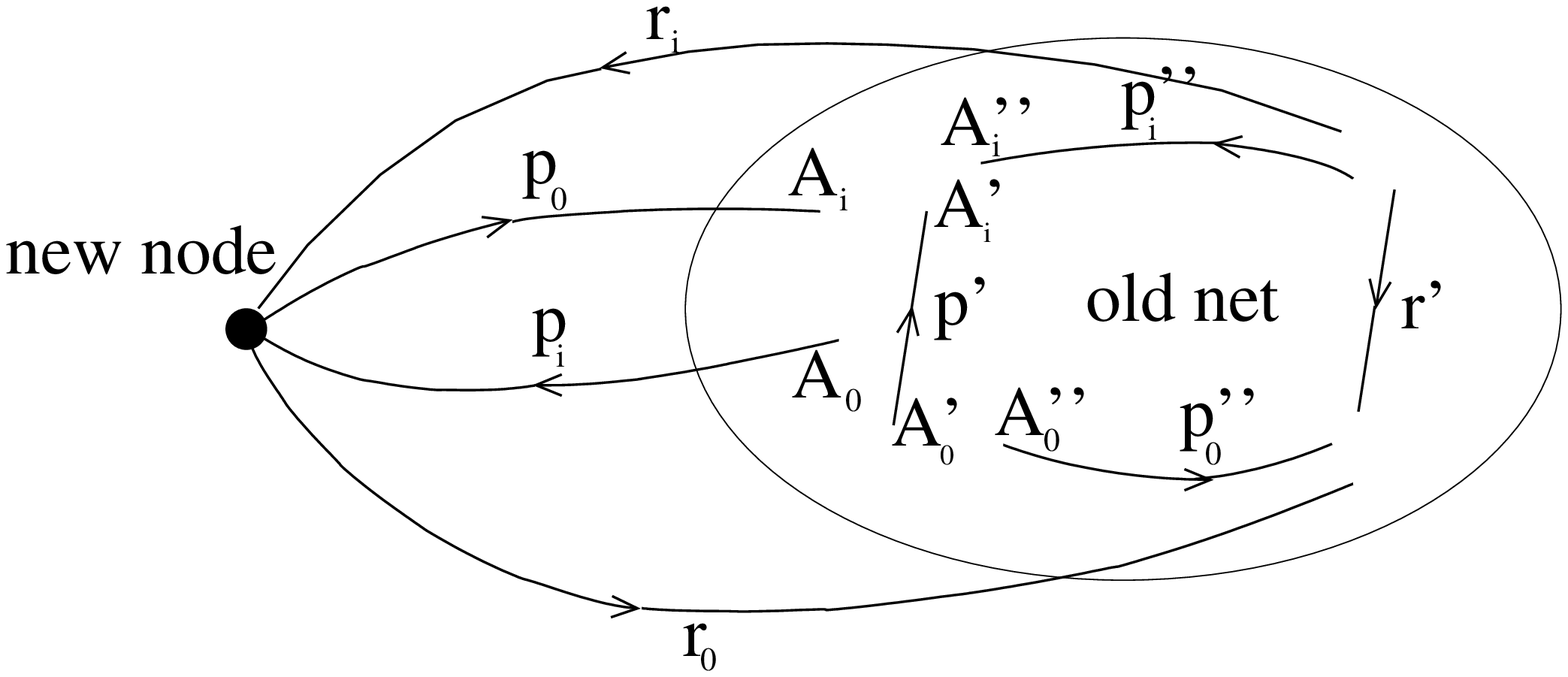}
\caption{
Scheme of the growth of the generalized model of a directed network.    
\ \ \ 
(i) At each time step, a new vertex is added. 
\ \ 
(ii) It has both outgoing and incoming edges. 
\ \ 
The target ends of $r_o$ outgoing edges are distributed randomly among old vertices. 
\ \ 
The source ends of $r_i$ incoming edges are distributed randomly among old vertices. 
\ \ 
The target ends of $p_o$ outgoing edges are distributed preferentially among old vertices 
with probability proportional to $k_{i}+A_i$. 
\ \ 
The source ends of $p_i$ incoming edges are distributed preferentially among old vertices 
with probability proportional to $k_{o}+A_o$. 
\ \ 
(iii) Simultaneously, $p^\prime$ edges are distributed preferentially between old vertices. 
\ \ 
Their target ends are distributed with the preference function, $k_{i}+A_i^\prime$ and 
their source ends -- with the preference function $k_{o}+A_o^\prime$. 
\ \ 
(iv) In addition, $r^\prime$ edges are distributed without preference among old vertices. 
\ \  
(v) In addition, $p_i^{\prime\prime}$ connections appear between old vertices 
with source ends being distributed without preference and with target ends -- 
with the preference function $k_{i}+A_i^{\prime\prime}$. 
\ \ 
Finally, $p_o^{\prime\prime}$ edges emerge between old vertices with 
target ends being distributed without preference and with source ones -- 
with the preference function $k_{i}+A_o^{\prime\prime}$. 
\ \  
Here $A_i$, $A_o$, $A_i^\prime$, $A_o^\prime$, $A_i^{\prime\prime}$, and $A_o^{\prime\prime}$ are constants.
\ \ 
The total number of connections that emerge at each increment of time is 
$n_t = p_o + p_i + r_o + r_i + r^\prime + p^\prime + p_i^{\prime\prime} + p_o^{\prime\prime}$. 
}
\label{f22}
\end{figure}


The number of possible channels is so large that we have to introduce new notations. 
The network grows by the rules described in the caption of Fig. \ref{f22}. We account for all combinations of linking without preference and linear preferential linking. Some new edges appear between new and old vertices, other connect pairs of old vertices. 
For different channels of linking, parameters of preferential linking differ from each 
other. Additional attractiveness takes different values for target and source ends of preferentially distributed edges. 
For brevity, we use the simplest preference functions of the form $k_{i}+A_{i}$ for distribution of target ends of links and of the form $k_{o}+A_{o}$ for distribution of source ends.  
In fact, this model generalizes the known models of networks with preferential linking of directed edges \cite{dms001,dm00e,krr00a}.

The above growing network is scale-free. Its exponents may be obtained in the continuum approach framework. 
Fortunately, part of parameters introduced in Fig. \ref{f22}, disappear from the final expressions for $\gamma_{i}$ and $\gamma_{o}$: 

\begin{eqnarray}
\gamma_{i} & = & 
1 + \left[ \frac{p_o}{n_t + A_i} + 
\frac{p^\prime}{n_t + A_i^\prime} + 
\frac{p_i^{\prime\prime}}{n_t + A_i^{\prime\prime}} \right]^{-1} 
, 
\nonumber 
\\[5pt]
\gamma_{o} & = & 
1 + \left[ \frac{p_i}{n_t + A_o} + 
\frac{p^\prime}{n_t + A_o^\prime} + 
\frac{p_o^{\prime\prime}}{n_t + A_o^{\prime\prime}} \right]^{-1}
\, .
\label{15-3}
\end{eqnarray} 
In principle, one must account for all above contributions.  
One may check that Eq. (\ref{15-2}) is a particular case of Eq. (\ref{15-3}). 
Twelve unknown parameters ($n_t,p_o,p_i,p^\prime,p_i^{\prime\prime},p_o^{\prime\prime},A_i,A_o,A_i^\prime,A_o^\prime,A_i^{\prime\prime},A_o^{\prime\prime}$) in Eq. (\ref{15-3}) make the problem of improving of the estimate of $\gamma_{i,o}$ (see (b) and (c)) hardly solvable. 


\begin{figure}
\epsfxsize=90mm
\epsffile{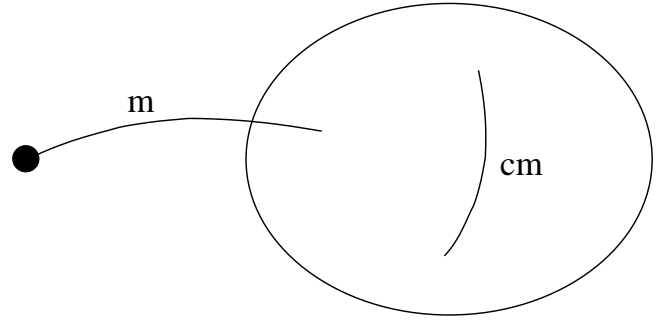}
\caption{
Scheme of growth of an undirected network with creation of connections between already existing vertices. At each time step, (i) a new vertex is added; (ii) it connects to $m$ preferentially chosen old vertices; (iii) $cm$ new edges connect pairs of preferentially chosen old vertices.
}
\label{f22a}
\end{figure}


A simpler case of a growing undirected network was considered in Ref. \cite{dm003}. At each time step, apart of $m$ new edges between a new vertex and old vertices, $mc$ new edges are created between the old vertices (see Fig. \ref{f22a}), so that the average degree of the network is $\overline{k}=2m(1+c)$. The connections to a new vertex are distributed among old vertices like in Barab\'asi-Albert model. The probability that a new edge is attached to existing vertices 
of degree $k^{(\mu)}$ and $k^{(\nu)}$ is proportional to $k^{(\mu)} k^{(\nu)}$. Here $\mu$ and $\nu$ are labels of the vertices. The resulting degree distribution is of a power-law form with the exponent 
 
\begin{equation}
\gamma = 2 + \frac{1}{1+2c} = 2 + \frac{m}{\overline{k}-m}
\, .  
\label{15-3a} 
\end{equation} 
Thus, $2<\gamma<3$. 
The same expression is valid if, at each time step, we delete $-mc>0$ randomly chosen edges (here $c<0$).  


\begin{figure}
\epsfxsize=90mm
\epsffile{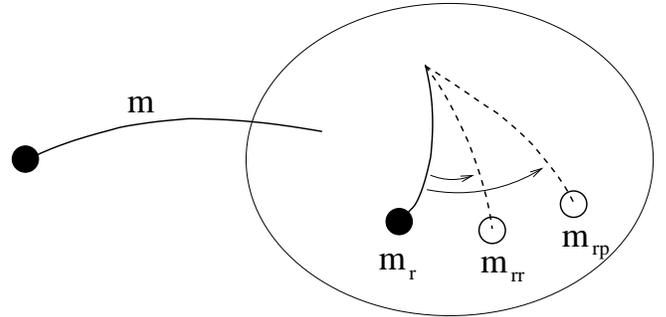}
\caption{
Scheme of the growth of an undirected network with the rewiring of connections in the old part of the network. At each time step, 
(i) a new vertex is added; 
(ii) it connects to $m$ preferentially chosen old vertices; 
(iii) $m_r$ old vertices are chosen at random, and, from each of these vertices,  one of edges is rewired to another vertex. In the $m_{rr}$ cases, the rewiring occurs to randomly chosen vertices. In the rest $m_{rp}=m_r-m_{rr}$ of cases, the rewired edge ends are attached to preferentially chosen vertices. 
}
\label{f22b}
\end{figure}


A very similar effect produces a rewiring of edges \cite{ab00a}. 
Now, instead of the creation of connections in the old part of an undirected growing network, 
at each time step, let each of $m_r$ randomly chosen vertices loose one of its connections (see Fig. \ref{f22b}). In $m_{rr}$ cases, a free end is attached to a random vertex. In the rest $m_{rp}=m_r-m_{rr}$ cases, a free end is attached to a preferentially chosen vertex.  
The continuum equation for the mean degree has the form 

\begin{equation}
\frac{\partial \overline{k}(s,t)}{\partial t} = 
(m+m_{rp}) \frac{\overline{k}(s,t)}{\int_0^t du \overline{k}(u,t)} 
+ \frac{m_{rr}-m_r}{t}
\,   
\label{15-3b}
\end{equation}   
with the boundary condition $\overline{k}(t,t)=m$. 
From this, one gets the following expression for the exponent of the degree distribution: 
 
\begin{equation}
\gamma = 2 + \frac{m-m_{rp}}{m+m_{rp}}
\, .  
\label{15-3c} 
\end{equation} 
Notice that here we did not account for the emergence of bare vertices, so the number of the rewirings $m_r$ has to be small enough. From Eq. (\ref{15-3c}), it follows that, as the number of the preferential rewirings grows, the $\gamma$ exponent decreases. Moreover, simulations in Ref. \cite{ab00a} demonstrated that, when the number of the rewirings is high enough, the degree distribution changes from a power law to an exponential one. 
  
One sees that the power-law in- and out-degree distributions arise from the power-law singularities $\overline{k}_{i}(s,t) \propto (s/t)^{-\beta_{i}}$ and 
$\overline{k}_{o}(s,t) \propto (s/t)^{-\beta_{o}}$ at the point $s=0$ (the oldest vertex). 
Therefore, the same vertices, as a rule, have the high values of both in- and out-degree.  
This means that the in- and out-degree of vertices {\em correlate}, and, of course, 
$P(k_{i},k_{o}) \neq P(k_{i})P(k_{o})$ (see discussion in the paper of Krapivsky, Rodgers, and Redner \cite{krr00a}). 
Moreover, even if we exclude the preferential linking from such network growth process, the rule ``the oldest is the richest'' is still valid for both in- and -out degree, and hence the correlation between $k_{i}$ and $k_{o}$ is again present.  

In Ref. \cite{krr00a}, the distribution $P(k_{i},k_{o})$ was analytically calculated for a model of this type. To get the exact result, the authors of this paper accounted for only two channels of the preferential attachment of new edges and made a number of simplifying assumptions. In their model, 
(i) a new edge may go out of a new vertex and, in this case, its target end is attached to some old vertex chosen with the probability proportional to $k_{i}+A_i$. (ii) Another possibility is connection of two old vertices $(\mu)$ and $(\nu)$ with the probability proportional to 
$(k_{i}^{(\mu)}+A_i^\prime)(k_{o}^{(\nu)}+A_o^\prime)$ (in Ref. \cite{krr00a}, $A_i^\prime=A_i$). Here
, $k_{i,o}^{(\mu)}$ and $k_{i,o}^{(\nu)}$ are the in- and out-degrees of these vertices. In addition, parameters of the model \cite{krr00a} are chosen in such a way that the exponents of the in- and out-degree distributions are equal, $\gamma_{i}=\gamma_{o}$. The resulting distribution has the following asymptotic form for large $k_{i}$ and $k_{o}$, 

\begin{equation}
P(k_{i},k_{o}) \propto 
\frac{k_{i}^{A_i-1}k_{\,o}^{A_o^\prime}}{(k_{i}+k_{o})^{2A_i+1}}
\, ,   
\label{15-4} 
\end{equation} 
which is very different from the product $P(k_{i})P(k_{o})$.

A model of growing directed networks with preferential linking was simulated in the paper \cite{t00b}. 
In- and out-degree distributions were observed to be of power-law form. The distribution of the sizes of connected clusters may be also interpreted as a power-law dependence in some range of the parameters of this model.


\subsection{Types of preference providing scale-free networks}\label{ss-types}

Many efforts were made to analyse different preference functions producing scale-free networks. 
The power-law preference function, $k^y$, does not produce power-law degree distributions if $y \neq 1$, see Sec. \ref{s-non-scale-free}. 
One can check that the necessary condition is a linear asymptotic form of the preference function at large values of degree \cite{krl00,kr00c}, so the function, in principle, may be nonlinear.  Nevertheless, main features can be understood if one consider linear preference functions. 
In general, the probability for a new link to be attached to a vertex $s$ at time $t$ is 
$p(s,t) = G(s,t)k(s,t) + A(s,t)$. The coefficient $G(s,t)$ may be called {\em fitness} of a vertex \cite{bb00b,bb00a}  $A(s,t)$ is {\em additional attractiveness}. 
As we have seen, $A$ can change the values of the exponents. Effect of the variation of $G$ may be even stronger. 

One can consider the following particular cases:

(i) $G=const$, $A=A(s)$. 
In this case, the additional attractiveness $A(s)$ may be treated as ascribed to individual vertices. 
A possible generalization is to make it a random quantity. One can check that the answers do not change crucially -- one only has to substitute the average value, $\overline{A}$, instead of $A$, into the previous 
expressions for the scaling exponents.

Note that $n$ and $m$ may also be made random, and this can be accounted for by the substitution of $\overline{n}$ and $\overline{m}$ into the expressions for the exponents. 

There exists a more interesting possibility -- to construct a direct generalization of the network considered in Sec. \ref{ss-estimations} where combination of the preferential and random linking was described.  
For this, we may ascribe the additional attractiveness not to vertices but to new edges and again make it a random quantity. In such an event, new edges play the role of fans with different passion for popularity of their idols, vertices. This is the case (ii), $G=const$, $A=A(t)$, where $A(t)$ is random. If the distribution function of $A$ is $P(A)$, the $\gamma$ exponent equals 

\begin{equation}
\gamma = 1 + 
\left[ 
\int \frac{dA\,P(A)}{1 + (\overline{n} + A)/\overline{m}} 
\right]^{-1}
\, ,  
\label{17-1} 
\end{equation} 
see Ref. \cite{dm00e}. The values of the exponent are again between $2$ and $\infty$. 

Let us pass to situations where $A=const$. 

(iii) $G = G(t)$. This case reduces to case (ii). 

\begin{figure}
\epsfxsize=78mm
\centerline{\epsffile{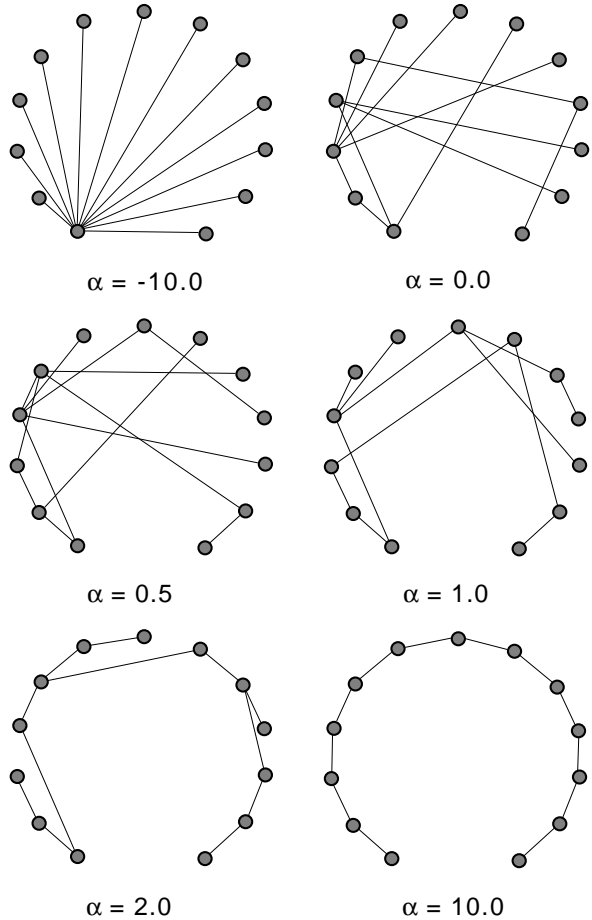}}
\caption{
Change of the structure of the network with aging of vertices with increase of the aging exponent $\alpha$. The aging is proportional to $\tau^{-\alpha}$, where $\tau$ is the age of a vertex. The network grows clockwise starting from the vertex below on the left. At each time step, a new vertex with one edge is added. 
}
\label{f23}
\end{figure}

(iv) $G(s,t) = f(t-s)$, aging of vertices. This case was considered in Refs. \cite{dm002,dm00e}. In particular, such a form of a preference function is quite reasonable in citation networks. Indeed, we rarely cite old papers. One may check that to keep the network scale-free, the function has to be of a power-law form, $G(s,t) = (t-s)^{-\alpha}$. In principle, the exponent $\alpha$ may be of any sign: $-\infty < \alpha < \infty$. 
Negative values of $\alpha$ are typical for very conservative citation networks (many references to Bible). 
Variation of the aging exponent $\alpha$ produces quite distinct networks, 
see Fig. \ref{f23}. If $\alpha$ is negative, links tends to be attached to the oldest vertices, 
if $\alpha$ is large, the network becomes a chain structure. 


\begin{figure}
\epsfxsize=70mm
\epsffile{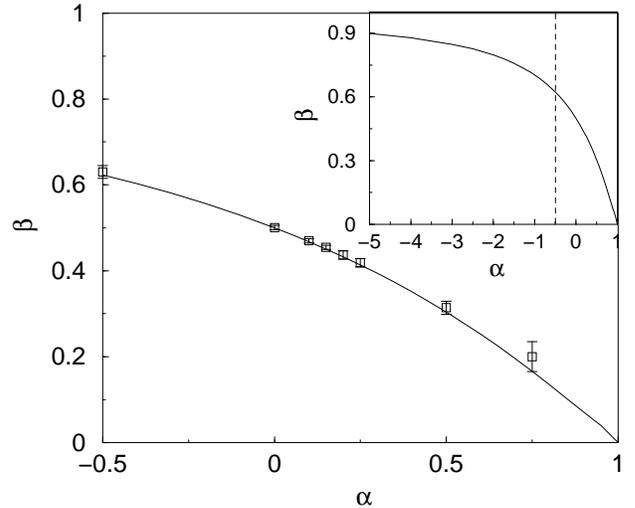}
\caption{
$\beta$ exponent of the average degree vs. the aging exponent $\alpha$ of the network with aging of vertices. 
The points are obtained from the simulations \protect\cite{dm002}. 
The line is the result of the calculations. 
The inset shows the analytical solution in the range $-5 < \alpha < 1$. 
Note that $\beta \to 1$ when $\alpha \to -\infty$.
}
\label{f24}
\end{figure}

\begin{figure}
\epsfxsize=67mm
\epsffile{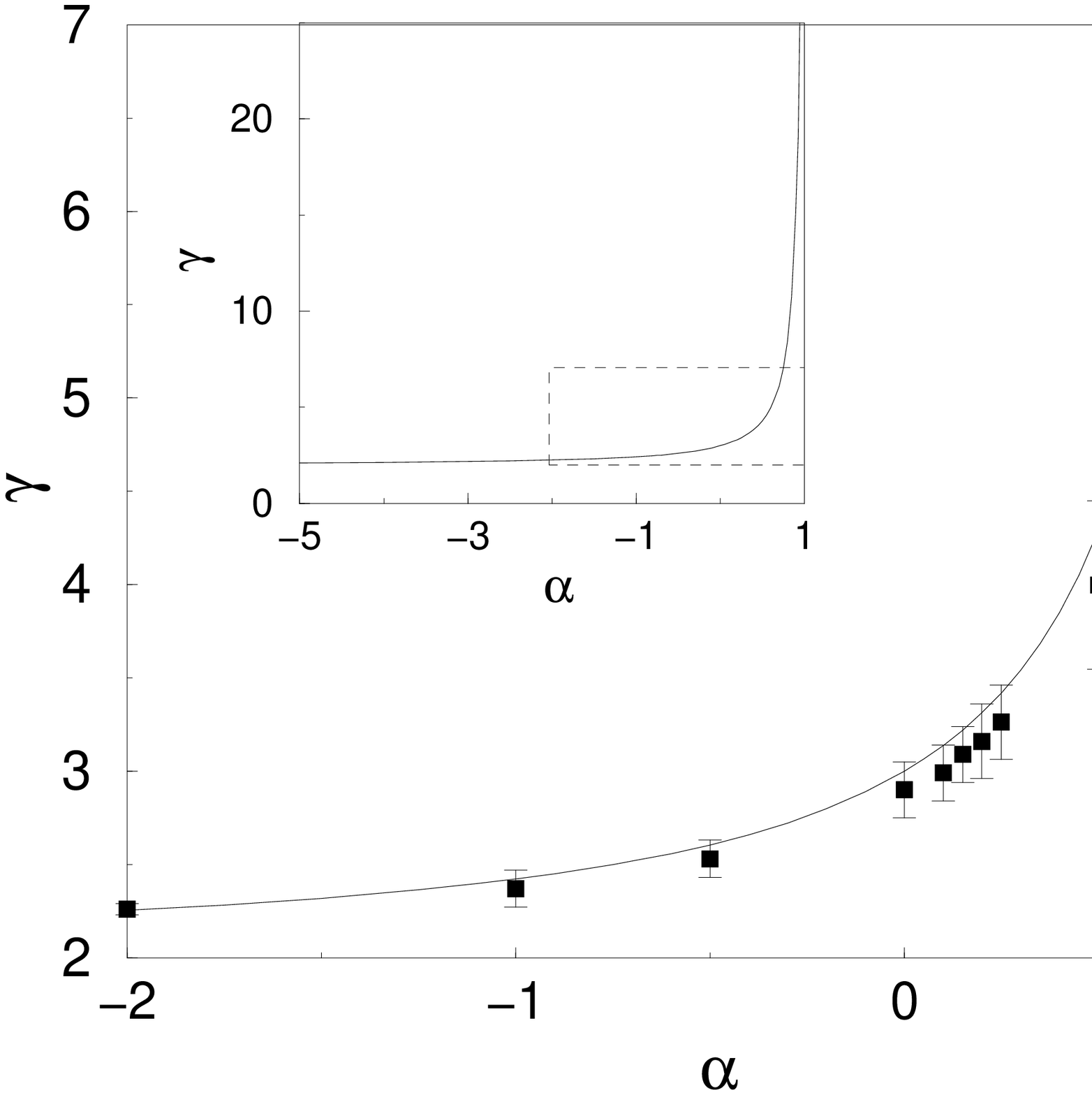}
\caption{
$\gamma$ exponent of the average degree vs. the aging exponent $\alpha$ of the network with aging of vertices. 
The points show the results of simulations \protect\cite{dm002}. 
The line is the analytical result. The inset depicts the analytical solution in the range $-5 < \alpha < 1$. 
}
\label{f25}
\end{figure}

Again it is possible to use the continuum approach. 
For the undirected network to which one vertex with one edge is added at each time step, after the introduction of the scaling variables, 
$\kappa(s/t) \equiv \overline{k}(s,t)$ and  $\xi \equiv  s/t$, 
one gets 
 
\begin{eqnarray}
-\xi (1-\xi)^\alpha \frac{d \ln \kappa(\xi)}{d \xi}  & = & 
\left[ \int_0^1 d\zeta \kappa(\zeta) (1-\zeta)^{-\alpha} \right]^{-1} = \beta 
, 
\nonumber \\[3pt]
\kappa(1) = 1
\, .
\label{17-2}
\end{eqnarray} 
From Eq. (\ref{17-2}), one obtains the solution $\kappa(\xi,\beta)$. Substituting it into the right equality in Eq. (\ref{17-2}) or, equivalently, into $\int_0^1 d\zeta \kappa(\zeta) = 2$, we get a transcendental equation for $\beta$. The resulting exponents, $0<\beta<1$ and 
$2<\gamma<\infty$, are shown in Figs. \ref{f24} and \ref{f25} \cite{dm002,dm00e}. 


(v) $G = G(s)$, a fluctuating function \cite{bb00a}.  

Let $G$ be a random variable having distribution $P(G)$. 
Then, in the particular case of the BA model, one gets the following equation for average degree 

\begin{equation}
\frac{\partial\overline{k}(s,t)}{\partial t} = 
\frac{G(s)\overline{k}(s,t)}{\int^t_0 ds\,G(s)\overline{k}(s,t)} = 
\beta(G(s))\frac{\overline{k}(s,t)}{t}  
\,   
\label{17-3} 
\end{equation} 
with the boundary condition $\overline{k}(t,t)=1$ (we set $m=1$ without loss of generality). 
In this case, $\overline{k}(s,t)=(s/t)^{-\beta(G(s))}$. Then, it is easy to check that $\beta(G) = cG$, where the constant $c$ can be obtained from the equation 

\begin{equation}
\int dG\,P(G) \frac{G}{1-cG} = 1/c  
\, .  
\label{17-4} 
\end{equation} 
Using Eq. (\ref{13-8}), one gets the distribution

\begin{equation}
P(k) \propto \int dG\,P(G) \left(1+\frac{1}{cG} \right) k^{-[1+1/(cG)]} 
\, .  
\label{17-5} 
\end{equation} 
If $P(G)=\delta(G-G_0)$, the network is the original BA model. 
If $P(G)$ is a distributed function, the answer changes. For instance, 
for $P(G)=\theta(G)\theta(1-G)$, i.e., when $G$ is homogeneously distributed in the range $(0,1)$, the distribution takes the following form: 

\begin{equation}
P(k) \propto \int_0^1 dG \left(1+\frac{1}{cG} \right) k^{-[1+1/(cG)]}  
\propto 
\frac{k^{-(1+1/c)}}{\ln k}
\, .  
\label{17-6} 
\end{equation}  
Here, constant $c$, which is the solution of Eq. (\ref{17-4}) with homogeneous distribution $P(G)$, equals $0.797\ldots$, so $\gamma=2.255\ldots$, i.e., it is smaller than the value $\gamma=3$ for the homogeneous BA model. 

We emphasize that $\gamma$ depends on a form of the distribution $P(G)$. 
In particular, results obtained with distribution $P(G)$, which consists of two delta-functions, are discussed in Sec. \ref{ss-capture}. 
The fluctuations of $G$ may also be introduced into models of growing networks from 
Secs. \ref{ss-masterequation}, \ref{ss-continuous}, and \ref{ss-estimations}. 
Results are similar to Eq. (\ref{17-6}).

A combination of fluctuating additional attractiveness $A(s)$ and fluctuating fitness $G(s)$ was considered in Ref. \cite{er01}. Calculations in this paper are very similar to the above derivation (an explicit rate-equation approach \cite{krl00,kr00c,krr00a} was used), but the result, namely the $\gamma$ exponent of the power-law dependence, corrected by a logarithmic denominator, 
depends on the form of the joint distribution $P(A,G)$. In Ref. \cite{er01} one may also find the results for the exponents of in- and out-degree distributions of a directed growing network with fluctuating fitness.  

One should note that most of the existing models of networks growing under mechanism of preferential linking produce the effect 
``the oldest is the richest''. (Here we do not dwell on situations when old vertices may die, divide into parts, or stop to attach new edges. The last possibility was studied in Refs. \cite{ke01b,ke01c}, and this is the case, where young vertices may have larger degrees than old vertices. Also, if vertices may divide into parts, it is hard to define the age of a vertex at large temporal scales.)  
Even in the case of fluctuating fitness $G$, older vertices are, with high probability, of larger degree than young vertices. 
Indeed, the power-law degree distributions have to be accompanied by strong singularities of the average degree of individual vertices 
$\overline{k}(s,t)$ at $s=0$. 
This follows from the derivations of scale-free degree distributions in the present section. The fluctuations of $G$ produce broadening of 
the degree distribution of individual vertices $p(k,s,t)$. 
In Sec. \ref{ss-capture} we will consider the situation in which the in-homogeneity of 
$G$ provides stronger effect than considered here.

It was stated in Ref. \cite{ah00a} that degree distribution of individual vertices (sites) of the Web practically does not depend on their age. This indicates inapplicability of the preferential linking concept. Authors of Ref. \cite{bajb00a} explained that these data are not sufficient to exclude the rule ``the oldest is the richest'' and that just the inhomogeneity of fitness $G$ hampers the observation of such an effect.


\subsection{``Condensation'' of edges}\label{ss-capture}

In the last of above situations, that is, in the case of inhomogeneous fitness $G$, the form of resulting degree distributions crucially depends on the form of the distribution $P(G)$. 
For some special forms of $P(G)$, a striking phenomenon occurs \cite{bb00b}. One or several the ``strongest'' vertices with the largest $G$ may capture a finite fraction of all edges. 
A related effect was considered in Ref. \cite{mh00}.
In Ref. \cite{bb00b} this intriguing effect was called ``Bose-Einstein condensation''. One can explain the essence of this phenomenon using a simple 
example \cite{dm00e}. 

Let us use the model of a growing network with directed edges introduced in Sec. \ref{ss-continuous}. To simplify the formulas, we set $A=0$ (one can see that this does not reduce the generality of the model which produces scaling exponents in the wide ranges of values, $2<\gamma<\infty$ and $0<\beta<1$). Let the rule of preference be the same as in Sec. \ref{ss-continuous}, 
i.e., the probability that an edge is attached to vertex $s$ is proportional to the in-degree $q_s$ of the vertex but with one exception --- one vertex, $\tilde{s}$, is ``stronger'' than others. This means that the   
probability that this vertex attracts an edge is higher. It has an additional factor, $g>1$, and proportional to $g q_{\tilde{s}}$. This means that 
$G_s = 1 + (g-1)\delta_{s,\tilde{s}}$.      
The equations for the average in-degree are 

\begin{eqnarray}
& & 
\frac{\partial \overline{q}_{\tilde{s}}(t)}{\partial t} = m \frac{g \overline{q}_{\tilde{s}}(t)}{(g-1) \overline{q}_{\tilde{s}}(t) +\int_0^t ds\,\overline{q}(s,t)} \, , \ \overline{q}_{\tilde{s}}(t=\tilde{s}) = q_i\, ,
\nonumber 
\\[5pt] 
& & 
\frac{\partial \overline{q}(s,t)}{\partial t} = 
m \frac{\overline{q}(s,t)}{(g-1) \overline{q}_{\tilde{s}}(t) +\int_0^t ds\,\overline{q}(s,t)}\, , \
\overline{q}(t,t) = n 
\, .
\label{19-1}
\end{eqnarray}
In the second of Eqs. (\ref{19-1}), $s \neq \tilde{s}$. Obviously, at long times, the total in-degree of the network is 
$\int_0^t ds\,\overline{q}(s,t) = (m+n)t + {\cal O}(1)$. 

At $t \gg \tilde{s}$, two situations are possible. In the first one, the in-degree $q_{\tilde{s}}(t)$ of the strongest vertex grows slower than $t$, and, at long times, the denominators are equal to $(m+n)t$, so that we get the exponents, $\beta = m/(m+n) \equiv \beta_0$ and 
$\gamma = 2 + n/m \equiv \gamma_0 = 1 + 1/\beta_0$, were $0<\beta_0<1, \ 2<\gamma_0<\infty$. 
Here, we introduce the exponents, $\gamma_0$ and $\beta_0$, of the network in which all vertices have equal ``strength'' (fitness), $g=1$.
The first line of Eq. (\ref{19-1}), in this case, looks as 
 
\begin{equation}
\frac{\partial \overline{q}_{\tilde{s}}(t)}{\partial t} = 
\frac{g m}{m+n} \frac{g \overline{q}_{\tilde{s}}(t)}{t}
\, .
\label{19-2}
\end{equation}
Hence, at long times, 
$\overline{q}_{\tilde{s}}(t) = const(q_i)\, t^{gm/(m+n)}$, and we see that the in-degree of the strong vertex  
does grow slower than $t$ only for
 
\begin{equation}
g < g_c \equiv 1 + \frac{n}{m} = \gamma_0 - 1 = \beta_0^{-1} > 1 
\, ,
\label{19-3}
\end{equation} 
so we obtain the natural threshold value. 

In the other situation, $g>g_c$, at long times, we have the only possibility, 
$q_{\tilde{s}}(t) = d\ t$, $d$ is some constant, $d < m+n$, since a more rapid  growth of $q_{\tilde{s}}(t)$ is impossible in principle. This means that, for $g>g_c$, a finite fraction of all preferentially distributed edges is 
captured by the strong vertex (in Ref. \cite{bb00b} just this situation is called the Bose-Einstein condensation). We see that a single strong vertex may produce a macroscopic effect.
In this case, Eq. (\ref{19-1}) takes the form, 

\begin{eqnarray}
& & 
\frac{\partial \overline{q}_{\tilde{s}}(t)}{\partial t} = 
\frac{gm}{(g-1)d + m + n}
\frac{\overline{q}_{\tilde{s}}(t)}{t}
\, ,
\nonumber 
\\[5pt] 
& & 
\frac{\partial \overline{q}(s,t)}{\partial t} = 
 \frac{m}{(g-1)d + m + n}
\frac{\overline{q}(s,t)}{t} 
\, ,
\label{19-4}
\end{eqnarray}  
where in the second of Eqs. (\ref{19-4}), $s \neq \tilde{s}$.  
Note that the coefficient in the first equation is always larger than the coefficient in the second equation, 
since $g_c>1$.  
From the first of Eqs. (\ref{19-4}), we get the condition 
 
\begin{equation}
\frac{gm}{(g-1)d + m + n} = 1 
\, ,
\label{19-4a}
\end{equation} 
so, for $g>g_c$, the following fraction of all edges in the network is captured by the strongest vertex: 
 
\begin{equation}
\frac{d}{m+n} = \frac{d}{m}\frac{m}{m+n} = \frac{1}{g_c}\frac{g-g_c}{g-1}
\, .
\label{19-5}
\end{equation} 
We have to emphasize that the resulting value of $d$ is independent on initial conditions! (Recall that we consider the long-time limit.) This ``condensation'' of edges on the ``strongest'' vertex leads to change of exponents. 
Using the condition Eq. (\ref{19-4a}), we readily get the following expressions for them, 
 
\begin{equation}
\beta = \frac{1}{g} < \beta_0 \, , \ \ \ \gamma = 1 + g > \gamma_0
\, .
\label{19-6}
\end{equation} 

\begin{figure}
\epsfxsize=63mm
\centerline{\epsffile{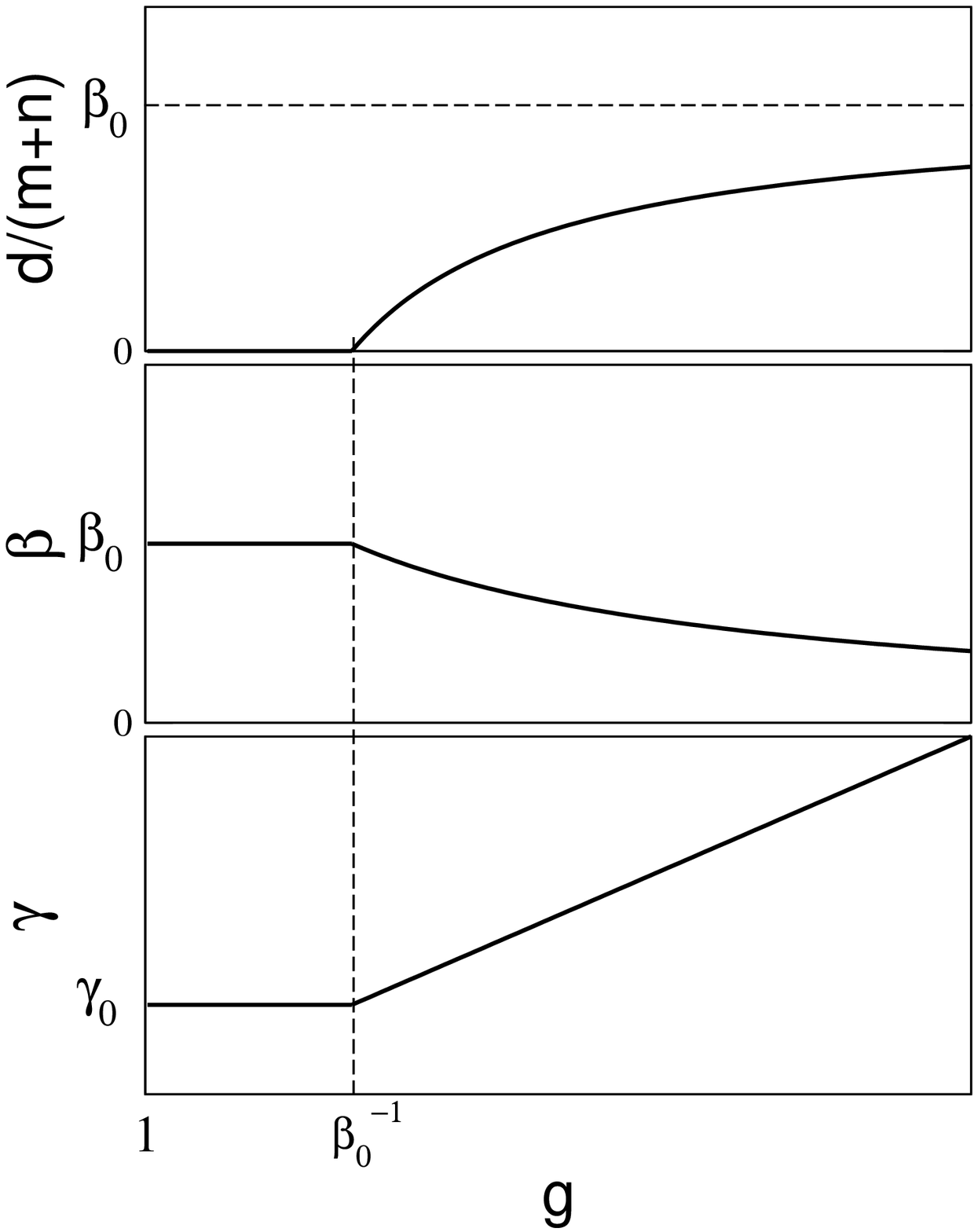}}
\caption{
``Condensation'' of edges. 
Fraction of all edges, $d/(m+n)$, captured by a single strong vertex at long times and the scaling 
exponents $\beta$ and $\gamma$ vs. relative fitness $g$ of the strong vertex \protect\cite{dm00e}. The network contains only one ``strong'' vertex. 
The condensation occurs above the threshold value $g_c = 1/\beta_0 = \gamma_0-1>1$. 
Here $\beta_0$ and $\gamma_0$ are the corresponding exponents for the network without a strong vertex. $d/(m+n)[g \to \infty] \to \beta_0$ and $\beta[g \to \infty] \to 0$. 
}
\label{f26}
\end{figure}

The fraction of all edges captured by the strongest vertex and the $\beta$ and $\gamma$ exponents vs. $g$ are shown in Fig. \ref{f26}.   
Note that the growth of $g$ increases the value of the $\gamma$ exponent. If the World is captured by Bill Gates or some czar, the distribution of wealth becomes more fair! 
One should note that the strong vertex does not take edges away from other vertices but only {\em intercepts} them.
The closer $\gamma_0$ is to $2$, the smaller $g$ is necessary to exceed the threshold. 
Above the threshold, the values of the exponents are determined only by the factor $g$. Nevertheless, the expression for $d/(m+n)$ contains $\gamma_0$, the exponent of the homogeneous network. (Recall that the threshold value is $g_c=\gamma_0-1)$. 

\vspace{37mm}$\phantom{x}$

\begin{figure}
\epsfxsize=85mm
\epsffile{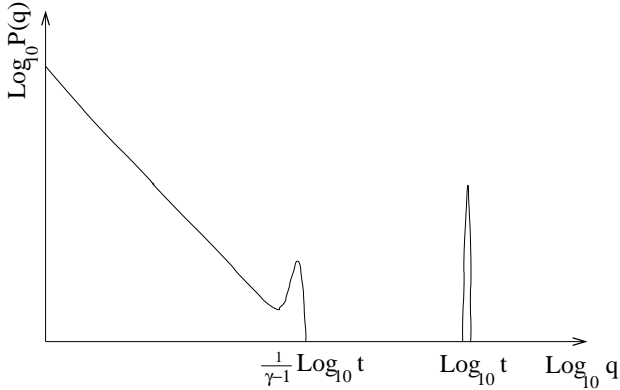}
\caption{
Schematic plot of the degree distribution of the network with one vertex, the fitness of which  
exceeds the threshold value \protect\cite{dm00e}. The peak is due to edges ``condensed'' on the strong vertex. A hump at the cutoff of the continuum part of the distribution is a trace of initial conditions (see Sec. 
\protect\ref{ss-relations} and Ref. \protect\cite{dms003}).
}
\label{f27}
\end{figure}


For $g>g_c$, in the edge condensation regime,  
the strongest vertex   
determines the evolution of the network. With increasing time, a gap between the in-degree of the strongest vertex and the maximal in-degree of all others grows (see Fig. \ref{f27}). A small peak at the end of the continuum part of the distribution is a trace of initial conditions, see Sec. \ref{ss-relations}. Note that the network remains scale-free even above the threshold, i.e., for $g>g_c$, although $\gamma$ grows with growing $g$. 

The above-described initial-condition-independent state is realized only in the limit of large networks. In the ``condensate phase'', relaxation to the final state is of a power-law kind \cite{dm00e}: 
 
\begin{equation}
\frac{\overline{q}_{\tilde{s}}(t) - d\,t}{(m+n)t} \propto t^{-(g-g_c)/g}
\, 
\label{19-7}
\end{equation} 
i.e.,  
the fraction of all edges captured by the strong vertex relaxes to the final 
value by a power law. Its exponent $(g-g_c)/g$ approaches zero 
at the  condensation point $g=g_c$. This behavior evokes strong associations with critical relaxation. 

\begin{figure}
\epsfxsize=65mm
\centerline{\epsffile{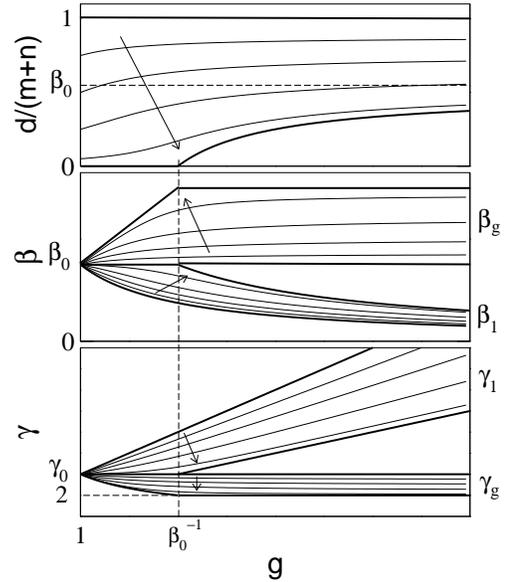}}
\caption{
Fraction of all edges, $d/(m+n)$, captured by the component of ``strong'' vertices, at long times and the scaling 
exponents $\beta$ and $\gamma$ vs. relative fitness, $g$, of ``strong'' vertices \protect\cite{dm002}. 
\ \ 
The network contains two kinds of vertices --- ``weak'' vertices and ``strong'' ones. 
We introduce two sets of exponents for the two components of the network, $\beta_1$ and $\gamma_1$ -- for the component consisting of vertices with the unit fitness (contains $(1-p)t$ vertices) and $\beta_g$ and $\gamma_g$ -- for the component consisting of vertices with the fitness $g$ (contains $pt$ vertices). 
Thin lines depict the dependences at fixed  values of $p$. 
Arrows show how these curves change when $p$ decreases from $1$ to $0$. At $p \to 0$, we obtain dependences shown in Fig. \protect\ref{f26} (a single strong vertex).  
At $p \to 1$, $d/(m+n) \to 1$, 
$\beta_g \to \beta_0$, 
$\beta_1 \to \beta_0/g$, $\gamma_1 \to 1+g/\beta_0$, 
$\gamma_g \to \gamma_0$. 
}
\label{f28}
\end{figure}

The threshold, that is, the ``condensation point'', can be easily smeared in the following way. 
Let vertices have, at random, two values of fitness, $1$ and $g>1$, the 
probability that a vertex has fitness $1$ is $1-p$, and, with probability $p$, a vertex has fitness $g$. The characteristics of such a network are shown in Fig. \ref{f28}. These are the fraction of all vertices captured by the component of the network consisting of ``strong vertices'' and the scaling exponents of both components as functions of $g$. One sees that the threshold is smeared, and the condensation phenomenon is absent.  

\vspace{38mm}$\phantom{x}$

For the observation of the condensation of edges, special distributions $P(G)$ are needed. 
If $P(G)$ is continuous, it must be of a specific form in the region of the largest fitness, $G_{max}$ \cite{bb00b}. 
The structure of the network for the continuous distribution $P(G)$ was discussed in Sec. \ref{ss-types}. We have already described the situation when the transcendental equation (\ref{17-4}) has a real root $c$. For some distributions,  
including the considered case of a single strong vertex with $g>g_c$,
this is impossible. This indicates the condensation phenomenon --- a finite fraction of edges condenses on a single vertex with the largest fitness.       

At this point we must make the following remark. 
All the growing networks that we consider in the present section have a general feature -- {\em each} of their vertices has a chance to get a new link. Only one circumstance prevents their enrichment -- seizure of this link by another vertex. In such kinetics of distribution of edges, there is no finite radius of ``interaction'' and there are no principal obstacles for capture of a great fraction of edges by some vertex. 

Bianconi and Barab\'asi \cite{bb00b} noticed that the form of Eq. (\ref{17-4}) is similar to the form of the well-known equation for the Bose gas. 
Edges were interpreted as Bose-particles. They are distributed among energy levels --- 
vertices. The energy of each level was related to the fitness $G$ of the corresponding vertex. The distribution of these levels can be obtained from $P(G)$. 
One may  indicate    
a set of the distributions $P(G)$ for which there is no solution of Eq. (\ref{17-4}) 
like in the classical phenomenon of Bose condensation. 
Using the analogy with this phenomenon, Bianconi and Barab\'asi demonstrated that, in this ``phase'', a finite fraction of edges (particles of the Bose-gas) condenses on the strongest vertex (the lowest energy level) and called this process ``Bose-Einstein condensation'' \cite{bb00b}. 

One can use various parameterizations of $P(G)$ but it is natural to study its variation mainly near $G_{max}$. E.g., in Ref. \cite{bb00b}, it was shown that $P(G) \propto (G_{max}-G)^\theta$ produces the condensation starting from some minimal value of the exponent $\theta_c$. 

In fact, in paper \cite{bb00b}, the equations describing the distribution of edges among vertices of the large network with inhomogeneous fitting are mapped to the equations for the Bose gas. The price of this mapping is the introduction of thermodynamic quantities such as temperature, etc. for the description of the network. Unfortunately, it is not easy to find an interpretation, e.g., for temperature in this situation. It is ``something'' related to the form of $P(G)$. Therefore, here, we prefer to consider the ``condensation'' effect without 
applying such analogies and the introduction of thermodynamic variables but directly using the distribution $P(G)$.


\subsection{Correlations and distribution of edges over network}\label{ss-distributionoflinks}

In the present section, we mainly studied degree distributions. We have to admit, however, that they provide rather incomplete description of a growing network. 
One can better imagine the network if the average elements of the adjacency matrix are known. Their values are easily calculated in the continuum approximation. In the simplest case of the 
citation graph, the average number of edges $\overline{b}(s,s^\prime,t)$ between vertices $s$ and $s^\prime$ at time $t$ ($s<s^\prime\leq t$) has a very convenient feature, 
$\overline{b}(s,s^\prime,t \geq s^\prime) = \overline{b}(s,s^\prime,s^\prime)$. 
This crucially simplifies the calculations, and the result for scale-free citation graphs is 
 
\begin{equation}
\overline{b}(s,s^\prime,t) = 
\frac{m}{t}(1-\beta) \left(\frac{s}{t}\right)^{-\beta} 
\left(\frac{s^\prime}{t}\right)^{\beta-1}
\, , 
\label{21-1}
\end{equation} 
where $m$ is the number of connections of each new vertex 
(see Ref. \cite{dm00e}). Recall that $\beta=1/(\gamma-1)$. 
This characteristic was obtained exactly for the model of Sec. \ref{ss-simplestscale-free}. One sees that, generally, the product does not factorize to $\overline{k}(s,t)\overline{k}(s^\prime,t)$. The only exception is the $\beta=1/2$ ($\gamma=3$) case. 
From Eq. (\ref{21-1}), in the scaling regime, we can estimate the average number of connections between ancestor vertices of degree $k$ and descendants with degree $k^\prime$. 
In the continuum approximation, this quantity is proportional to the probability $P(k,k^\prime)$ that 
vertices of degree $k$ (ancestor) and $k^\prime$ (descendant) are connected: 
 
\begin{equation}
P(k,k^\prime) \propto k^{-1/\beta}k^{\prime-2} = k^{-(\gamma-1)}k^{\prime-2}
\, . 
\label{21-1a}
\end{equation}
The origin of the factor $k^{-(\gamma-1)}$ on the left-hand side of the equation is clear: new vertices are attached to old ones with probability $\sim kP(k)$, where $P(k)$ is the degree distribution. Meanwhile, degrees of nearest neighbors of a vertex in equilibrium scale-free networks are also distributed as $kP(k)$. Indeed, in equilibrium networks with statistically uncorrelated vertices, this degree distribution coincides with that for an end vertex (either of the two ones) of a randomly chosen edge, which is proportional to $kP(k)$ 
(see Sec. \ref{ss-theory}). 
Then, in equilibrium networks, the probability that a randomly chosen edge connects vertices of degrees $k$ and $k^\prime$ is 
$P(k,k^\prime) = k P(k) k^\prime P(k^\prime)/[\sum_k kP(k)]^2$, that differs sharply from Eq. (\ref{21-1a}). 
The factor $k^{\prime-2}$ in Eq. (\ref{21-1a}) is, in particular, the degree distribution of the nearest neighbors of the oldest (the richest) vertex (compare with the degree distribution of the nearest neighbors of a new vertex, $k^{-(\gamma-1)}$). 

The distribution $P(k,k^\prime)$ was originally obtained by Krapivsky and Redner \cite{kr00c} in the framework of the rate equation 
approach \cite{krl00,kr00c,krr00a} which is similar to the master 
equation one, which was discussed in Sec. \ref{ss-masterequation}, 
so we do not present their details here. The main statement is that 
this probability does not factorize. 
This means that {\em degrees of neighboring 
vertices in growing networks are correlated}. 


If one keeps fixed the large degree $k$ of an ancestor vertex, then, the most probable linking is with a descendant vertex of the smallest degree $k^\prime\sim1$. 
If the large degree $k^\prime$ of a descendant vertex is fixed, the above probability has a maximum at some 
$k$ which is smaller than $k^\prime$ but of the order of it.  

This absence of the factorization (the correlations) indicates a sharp difference of {\em growing} networks from  equilibrium graphs with statistically independent vertices. The reason is the obvious absence of time-reversal symmetry --- quite natural asymmetry between parents and children. Therefore, these correlations are present even for networks growing without preferential linking. 

Reliable measurements of the joint degree distribution of neighboring vertices, $P(k,k^\prime)$, are difficult because of poor statistics. Nevertheless, one can easily obtain information about the correlations in a network measuring the dependence of the average degree of nearest neighbors of a vertex on its degree, $\overline{k}_{nn}(k)$ (see discussion of the empirical data \cite{pvv01a} in Sec. \ref{sss-internet}). This is an important characteristic of correlations in growing network, so let us discuss it briefly. 

For a scale-free citation graph with the degree distribution exponent $\gamma<3$, from Eq. (\ref{21-1}), one may easily obtain the dependence $\overline{k}_{nn}(k) \propto k^{-(3-\gamma)}$ for small enough degrees $k$. For larger $k$, the dependence is a slow, logarithm-like function. 

If, in addition, new connections in a growing scale-free network also emerge between old vertices, the form of $\overline{k}_{nn}(k)$ depends on the details of the linking procedure. For example, 
(a) if new connections in the old part of the network emerge without any preference, the power-law singularity in $\overline{k}_{nn}(k)$ is retained; 
(b) if these edges connect randomly chosen old vertices with old vertices which are chosen preferentially, the power-law singularity is retained; 
but (c) if these edges connect pairs preferentially chosen old vertices, the singularity disappears as the number of such connections increases.  

For directed networks,
it is easy to introduce the notions of in- and out-components with respect to any vertex. 
One can define the out-component as the set of all ``ancestors'' of the vertex {\em plus itself}, i.e., all the vertices that can be reached if one starts from this vertex \cite{kr00c}. 
The in-component of the vertex contains all the vertices from which it can be reached, i.e., all its ``descendants'' {\em plus itself}. 


The distribution of the sizes of the in- and out-components of citation graphs (which are, in fact, directed networks) and their other characteristics were calculated in Ref. \cite{kr00c}. These results provide information about the topology of these networks. 
For the citation networks with $t$ vertices, the distribution of the in-component sizes $s$ was found to be proportional to $t/s^2$ for $s \gg 1$. This relation is valid for a wide variety of preference functions, including even the absence of any preference. 
For such a form of the in-component size distribution, the following condition is necessary: the power $y$ in the preference function $k^y$ should not exceed $1$. 

In Ref. \cite{kr00c}, the out-components of scale-free citation graphs were studied. 
E.g., for the BA model, that is, for the citation graph with $\gamma=3$, the out-component size distribution is $\ln^{s-1}(t+1)/[(t+1)(s-1)!]$. 
Here, $s$ is the out-component size.
This form is valid for the network with one edge (and, as usually, one vertex) added per unit of time. The distribution has a maximum at $s-1 = \ln (t+1)$ and quickly decays at larger $s$. Hence, the typical size of the out-component is of the order of $\ln t$, i.e. of the order of the typical shortest-path length in classical random graphs (see Sec. \ref{ss-shortest}). Similar results were obtained for all scale-free citation graphs \cite{kr00c}. 
The relation for typical size of the out-component is also valid for any citation graph with power $y$ of the preference function $k^y$ less or equal $1$. In this respect, these networks are similar to the classical random graphs.


\subsection{Accelerated growth of networks}\label{ss-accelerating}

The linear growth, when the total number of edges in the network is a linear function of its size (the total number of vertices), is only a particular case of the network evolution. For instance, data on the WWW growth \cite{dm00e} 
(see Sec. \ref{sss-www}), for the Internet \cite{fff99,pvv01a,gkk01e} (see Sec. \ref{sss-internet}), for networks of citations in scientific literature \cite{v01a} (see Sec. \ref{ss-citations}), and for collaboration networks \cite{bjnr01a,jnb01a} 
(see Sec. \ref{ss-collaborations}) demonstrate that the total numbers of edges in these networks grow faster than the total numbers of vertices, and one can say that the growth is {\em accelerated} \cite{dm004}, that is, nonlinear. 

One can show that a power-law dependence of the input flow of links may produce scale-free networks \cite{dm004}, and non-stationary degree distributions may emerge.  
In such a case, the scaling relations of Sec. \ref{ss-relations} are easily generalized. In the limit of the large network size, in general, one can write 

\begin{equation}
P(k,t) \propto t^z k^{-\gamma}
\, ,  
\label{23-1}
\end{equation} 
and

\begin{equation}
\overline{k}(s,t) \propto t^\delta \left( \frac{s}{t} \right)^{-\beta}
\, .   
\label{23-2}
\end{equation} 
General relations in the present subsection are valid for degree-, in-degree, and out-degree distributions, so that 
here $k$ denotes not only degree but also in- and out-degree. 
One can show that the exponents $z$, $\delta$, and $\beta$ are coupled by the relation, 
$z=\delta/\beta$, and the old relation (\ref{8-14}) is valid.
The distribution for individual vertices now is of the form 

\begin{equation}
p(k,s,t) = \frac{s^{1/(\gamma-1)}}{t^{(1+z)/(\gamma-1)}} 
f\left( k\,\frac{s^{1/(\gamma-1)}}{t^{(1+z)/(\gamma-1)}}  \right)
\, .   
\label{23-3}
\end{equation} 
Also, 

\begin{equation}
P(k,t) =  t^z k^{-\gamma} F(kt^{-(1+z)\beta}) =
t^z k^{-\gamma} F(kt^{-(1+z)/(\gamma-1)})
\, ,
\label{23-4}
\end{equation} 
and the distribution has a cut-off at $k_{cut} \sim t^{(1+z)/(\gamma-1)}$. 
We emphasize that Eqs. (\ref{23-3}) and (\ref{23-4}) are quite general relations obtained from the assumption of a power-law dependence of the total number of edges on the total number of vertices in the network. 

As demonstrating examples, 
in Ref. \cite{dm004}, two models for accelerating growth of networks with preferential 
linking were studied. In particular, the in-degree distributions of directed networks were considered. 
The input flow of edges is suggested to grow as $t^a$, where $a$ is a new exponent, $a<1$. 
The restriction is introduced to avoid multiple edges in the network at long times.    

\begin{figure}
\epsfxsize=75mm
\centerline{\epsffile{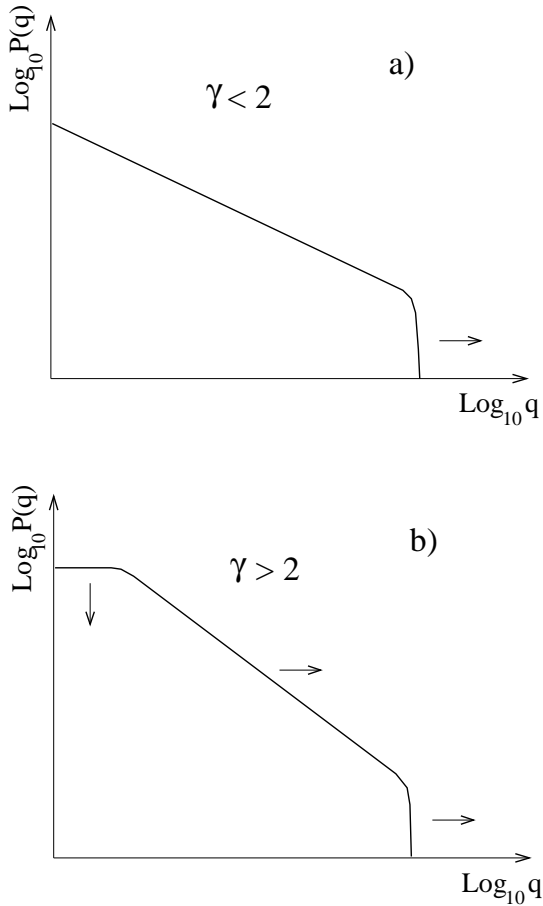}}
\caption{
Schematic log-log plots of degree distributions in the two models for accelerating growth of networks which are discussed in Sec. \ref{ss-accelerating}. 
The first model produces the stationary degree distribution with the exponent $\gamma<2$ (a) at long times. 
The degree distribution of the second model (b) is non-stationary, $\gamma>2$. 
The arrows indicate changes of the distributions as the networks grow.    
}
\label{f28a}
\end{figure}

In the first model, the additional attractiveness is constant. 
In this case, $\beta=1+a$, $\gamma=1+1/(1+a)$, $\delta=z=0$, and the degree-distribution is stationary (if we ignore its time-dependent cutoff), see Fig. \ref{f28a}, a. Therefore, the degree distribution 
of a non-linearly growing scale-free network may have the $\gamma$ exponent less than 2. When the input flow of edges grows proportionally to $t$, $\gamma=3/2$.

The second model is also based on the model of Sec. \ref{ss-continuous} but, in this case, 
the additional attractiveness $A$ (or the number of incoming edges $n$ of new nodes) 
grows with increasing time. For instance, let the additional attractiveness be proportional to 
the average in-degree of the network with a constant factor, $A(t)=B\overline{q}(t)=Bc_0t^a/(1+a)$. 
In this case, the $\gamma$ exponent exceeds $2$ and the distribution is non-stationary: $\gamma = 2 + B(1+a)/(1-Ba)$, $\beta = (1-Ba)/(1+B)$, $\delta=a$, and $z = a(1+B)/(1-Ba)$ (see Fig. \ref{f28a}, b). 
In such an event, the scaling regime is realized only if $Ba<1$. 


The same results are valid for degree distributions of undirected networks which grow nonlinearly. 

In general, assuming a power-law dependence of the input flow of edges on the network size ($a>0$), it is easy to obtain the following relations for the exponents $\gamma$, $z$, and $a$ \cite{dm00e,dm004}. If one assumes that $1<\gamma<2$, then 

\begin{equation}
\gamma = 1 + \frac{1+z}{1+a}
\, ,  
\label{23-5}
\end{equation} 
so $z$ should be smaller than $a$. 
This situation is realized in the first model above (see Fig. \ref{f28a}, a). 
At long times, the degree distribution is stationary. One may show that the cutoff $k_{cut} \sim t^{a+1}$, that is, of the order of the total degree of the network, so that the cutoff is in fact absent. 

If one assumes $\gamma>2$, the relation 

\begin{equation}
\gamma = 1 + \frac{z}{a}
\,   
\label{23-6}
\end{equation} 
is valid, so that one has to have  $z>a$. The distribution is non-stationary (see Fig. \ref{f28a}, b), and this is the case for the latter model. 

In both above models, the input flow was preset, that is, we suggest its power-law time dependence. However, such a power-law growth may arise quite naturally. Let us consider an illustrating example. An undirected citation graph grows according the following rules: 

(i) one new vertex is added to the network in unit time; 

(ii) with a probability $1-p$ it connects to a randomly chosen vertex or, 
with complementary probability $p$, not only to this vertex but also to all its nearest neighbors. 

If one is interested only in exponents, 
the same result 
is valid if a new vertex connects to 
a randomly chosen vertex plus to some of its nearest neighbors, each being chosen with the probability $p$. 
A new vertex actually copies (inherits) a fraction of connections of its ancestor (compare with a network growth process which leads to multifractal distributions \cite{dms009,dms01i}, see Sec. \ref{s-non-scale-free}). Such copying processes may be realized, for example, in networks of protein-protein interactions (see discussion in Ref. \cite{vfmv01a}).

These growth rules lead to an effective linear preferential attachment of edges to vertices with a large number of connections.  
One can easily see that the average degree (and the input flow of vertices) of this graph grows as $t^{2p-1}$ when $p>1/2$. For $p<1/2$, the mean degree approaches the constant value $2/(1-2p)$ at long times, and the degree distribution is stationary with exponent $\gamma=1+1/p>3$. 
When $p>1/2$, the degree distribution is non-stationary, like in the latter model, and $\gamma=1+1/(1-p)>3$. 

A model of a directed growing network, whose mean degree logarithmically grows with $t$, was proposed by V\'azquez \cite{v00}. The preferential linking arises in this model dynamically. Simulations and heuristic arguments in Ref. \cite{v00} have showed that the exponent $\gamma$ of this network is equal or close to $2$.   


Above models have input flow of edges growing proportionally to $t^a$. The situation when it grows as $t^a + const$ is also very interesting. Of course, at long times, such growth yields the above distributions. However, for real finite networks, this limit may not be approached. An intriguing application of these ideas can be proposed for the word web having been constructed in Ref. \cite{fs01} (see Sec. \ref{sss-word_web}). 

Recall that the empirical degree distribution of the word web has a complex form (see Fig. \ref{f6a}) with two distinct regions. A high quality of the empirical degree distribution due to the large number of vertices in the word web, $t\approx 470\,000$, and its high mean degree $\overline{k}(t) \approx 72$ together with the specific complex form gives a real chance to explain convincingly the structure of the word web. 


\begin{figure}
\epsfxsize=75mm
\centerline{\epsffile{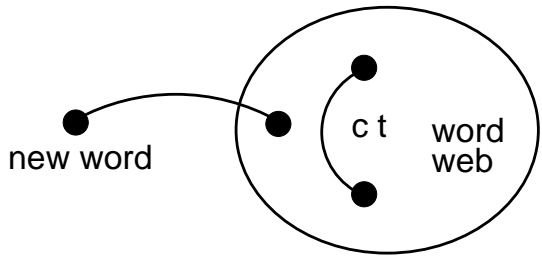}}
\caption{
Scheme of the word web growth \protect\cite{dm01d}. 
At each time step a new word appears, so $t$ is the total number of words. It connects to some preferentially chosen old word. Simultaneously, $ct$ new undirected edges emerge between pairs of preferentially chosen old words. 
All the edges are undirected. 
We use the simplest kind of the preferential attachment when a vertex is chosen with the probability proportional to the number of its connections.  
}
\label{f28b}
\end{figure}


Let us consider the minimal model for the evolving word web \cite{dm01d}
(see the discussion of  
practically the same model for networks of collaborations in Ref. \cite{bjnr01a}). 
We use the following rules of the growth of this undirected network (see Fig. \ref{f28b}). At each time step, a new vertex (word) is added to the network, and the total number of vertices, $t$, plays the role of time. At 
its birth, the new word connects to several old ones. We do not know the original number of connections. We only know that it is of the order of $1$. It would be unfair to play with an unknown parameter to fit the experimental data, so we set this number to $1$ (one can check that the introduction of this parameter does not change the degree distribution of the word web noticeably). 
We use the simplest natural version of the preferential linking, so a new word is connected to some old one $\mu$ with the probability proportional to its degree $k_\mu$, like in the Barab\'asi-Albert model \cite{ba99}. In addition, at each increment of time, $ct$ new edges emerge between old words, where $c$ is a constant coefficient that characterizes a particular network. 
The linear dependence appears if each vertex makes new connections at a constant rate, and we choose it as the most simple and natural. These new edges emerge between old words $\mu$ and $\nu$ with the probability proportional to the product of their degrees 
$k_\mu k_\nu$ \cite{ab00a,dm003} (see Sec. \ref{ss-estimations}). 

The mean degree of the network is equal to $\overline{k}(t)=2+ct$. 
According to the preceding analysis, this yields the stationary degree distribution with the exponent $\gamma=3/2$ at long times. The additional constant, equal to 2, is important 
if we are interested in 
how the stationary distribution is approached. 
Simple calculations \cite{dm01d} in the framework of the continuum approach yield the following degree distribution 

\begin{equation}
P(k,t) = \frac{1}{ct} \frac{cs(2+cs)}{1+2cs} \frac{1}{k}
\, ,  
\label{23-7}
\end{equation}  
where $s=s(k,t)$ is the solution of the equation 

\begin{equation}
k(s,t) = 
\left(\frac{ct}{cs}\right)^{1/2} \left(\frac{2+ct}{2+cs}\right)^{3/2}
\, .  
\label{23-8}
\end{equation} 

This distribution has two distinct regions separated by the crossover point $k_{cross} \approx \sqrt{ct}(2+ct)^{3/2}$. The crossover moves in the direction of large degrees while the network grows. 
Below this point, the degree distribution is stationary, 
$P(k) \cong \frac{1}{2}k^{-3/2}$ (we use the fact that in the word web $ct \gg 1$). Above the crossover point, we obtain 
$P(k,t) \cong \frac{1}{4}(ct)^3 k^{-3}$, so that the degree distribution is non-stationary in this region. 

At first sight, contribution to the average degree of the network (or the input flow of new links) from connections to new words seems negligible when compared with links which emerge between old words ($2 \ll ct \approx 70$). 
Nevertheless, as we see, this small contribution produces an observable effect.   

The position of the cutoff produced by finite-size effect (see Sec. \ref{ss-relations} for the relation  
$t\int_{k_{cut}}^\infty dk\,P(k) \sim 1$), 
is $k_{cut} \sim \sqrt{t/8}(ct)^{3/2}$. 
In Fig. \ref{f6a}, we plot the degree distribution of the model  
(the solid line). To obtain the theoretical curve, known parameters of the word web were used, $t$ and $\overline{k}(t)$. 
The deviations from the continuum approximation are accounted for in the small $k$ region, $k<10$. One sees that agreement with the empirical distribution is excellent. Positions of theoretical crossover and cutoff are also perfect.  Note that no fitting was made. 
For a better comparison, in Fig. \ref{f6a}, the theoretical curve is displaced upward to exclude two experimental points with the smallest $k$ since these points are dependent on the method of the construction of the word web, and any comparison in this region is meaningless in principle. 

Note that few words are in the region above the crossover point $k_{cross} \approx 5\times 10^3$. As language grows, $k_{cross}$ increases rapidly but, as it follows from above relations, the total number of words of degree greater than $k_{cross}$ does not change. It is a constant of the order of 
$1/(8c) \approx t/(8\overline{k}) \sim 10^3$, that is, of the order of the size of 
a small set of words forming the 
kernel lexicon of the British English which was estimated as $5\,000$ words \cite{fs00}. Thus, the size of the core of language does not vary as language evolves.


\subsection{Decaying networks}\label{ss-decay}

\begin{figure}
\epsfxsize=81mm
\epsffile{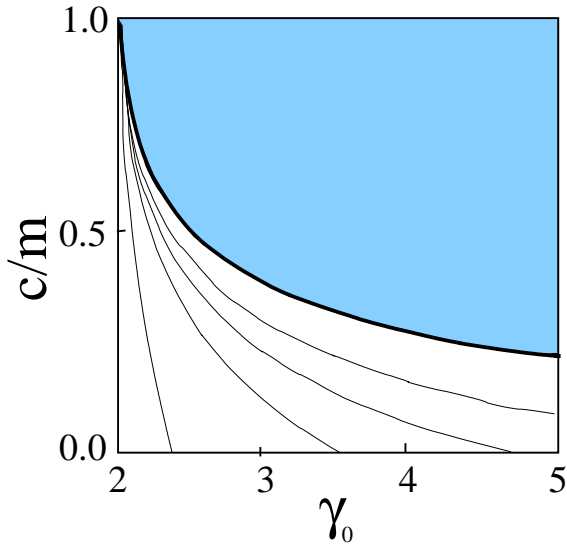}
\caption{
Phase diagram of the directed network growing under condition of permanent random damage -- 
the permanent deletion of random edges --- on axis $\gamma_0$, $c/m$. 
At each time step, $m$ new edges are added and 
$c$ random edges are deleted (see the text). 
$\gamma_0$ is the scaling exponent of the corresponding network growing without deletion 
of edges. Curves in the plot are lines of constant $\gamma_i$. 
$\gamma_i=\infty$ on the line $\gamma_0 = (c/m)+(c/m)^{-1}$. 
In the dashed region, the network is out of the class of scale-free nets. 
}
\label{f29}
\end{figure}

In Sec. \ref{ss-estimations} we have described a wide spectrum of possibilities to add edges to a network. Results for various cases may be found in Refs. \cite{ba99,baj99,krl00,kr00c,dms003,dms001,dm00e,krr00a,t00b,bb00b,bb00a,bajb00a}. Here, we discuss the opposite situation, namely, a fraction of edges may disappear during the network growth. This situation, an additional {\em permanent} 
deletion of edges, is considered in Refs. \cite{ab00a,dm003,dm00e}. Here we write down the result for a typical case of the model of directed growing network from Sec. \ref{ss-estimations} in which we, for brevity, set $n=0$. Each time a vertex and $m$ edges are added, $c$ old randomly chosen connections disappear. 
If we define $\gamma_i(c=0) \equiv \gamma_0$, the $\gamma_i$ exponent is 
 
\begin{equation}
\gamma_i = 2 + \frac{\gamma_0-2}{1-\gamma_0 c/m+(c/m)^2}
\, .
\label{25-1}
\end{equation} 
The resulting phase diagram, $c/m$ vs. $\gamma_0$ is shown in Fig. \ref{f29}.
One sees that random removal of edges increases the $\gamma_i$ exponent which
grows monotonously with increasing $c/m$ until it becomes infinite on the line  
$\gamma_0 = (c/m) + 1/(c/m)$.  
In the dashed region of Fig. \ref{f29}, 
the network is out of the class of scale-free nets. 
Note that, for large enough $c/m$, the network may decay to a set of uncoupled clusters. 


The effect of permanent random damage (the removal of edges) on undirected growing networks, in fact, has been considered in Sec. \ref{ss-estimations} (see the discussion of Eq. (\ref{15-3a})).   

The permanent random removal of vertices produces a different effect on the growing networks with preferential linking \cite{dm00e}. In this case, the $\gamma$ exponent does not change. Nevertheless, the exponent $\beta$ varies. Let a randomly chosen vertex be deleted with probability
$c$ each time a new vertex is added to the network. If, again we introduce the notations $\gamma(c=0) \equiv \gamma_0$ and $\beta(c=0)\equiv \beta_0$, then one can show that $\beta=\beta_0/(1-c)$ and 
$\gamma = 1 + 1/[\beta(1-c)] = 1 + 1/\beta_0 = \gamma_0$. We see that the scaling relation (\ref{8-14}) is violated in this situation. The reason of this violation and of the change of $\beta$ is an 
effective re-normalization of the $s$ variable due to the removal of vertices. 
In such an event, scaling forms of the degree-distributions for individual vertices and of the total degree-distribution are $p(k,s,t) = (s/t)^\beta f[ k (s/t)^\beta ]$ and $P(k) = k^{-\{1+1/[\beta(1-c)] \}} F(k/t^\beta)$. 

One can show that the permanent deletion of a fraction of vertices with the largest values of degree, that is an analogy of 
intentional damage (attack) \cite{ba00a,ceah00b,cnsw00}, destroys 
the scaling behavior of the network \cite{dm00e}.


\subsection{Eigenvalue spectrum of the adjacency matrix}\label{ss-spectrum}

The structure of the adjacency matrix and, therefore, of the network itself, can be characterized by its eigenvalue spectrum $G(\lambda)$. The eigenvalue spectra of classical random graphs are well studied. For the undirected infinite random graph with the Poisson degree distribution, the (re-scaled) eigenvalue spectrum has a semi-circle shape (here, $G(\lambda)=G(-\lambda)$) \cite{mbook91,cdsbook79}. If such a graph is large but finite, the tail of the distribution 
decreases exponentially with growing $\lambda$. 

In the recent paper \cite{fdbv01}, the eigenvalue spectrum of the BA model, that is, of the growing scale-free network with $\gamma=3$, was studied numerically (see also Ref. \cite{gkk01}). 
A sharp difference from classical random graphs was observed. 
It was found that its shape is very far from a semi-circle, and the tail of the spectrum is of a power law form (compare with the observations for the Internet \cite{fff99}, see Sec. \ref{sss-www}).


\subsection{Scale-free trees}\label{ss-trees}

Above we discussed two main construction procedures producing scale-free networks: the preferential linking mechanism for growing networks and a rather heuristic procedure of Molloy and Reed for equilibrium networks. In Ref. \cite{bck01}, the (grand canonical) statistical ensemble of tree-like graphs was constructed, that is, an equilibrium random tree-like graph. In the particular case 
of mean degree $\overline{k}=2$, the procedure \cite{bck01} yields scale-free trees with a $\gamma$ exponent which takes values in the range $(2, \infty)$. Cutoffs of the degree distributions of these equilibrium networks were found to be at the same point $k_{cut} \sim N^{1/(\gamma-1)}$ as for growing scale-free nets (see Eq. (\ref{11-5}) in Sec. \ref{ss-relations}). Here $N$ is the network size. 

In sharp contrast to the standard logarithmic dependence of the average shortest-path length $\overline{\ell}$ on the network size,     
the constructed trees were found to have a quite different geometry with a power-law dependence $\overline{\ell} \propto N^{1/d_H}$. Here the fractal dimension 
$d_H=2$ for $\gamma>3$ and $d_H = (\gamma-1)/(\gamma-2)$ when  $2 < \gamma < 3$.


\section{Non-scale-free networks 
with preferential linking}\label{s-non-scale-free}

A linear form of the preference function discussed in Sec. \ref{s-scale-free} is only a very particular case. 
It is hard to believe that just this case is the most wide-spread in Nature. 
Moreover, recent empirical studies of collaboration networks in the scientific literature \cite{bjnr01a,jnb01a} suggest that the preference function may be of a power-law form (although in Ref. \cite{n01b}, a linear attachment was observed in such networks). 

One can show that not only a linear of the preference function produces scale-free networks but all the preference functions that have linear asymptotes in the range of large values of degree \cite{krl00,kr00c}. Other preference functions 
do not provide scale-free networks. The case of a power-law preference function was explicitly considered in Refs. \cite{krl00,kr00c}. The continuum approach arguments for this situation can be found in Ref. \cite{dm00e}. 

In Refs. \cite{krl00,kr00c}, the extension of the BA model to the case of the preference function proportional to $k^y$ was studied. The results are the following. The situations with 
$0<y<1$ and $y>1$ differs sharply one from each other. The case $0<y<1$ is, in fact, describes crossover from the BA model (linear preferential linking) to the linking without preference ($y=0$) that produces exponential degree-distributions (see Sec. \ref{s-exponential}). The exact result for the stationary degree distribution\cite{krl00,kr00c} is 
 
\begin{equation}
P(k) \propto \left\{ 
\begin{array}{ll} 
k^{-y}\exp\left[-\mu \frac{k^{1-y} - 2^{1-y}}{1-y} \right]  
& \frac{1}{2} < y < 1 \, , 
\\[7pt]
k^{(\mu^2-1)/2}\exp\left[-2\mu\sqrt{k} \right]  & y = \frac{1}{2} \, ,
\\[7pt]
k^{-y}\exp\left[-\mu \frac{k^{1-y}}{1-y} + \frac{\mu^2}{2}\frac{k^{1-2y}}{1-2y} \right]  
& \frac{1}{3} < y < \frac{1}{2} \, ,
\\[7pt]
\ldots & 
\end {array} 
\right.
\label{27-1}
\end{equation} 
Here, $\mu$ depends on $y$ and varies from $1$ when $y=0$ to $2$ for $y=1$. Near these points, $\mu(y)$ is linear: 
$\mu(y) - 1 \cong 0.5078\,y$ and $2-\mu(y) \cong 2.407(1-y)$. 

In the case of $y>1$, most of connections come to the oldest vertex. Furthermore, for $y>2$, there is a finite probability that it is connected to {\em all} other vertices. For simplicity, let, at each increment of time, one vertex with one edge be added. Then, probability ${\cal P}(t)$ that the oldest vertex captures all edges satisfies the relation: 
${\cal P}(t+1) = {\cal P}(t) t^y/[t\cdot 1^y + t^y]$. Then,
 
\begin{equation}
{\cal P}(t \to \infty) = \prod_{t=1}^\infty \frac{1}{1+t^{1-y}}
\, .
\label{27-2}
\end{equation} 
This probability is indeed nonzero when $y>2$. 

Applying the master equation approach to this network one can obtain the following results \cite{krl00,kr00c}. If $y>2$, all but a finite number of vertices are connected with the oldest vertex. 
For $1<y<2$, the oldest vertex is connected to almost every other vertex but various situations are possible for the distribution of edges. For $(j+1)/j < y < j/(j-1)$, the number of vertices of degree $k>j$ 
(this number is equal to $tP(k>j)$) is finite and grows as $t^{k-(k-1)y} < t^1$ for $k \leq j$. This just means that $P(k=1,t \to \infty) \to 1$, so practically all the vertices are of unit degree and almost all the connections are with the oldest vertex. 

It would be a mistake to presume that linear preferential linking always provides scale-free networks networks (or, more precisely, networks with power-law degree distributions). We have to repeat that the growth producing power-law distributions is only a very particular situation. In
 Refs. \cite{dms009,dms01i}, the idea of preferential linking \cite{ba99} was combined with {\em partial inheritance (partial copying)} of degree of individual vertices by new ones \cite{sk99,sk00}. 
(The papers \cite{sk99,sk00} as well as Refs. \cite{cdks98,br00,jk98,jk00,gh98} are devoted to more complex models generically related to the biological evolution processes.) 

As an illustrating example, the directed network in which the growth is governed by the same rule of preferential linking as in the BA model, was considered analytically. At each time step, 
apart from $m$ new edges being distributed preferentially, some additional connections emerge. This additional new edges are attached to a new vertex. This one is born with a random number of incoming edges which is distributed according to some distribution function $P_c(q,t)$ depending on the state of the network at the birth time. In this rather general situation, the master equation looks like 

\begin{eqnarray}
& & t\frac{\partial P(q,t) }{\partial t} + P(q,t) + 
\frac m{\overline{q}(t) } [ q P(q,t) - (q-1) P(q-1,t) ] 
\nonumber 
\\[5pt]
& & = P_c(q,t) 
\, ,
\label{27-3}
\end{eqnarray} 
where $\overline{q}(t)$ is the average in-degree of the net at time $t$. 
One sees that the term $\delta_{q,0}$ on the right hand side of Eq. (\ref{8-4}) from Sec. \ref{ss-masterequation} is substituted by 
$P_c(q_t)$ in Eq. (\ref{27-3}), so Eq. (\ref{27-3}) is the direct generalization of Eq. (\ref{8-4}). If every new vertex is born by some randomly chosen old one, and at the moment of birth, it {\em ``inherits'' (copies)} on average, a fraction $c$ of its parent's connectivity, the distribution $P_c(t,q)$ takes a specific form which allows us to solve the problem explicitly. 
More precisely, with probability $c$, each of $q$ incoming edges of a ``parent'' creates an edge attached to its heir, so that this copying is only {\em partial}. 
In particular, such type of ``inheritance'' (copying) produces nodes of zero degree which cannot get new connections. Hence, it is worth to consider only ``active'' nodes of non-zero degree.

The resulting network is not scale-free. One may check that its in-degree distribution is of a {\em multifractal} type. That means that moments of the distribution depend on network size in the following way: 
$M_n(t) \sim t^{\tau(n)}$ where $\tau(n)$ is a non-linear function of $n$. 
Therefore, special attention must be paid to temporal evolution of the in-degree distribution. 

If, for example, $c$ is a random number distributed homogeneously within the interval $(0,1)$, and the evolution of the network starts from the distribution $P(q,t_0 \gg 1)=\delta_{q,q_0}$, then 
in-degree distribution $P_1(q,t)$ of the vertices of non-zero degree is of the form 

\begin{equation}
P_1(t,q) = d_1 t^{-\sqrt{2}}\ln (d_2 q) \exp [\sqrt{2\ln t\ln (t/q^2) }] 
\, . 
\label{27-4}
\end{equation} 
Here, $1 \ll q \ll q_0\sqrt{t}$, $d_1=0.174\ldots$, and $d_2=0.840\ldots$. For $t \to \infty$, Eq. (\ref{27-3}) takes the stationary form 

\begin{equation}
P_1(q) =\frac{d_1}{q^{\sqrt{2}}}\ln (d_2q) 
\, .
\label{27-5}
\end{equation} 
One may check that, in this case, $\tau(n)$ is indeed nonlinear, $\tau(n) = n/2 - n(n+1)$, and the distribution is multifractal. 

One should note that the nature of the new term in Eq. (\ref{27-3}) is rather general, and such effects should exist in various real networks. Unfortunately, 
as far as we know, no checks for multifractality of real degree distributions were made yet. 
In fact, the quality of the existing experimental material (see Sec. \ref{s-nature}) does not let one to 
separate power-law and multifractal behaviors. 
It is quite possible, that what is often reported as a power-law degree distribution is in fact a multifractal one. 
The situation may be similar to the one in the field of the self-organized criticality where numerous distributions first perceived as pure power-law dependences, now are treated as multifractal functions. 

Recently, a multifractal degree distribution was obtained in a model describing the evolution of protein-protein interaction networks \cite{vfmv01a}. 


\section{Percolation on networks}\label{s-percolative}

Rigorously speaking, percolation is a phenomenon determined for structures with well-defined metric structure, like, e.g., regular lattices. In case of networks, where it is hard to introduce metric coordinates, one can speak about the emergence of a giant component. In physical literature, a 
phenomenon related to the emergence of a giant component in networks is usually called percolation, and the phase transition of the emergence of a giant component is called a percolation threshold \cite{nsw00,mr95,mr98}. We follow this tradition. 

If the giant component is absent, the network is only a set of small clusters, so that the study of this characteristic is of primary importance. 
For regular lattices, to observe the percolation phenomenon, one must remove a 
fraction of sites or bonds.
In case of networks it is not necessary to delete vertices or edges to eliminate their giant components. For instance, one can approach the percolation threshold changing the degree distribution of a network. 

One should note that percolation phenomena in equilibrium and evolving (growing) networks are of different nature, so hereafter we consider them separately. 
Furthermore, the existing percolation theory for equilibrium 
networks \cite{nsw00} and its generalizations are valid only for specific graphs constructed by the Molloy-Reed procedure (see Sec. \ref{s-notions}). 
Also, inasmuch as giant components are discussed, we stress that networks must be large.


\subsection{Theory of percolation on undirected equilibrium networks}\label{ss-theory}

Very important results on percolation on random networks with arbitrary degree distributions and random connections are due to Molloy and 
Reed \cite{mr95,mr98}. 
They were subsequently developed in papers \cite{nsw00,cnsw00,mn00,mn00h}, and the problem was brought to the level of physical clarity. Here, 
we dwell on the latter efficient approach for equilibrium networks constructed by 
the Molloy-Reed procedure. 

The generating function (or the $Z$-transform) apparatus is used extensively in modern graph theory \cite{jtrbook00}. The $Z$-transform 
of the degree distribution is defined as 

\begin{equation}
\Phi(y) = \sum_{k=0}^\infty P(k)y^k 
\, ,
\label{29-1}
\end{equation} 
where $|y| \leq 1$. Obviously, $\Phi(1)=1$. 
For example, for the Poisson distribution (see Eq. (\ref{3-2}) in Sec. \ref{s-classical}), this yields the $Z$-transform $\Phi(y) = \exp[z_1(y-1)]$, 
where $z_1\equiv\overline{k}$ is the average degree of a vertex, i.e., the average number of the nearest neighbors.  
The inverse $Z$-transform is 

\begin{equation}
P(k) = \left.\frac{1}{k!}\frac{d^k \Phi(y)}{dy^k}\right|_{y=0} = 
\frac{1}{2\pi i}\oint\limits_C dy \,\frac{\Phi(y)}{y^{k+1}}  
\, ,
\label{29-2}
\end{equation} 
where $C$ is a contour around $0$ which does not enclose singularities of $\Phi(y)$. Moments of the distribution can be easily obtained from the $Z$-transform: 

\begin{equation}
\overline{k^n} = \left[\left(y\frac{d}{dy}\right)^n \Phi(y) \right]_{y=1} 
\, .
\label{29-3}
\end{equation} 
In particular, the number of the nearest neighbors is $z_1 \equiv \overline{k} = \Phi^\prime(1)$ (naturally, $z_0=1$).

This technique is especially convenient for the description of branching processes and trees. Let us outline the key points of the calculations. 
One can study the percolation by ``infecting'' a random vertex and considering the process of the infection spreading step by step. 
Let this vertex belong to some connected component. 
At the first step, the nearest neighbors will be infected, at the second -- the second neighbors, etc. until the step at which all the connected component will be infected. 

The first thing we should know is the following.     
Suppose that we randomly choose an edge in the network. 
It connects two vertices. Each of them may have some extra edges being attached.
How do edges breed (multiply) at ends of the randomly chosen edge? To know this, one should calculate the degree distribution for an end vertex (either of the two) of the  
edge. 
This distribution is equal to $kP(k)/\sum_k kP(k)$. Indeed, the edge is attached to a vertex with probability proportional to its degree $k$, and the degree distribution of vertices is $P(k)$. The denominator ensures proper normalization. $Z$-transform of the resulting distribution is 

\begin{equation}
\frac{\sum_k kP(k) y^k}{\sum_k kP(k)} = 
y \frac{\Phi^\prime(y)}{\Phi^\prime(1)} = y \frac{\Phi^\prime(y)}{z_1} \equiv y\Phi_1(y)
\, .
\label{29-4}
\end{equation} 

Meanwhile, one sees that the probability that a randomly chosen edge of such a graph connects vertices of degrees $k$ and $k^\prime$ is 

\begin{equation}
P(k,k^\prime) = \frac{kP(k) k^\prime P(k^\prime)}{[\sum_k kP(k)]^2}
\, 
\label{29-4a}
\end{equation} 
(compare with the corresponding distribution (\ref{21-1a}) for growing networks).

Actually, one needs a slightly different distribution than Eq. (\ref{29-4}). We have to know the distribution of the number of connections {\em minus one} for either of the two end vertices of a randomly chosen edge since we do not want to account for the original edge itself. This probability equals 
$(k+1)P(k+1)/\sum_k (k+1)P(k+1)$, so one immediately gets the corresponding $Z$-transform:

\begin{equation} 
\Phi_1(y) = 
\frac{\Phi^\prime(y)}{\Phi^\prime(1)} = \frac{\Phi^\prime(y)}{z_1}
\, .
\label{29-5}
\end{equation} 
Note that $\Phi_1(1) = 1$.

Let us start from a randomly chosen vertex and look how the numbers of its second-nearest neighbors are distributed.  
We recall that the network is large and its vertices are statistically 
uncorrelated (i.e. connections are random), so, in particular, one can 
neglect connections between the nearest neighbors. Moreover, one should state 
that if such a network is infinitely large, then almost each of 
its connected components has a tree-like structure. The results that we discuss are based on this key statement. Then, one can see that the $Z$-transform of the number of the second neighbors of a vertex is

\begin{equation} 
\sum_k P(k)[\Phi_1(y)]^k = \Phi(\Phi_1(y))
\, .
\label{29-6}
\end{equation} 
Indeed, the reason to write $P(k)$ in the sum is obvious: it is the probability that the original vertex has $k$ edges. To understand the $[\Phi_1(y)]^k$ factor, one should know the following property of $Z$-transform. If $\Psi(y)$ is the $Z$-transform of the distribution of values of some 
quantity $X$ of a system, then $[\Psi(y)]^m$ is the $Z$-transform of the distribution of the sum $\sum_{i=1}^m X_i$ of the values of this quantity observed in $m$ independent realizations of the system. 
At this point we use the basic assumption that vertices of the network are statistically uncorrelated.  
From this and Eq. (\ref{29-5}), the form of Eq. (\ref{29-6}) follows immediately. Proceeding in this way, one gets the $Z$-transform of the distribution of numbers of third-nearest neighbors, $\Phi(\Phi_1(\Phi_1(y)))$ (we have again used the tree-like structure of the network), etc. 

Eqs. (\ref{29-4})-(\ref{29-6}) are the basic relations of this approach. From 
Eqs. (\ref{29-3}) and (\ref{29-6}), one gets the average number of second-nearest neighbors of a vertex, 

\begin{equation} 
z_2 = \Phi^\prime(1)\Phi_1^\prime(1) = z_1 \Phi_1^\prime(1) = 
\sum_k k(k-1)P(k)
\, .
\label{29-7}
\end{equation} 
Using the above $Z$-transform of the distribution of number of $m$-th-nearest neighbors and Eq. (\ref{29-3}), one readily obtains  

\begin{equation} 
z_m = [\Phi^\prime(1)]^{m-1}\Phi_1^\prime(1) = \left[ \frac{z_2}{z_1} \right]^{m-1} z_1
\, .
\label{29-8}
\end{equation} 
We see that $z_m$ is completely determined by $z_1$ and $z_2$. 
If the giant connected component spans almost surely all of the network, the typical shortest path between a pair of randomly chosen vertices may be estimated from the 
condition,  
$\sum_{m=0}^{\overline{\ell}} z_m \sim N$. Substituting Eq. (\ref{29-8}) into this condition and assuming $N \gg z_1,z_2$, one obtains \cite{nsw00} 

\begin{equation} 
\overline{\ell} \approx \frac{\ln(N/z_1) + \ln[(z_2-z_1)/z_1]}{\ln(z_2/z_1)}
\, .
\label{29-9}
\end{equation} 
This relation improves on the classical estimate of the typical shortest path written in Sec. \ref{ss-shortest}. If the fraction $S$ of the 
network occupied by the giant connected component is less than one, one may 
try to improve the estimation by replacing $N \to NS$ in Eq. (\ref{29-9}) \cite{nsw00}. 
Note that Eq. (\ref{29-9}) is of a general nature. It contains only the local characteristics of the network, the numbers of the first and second nearest neighbors, $z_1$ and $z_2$. 

One should stress that Eq. (\ref{29-9}) is an estimate. Furthermore, an obvious problem occurs when $\gamma \leq 3$, when the mean number 
of second neighbors, $z_2$, diverges (see Eq. (\ref{29-7})) as $N\to\infty$. 
Indeed, let us try to estimate $\overline{\ell}$ from below in this ``dangerous'' region just using Eq. (\ref{29-9}). When $2<\gamma<3$, 
accounting for the degree-distribution cut-off position $k_{cut} \sim N^{1/(\gamma-1)}$ (see Sec. \ref{ss-relations}), we get $z_2 \sim \int^{k_{cut}}k^2 k^{-\gamma} \sim N^{(3-\gamma)/(\gamma-1)}$. 
Substituting this relation into Eq. (\ref{29-9}) gives the finite value
$\overline{\ell} \gtrsim  (\gamma-1)/(3-\gamma) + 1 = 2/(3-\gamma)$ 
in the large network limit (note that another estimate was recently made in Ref. 
\cite{pl01a}). 
The last equation demonstrates that the above approach (the tree ansatz) fails when $\gamma < 3$ but may also be considered as a very rough estimate from below for $\overline{\ell}$.

Let us consider the distribution of the sizes of connected components of the networks. 
It is convenient to introduce the distribution of the sizes of components which are 
reachable if we start from a randomly chosen edge and move through one of its ends. Let its $Z$-transform be $H_1(y)$. In Refs. \cite{nsw00}, following important equation was obtained for it, 

\begin{equation} 
H_1(y) = y \Phi_1(H_1(y))
\, .
\label{29-10}
\end{equation} 
The tree-like structure of large networks under consideration was again used. In this situation, the probability to reach some connected component moving in such a way is equal to the sum of probabilities 
(i) that there is only a single vertex, i.e., the dead end, 
(ii) that this vertex has one extra edge leading to another component,  
(iii) that it has two extra edges leading to two other components, and so on. Accounting for this structure and for the already used property of powers of $Z$-transform, one gets Eq. (\ref{29-10}) (compare it with the basic Eq. (\ref{29-4})). 
Now one can easily write the expression for the $Z$-transform of the distribution of sizes of connected components, that is, the components reachable starting from a randomly chosen vertex. Practically repeating the derivation of Eq. (\ref{29-6}), one obtains \cite{nsw00}

\begin{equation} 
H(y) = y \Phi(H_1(y))
\, .
\label{29-11}
\end{equation} 
The factor $y$ appears here, since the starting vertex also belongs to the connected component. 

From Eqs. (\ref{29-10}) and (\ref{29-11}), using Eqs. (\ref{29-1}) and (\ref{29-2}), we can find the distribution that we discuss. It is easy to find the average connected component size $\overline{s}$. From Eq. (\ref{29-11}), it follows that 

\begin{equation} 
\overline{s} = H^\prime(1) = 1 + \Phi^\prime(1)H_1^\prime(1)
\, .
\label{29-12}
\end{equation} 
$H_1^\prime(1)$ can be obtained from Eq. (\ref{29-10}), 

\begin{equation} 
H_1^\prime(1) = 1 + \Phi_1^\prime(1)H_1^\prime(1)
\, ,
\label{29-13}
\end{equation}  
so 

\begin{equation} 
\overline{s} = 1 + \frac{\Phi^\prime(1)}{1 - \Phi_1^\prime(1)}
\, .
\label{29-14}
\end{equation} 
From this, one sees that the giant connected component exists when 

\begin{equation} 
\Phi_1^\prime(1) > 1
\, ,
\label{29-15}
\end{equation} 
that is, when
 
\begin{equation} 
\Phi^{\prime\prime}(1) - \Phi^\prime(1) > 0
\, ,
\label{29-16}
\end{equation}   
or, equivalently, when 
 
\begin{equation} 
\sum_k k(k-2)P(k) = \overline{k^2} - 2\overline{k} > 0
\, .
\label{29-17}
\end{equation}  
The average size of connected components turns out to be infinite and the giant connected component emerges when $\sum_k k(k-2)P(k) = 0$. 
Accounting for expression (\ref{29-7}) for $z_2$ we see that 
{\em the giant connected component is present when average number of second nearest neighbors is greater than the average number of nearest neighbors}, 
 
\begin{equation} 
z_2 > z_1
\, .
\label{29-17a}
\end{equation}  

This strong result is due to Molloy and Reed \cite{mr95,mr98} and derived above following Refs. \cite{nsw00,mn00h} (heuristic arguments leading to Eq. (\ref{29-17}) may be found in Ref. \cite{ceah00a}). 
It has several important consequences. 
In particular, from Eq. (\ref{29-17}), it follows that the giant connected component is present when $\sum_{k=3} k(k-2)P(k) > P(1)$ (the case $P(2)=1$ is special since just in this situation the network has no tree-like structure).
Isolated vertices do not influence the existence of the giant connected component. 
Then, if
$P(1)=0$, the giant component exists when $\sum_{k\geq3}P(k)>0$. We see that dead ends are of primary importance for the existence of the giant component.  
Indeed, only the term with $P(1)$ in Eq. (\ref{29-17}) prevents the giant connected component.

If the giant component exists, the resulting relations are the same, equations (\ref{29-10}) and (\ref{29-11}), but now $H(y)$ corresponds to the distribution of the sizes of connected components of the network {\em excluding} the giant component, so $H(1)=1-W$, where $W$ is the relative size of the giant connected component. Then, from Eqs. (\ref{29-10}) and (\ref{29-11}), one sees \cite{nsw00} that 
 
\begin{equation} 
1 - W = \Phi(t_c)
\, ,
\label{29-18}
\end{equation}  
where $t_c$ is the smallest real non-negative solution of  
 
\begin{equation} 
t_c = \Phi_1(t_c)
\, .
\label{29-19}
\end{equation} 

Notice that the effect of isolated vertices is trivial. They produce the natural addendum $P(0)$ on the right hand part of Eq. (\ref{29-18}), and this is all. Also, the isolated vertices do not influence the existence of the giant component. 
Therefore, we can exclude these nodes from consideration and set $P(0)=0$. 
Then, from Eqs. (\ref{29-18}) and (\ref{29-19}), we immediately see that the absence of dead ends, i.e., $P(1)=0$, is sufficient for $W=1$. (Note again that it should be $P(2)<1$.) Indeed, in this case, 
$\Phi_1(0) \propto \Phi^\prime(0)=0$, so $t_c=0$ and $W=1$. 
If $P(1)>0$, usually $W<1$. 
It seems,  
there exists a situation \cite{nsw00}, in which $W=1$, even if $P(1)>0$. 
Let the degree distribution be of a power-law form, $P(k) \propto k^{-\gamma}, \ k\geq 1$ with exponent $\gamma \leq 2$.
 Then its first moment   
$\overline{k} \equiv z_1 = \Phi^\prime(1) = \sum_{k=1}^\infty k k^{-\gamma}$ 
diverges. This means that 
$\Phi_1(y<1) = 0$,
and Eq. (\ref{29-19}) has the solution $t_c=0$. Therefore, this case provides $W=1$. 
We should warn, however, that, formally speaking, the tree ansatz, which is the basis for the above conclusions, is not applicable in this situation. 
Notice that degree distributions with 
$\gamma\leq2$ lead to the divergence of the first moment, so the number of edges in such nets grows faster than the number of vertices as the network size increases (see Sec. \ref{ss-accelerating}).

From Eqs. (\ref{29-18}) and (\ref{29-19}), the size of the giant connected component of the classical random graph of Erd\"{o}s and R\'{e}nyi can be easily found. 
We have seen that its Poisson degree distribution produces $\Phi(y)=e^{z_1(y-1)}$. Then 
$1-W = e^{-z_1 W}$. Hence the giant connected component of such a network exists if its average degree exceeds one \cite{bbook85}.

From Eqs. (\ref{29-10}) and (\ref{29-11}), one can understand the analytical properties of $H(y)$ and $H_1(y)$ near the percolation threshold. 
The analytical structure of $H(y)$, in turn, determines the asymptotic form of the size distribution of connected components.  
Substituting the inverse of $r=H_1(y)$, i.e., $y=H_1^{-1}(r)$, into Eq. (\ref{29-10}), we get 
 
\begin{equation} 
y = r/\Phi_1(r)
\, .
\label{29-21}
\end{equation} 
The derivative of $dy(r)/dr$ is zero at the point of singularity of $H_1(y)$, 
$y^\ast$. This one is, as one can find from Eq. (\ref{29-21}), at the point $r^\ast$ which is determined from the equation 
 
\begin{equation} 
\Phi(r^\ast) - r^\ast \Phi_1(r^\ast) = 0
\, .
\label{29-22}
\end{equation} 
At the percolation threshold, where the giant connected component emerges, 
$\Phi_1(1) = 1$. Accounting, in addition, for the equality $\Phi(1)=1$, 
one sees that, at the percolation threshold, $r^\ast=1$. Then, it follows from 
Eq. (\ref{29-21}) that, in this situation, the singularity of $H_1(y)$ reaches 
$y^\ast=1$. 

At the percolation threshold, it is easy to expand $y(r)$ about $r^\ast=1$ using 
Eq. (\ref{29-21}). This gives 
$y[1+(r-1)] \cong 1 + 0 - \Phi^{\prime\prime}(1)(r-1)^2/2$. 
If $\Phi^{\prime\prime}(1)\neq0$, that is not true only for very special distributions, one gets $r = H_1(y) \cong 1 + (1-y)^{1/2}$ near $y^\ast=1$.

Equation (\ref{29-11}) shows that this singularity in $H_1(y)$ coincides with the one in $H(y)$ since $\Phi_1(y)$ has no singularities for $y \leq 1$. 
Then, at the percolation threshold, near $y^\ast=1$, $H(y)$ looks as 
 
\begin{equation} 
H(y) \cong C_1 + C_2 (1-y)^{1/2}
\, ,
\label{29-23}
\end{equation} 
where $C_1$ and $C_2$ are constants. Knowing the analytical structure of 
$H(y)$ at the percolation threshold, and using the properties of the $Z$-transform 
(recall that a power-law function $w^{-b}$, as $w \to \infty$, yields the Z-transform of the form 
$$
C_3 + C_4\, (1-y)^{b-1} + \mbox{analytical terms}
$$ 
near $y=1$, where $C_3$ and $C_4$ are constants), 
one can restore the structure of the distribution ${\cal P}_s(w)$ of sizes of the connected components in the network near this point. It looks like \cite{nsw00}
 
\begin{equation} 
{\cal P}_s(w) \sim w^{-3/2} e^{-w/w^\ast}
\, ,
\label{29-24}
\end{equation} 
where $w^\ast=1/\ln|y^\ast|$, $y^\ast$ is the point of the singularity in $H(y)$ closest to the origin that is just the singularity that we discussed above. 
Near the percolation threshold, $y^\ast$ is close to $1$. 
The values of $w^\ast$ and $y^\ast$ depend on a particular form of the 
degree-distribution. 
The exponent $3/2$ is the same for all reasonable degree distributions. 
This value is quite natural. Indeed, at the threshold point the average size of connected components diverges, so this exponent cannot be greater than $2$. 
We emphasize that Eq. (\ref{29-24}) is valid only near the percolation threshold. 

The power-law form ${\cal P}_s(w) \sim w^{-3/2}$ of the size 
distribution for connected components at the percolation threshold point corresponds 
to the form ${\cal P}(w) \sim w^{-5/2}$ for the probability that randomly 
chosen vertex belongs to a finite connected component of size $w$. The latter 
probability is a basic quantity in percolation theory. From this form, using 
standard arguments of percolation theory, one finds that if the 
giant component is absent, the largest connected component has size of the order of $N^{2/3}$ (near the percolation threshold). Here $N$ is the size of the network. In this situation, the size of the second largest connected component is of the order of $\ln N$. Also, 
one can prove that, if the giant connected component exists and $\gamma>2$, the sizes of all other connected components are of the order of $\ln N$ or less \cite{acl00}. 

In principle, the outlined theory, allows one to compute the main statistical properties of these equilibrium networks. Analytical calculations are possible only for the simplest degree distributions but numerics is easily applicable \cite{nsw00}. The results may be also checked by simulation using, e.g., efficient algorithm for percolation problems \cite{mz01}.  
 
In the same paper \cite{nsw00}, one can find another generalization of this theory to the case of undirected bipartite graphs (see Fig. \ref{f2} in Sec. \ref{ss-collaborations}). 
The bipartite graphs in which connections are present only between vertices of different kinds were considered. Such networks, in particular, describe collaborations (see Sec. \ref{ss-collaborations}). 
The proposed theory describes percolation on these networks. In 
addition, it allows one to calculate the degree distribution of the one-mode projection of the bipartite graph from two degree distributions of its vertices of different kinds. These relations were checked using data on real collaboration graphs, of the Fortune 1000, movie actors, collaborations in physics, etc. 
For example, from the data 
for the network of the members of the boards of directors of the Fortune 1000 companies, the distribution of the numbers of boards on which a director sits and the distribution of the numbers of directors on boards were 
extracted. From these two distributions, the distribution of the total numbers of co-directors for each director was calculated. The result turned to be very close to the corresponding empirical distribution of the co-directors.


\subsection{Percolation on directed equilibrium networks}\label{ss-directed}

This theory can be generalized to equilibrium directed networks with statistically uncorrelated vertices \cite{nsw00,dms01a} (see Fig. \ref{f5} and Sec. \ref{sss-www}, where the definitions of a giant strongly connected component (GSCC), a giant weakly connected component (GWCC), and giant in- and out-components (GIN and GOUT) were given). 
In this case, it is natural to consider the joint in- and out-degree distribution $P(k_{i},k_{o})$ and its Z-transform 

\begin{equation}
\Phi(x,y) \equiv \sum_{k_{i},k_{o}}P(k_{i},k_{o})x^{k_{i}}y^{k_{o}}
\, .  
\label{29-24a}
\end{equation}  
If all the connections are inside of the network  
the average in- and out-degrees are equal (see Eq. (\ref{0-2}) in Sec. \ref{ss-degree}): 
$\partial_x\Phi(x,1)\left.\right|_{x=1} = \partial_y\Phi(1,y)\left.\right|_{y=1} \equiv z^{(d)}=z_1/2$, where $z_1$ is the mean degree of the network. 
Z-transform of the degree distribution is $\Phi^{(w)}(x)=\Phi(x,x)$. 
Using Eqs. (\ref{29-18}) and (\ref{29-19}), from $\Phi^{(w)}(x)$, 
one gets the relative size $W$ of the  
GWCC. 

The sizes of the GIN and GOUT can be obtained in the framework of the following rigorous procedure \cite{nsw00}. One introduces the $Z$-transform of the out-degree distribution of the vertex approachable by following a 
randomly chosen edge when one moves 
{\em along} the edge direction, $\Phi_1^{(o)}(y) \equiv \partial_x\Phi(x,y)\left.\right|_{x=1}/z^{(d)}$. Also, 
$\Phi_1^{(i)}(x) \equiv \partial_y\Phi(x,y)\left.\right|_{y=1}/z^{(d)}$ corresponds to the in-degree distribution of the vertex which one can 
approach moving {\em against} the edge direction. The GIN and GOUT are present if 
$\Phi_1^{(i)\,\prime}(1) = \Phi_1^{(o)\,\prime}(1) = \partial^2_{xy}\Phi(x,y)\left.\right|_{x=1,y=1}/z^{(d)} > 1$, i.e. when  
 
\begin{eqnarray}  
\sum_{k_{i},k_{o}}
(2k_{i}k_{o} - k_{i} - k_{o}) P(k_{i},k_{o}) & = & 
\nonumber
\\[5pt]  
2\sum_{k_{i},k_{o}} 
k_{i}(k_{o} - 1) P(k_{i},k_{o}) & = & 
\nonumber
\\[5pt]
2\sum_{k_{i},k_{o}} 
k_{o}(k_{i} - 1) P(k_{i},k_{o}) & > & 0
\, 
\label{29-25}
\end{eqnarray}  
--- the generalization of Eq. (\ref{29-17}). This is also the condition for the existence of the GSCC. 
In this case, there exist non-trivial solutions of the equations 

\begin{equation}
x_c = \Phi_1^{(i)}(x_c) \  , \ \ \ \ y_c = \Phi_1^{(o)}(y_c)
\, .  
\label{29-26}
\end{equation} 
They have the following meaning. 
$x_c<1$ is the probability that the connected component, obtained by moving {\em against} the edge directions starting from a randomly chosen edge, is finite. 
$y_c<1$ is the probability that the connected component, obtained by moving {\em along} the edge directions starting from a randomly chosen edge, is finite. 
The expressions for the relative sizes of the GIN, $I$, and GOUT, $O$, have the form \cite{nsw00}  

\begin{equation}
I = 1 - \Phi(x_c,1) \  , \ \ \ \ O = 1 - \Phi(1,y_c)
\, .  
\label{29-27}
\end{equation} 

Recall that, in our definition, the GSCC is the interception of GIN and GOUT. 
Accounting for the meaning of $x_c$ and $y_c$, we can find exactly the relative size of the GSCC (see the derivation in Ref. \cite{dms01a}): 

\begin{eqnarray}
S & = &
\sum_{k_i,k_o} P(k_i,k_o)(1 - x_c^{k_i})(1 - y_c^{k_o})
=
\nonumber
\\[5pt]
& & 
1 - \Phi(x_c,1) - \Phi(1,y_c) + \Phi(x_c,y_c)
\, .
\label{29-28}
\end{eqnarray} 
Furthermore, knowing $W$, $S$, $I$, and $O$, it is easy to obtain the relative size of 
{\em tendrils},

\begin{equation}
T = W + S - I - O
\, .  
\label{29-29}
\end{equation} 

Equations (\ref{29-18}), (\ref{29-19}), and (\ref{29-26})-(\ref{29-29}) allow us to obtain exactly the relative sizes of all the giant components of equilibrium directed networks with arbitrary joint in- and out-degree distributions, if their vertices are statistically uncorrelated. 
It is useful to rewrite Eq. (\ref{29-28}) in the form

\begin{equation}
S = I O + \Phi(x_c,y_c) - \Phi(x_c,1)\Phi(1,y_c)
\, .  
\label{29-30}
\end{equation} 
If the joint distribution of in- and out-degrees factorizes, 
$P(k_i,k_o)=P_i(k_i)P_o(k_o)$, Eq. (\ref{29-30}) takes the simple form $S=IO$. 
Here, $P_i(k_i)$ is the in-degree distribution of the network, and $P_o(k_o)$ is the out-degree distribution. 
Otherwise, such factorization of $S$ is impossible. 
At the point of the emergence of GIN, GOUT, and GSCC, $x_c=y_c=1$, and $I$, $O$, and $S$ simultaneously approach zero.


\begin{figure}
\epsfxsize=90mm
\epsffile{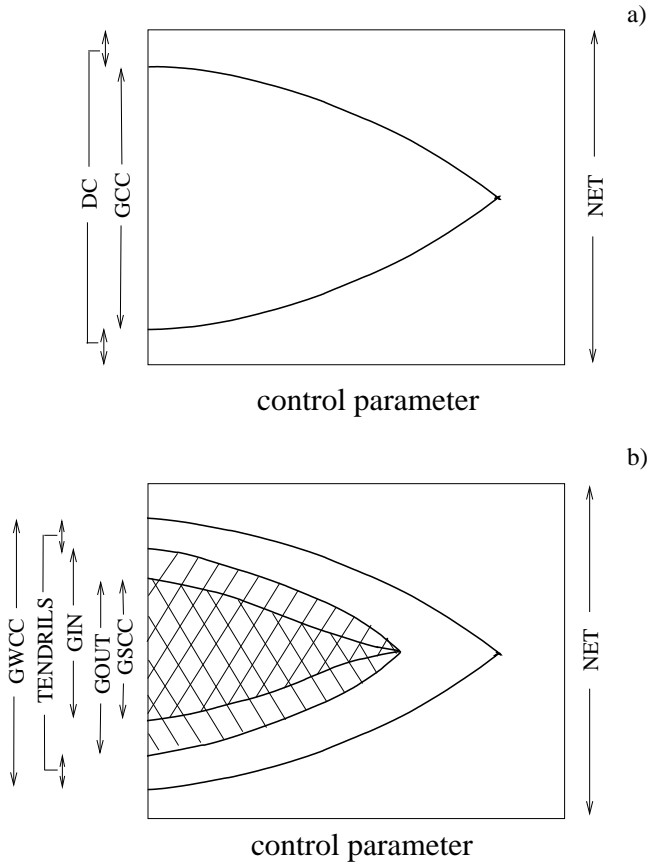}
\caption{
Schematic plots of the variations of all the giant components vs. some control parameter for the undirected equilibrium network $(a)$ and for the directed equilibrium one $(b)$. In the undirected graph, the meanings of the giant connected component (GCC), i.e., its percolating cluster, and the GWCC coincide. 
Near the point of the emergence of the GWCC, its size is a linear 
function of the control parameter. 
Near the point of the emergence of the GSCC, GIN, and GOUT, the sizes of the giant in- and out-components linearly varies with the control parameter, 
and the size of the giant strongly connected component is a quadratic function of the deviation of the control parameter from this point. 
}
\label{f29a}
\end{figure}


Variations of the giant components of equilibrium directed networks with some control parameter were studied in Ref. \cite{dms01a} (see Fig. \ref{f29a}). 
$I$ and $O$ approach the point of the emergence of the GSCC in a 
linear fashion, and $S$ a quadratic function of the deviation of the control parameter from its critical value.
 
Not pretending to apply this theory for equilibrium networks to the 
WWW, which is certainly non-equilibrium (growing), we recall the relative sizes of the giant components of the WWW. From the data of Ref. \cite{bkm00} (see Sec. \ref{sss-www}), in the WWW, $I \approx O \approx 0.490$, so $IO \approx 0.240$, that is, less than that measured in Ref. \cite{bkm00} $S \approx 0.277$ but is not far from it. 

An attempt was made to model the WWW using the measured in- and out-degree distributions and estimate the sizes of these components \cite{nsw00} but the result turned out to be far from reality. 
The main point of the discrepancy was the following. 
The reasonable values of parameters of the model network, used in the calculation, (in particular, the reasonable fraction of vertices with zero out-degree) produce a huge difference between the sizes of the giant in- and out-components unlike nearly equal in- and out-components of the WWW (see Sec. \ref{sss-www}). 
The authors of the paper \cite{nsw00} ascribed this discrepancy to the approximation 
$P(k_{i},k_{o}) \approx P_i(k_{i})P_o(k_{o})$ 
which they used in their calculations, so the correlations between 
in- and out-degree of vertices discussed in Sec. \ref{ss-estimations} were not accounted for. We may add that the equilibrium nets with statistically uncorrelated vertices, which we consider in this section, are far from the WWW whose growth 
produces strong correlations between its vertices.


\subsection{Failures and attacks}\label{ss-failure}

The effect of random damage and attack on communications networks (WWW and Internet) was simulated by Albert, Jeong, and Barab\'{a}si \cite{ba00a}. 
In their simulations, they used: 
 
(i) the real sample of the WWW containing 
$325\,729$ vertices and $1\,498\,353$ links, 
 
(ii) the existing map of the Internet containing $6\,209$ vertices and $24\,401$ links, 
 
(iii) the model for a scale-free network with the $\gamma$ exponent equal $3$ (the BA model), and 
 
(iv) for comparison, the exponential growing network \cite{ba99,baj99} (see Sec. \ref{s-exponential}). 

One should again stress that all these {\em growing} networks 
differ from the equilibrium networks considered in Secs. \ref{ss-theory} and \ref{ss-directed}. Edges between their vertices are distributed in a different manner because of their growth (see Sec. \ref{ss-distributionoflinks}). 
Recall that the networks (i) and (ii) have the $\gamma$ exponents in the range between 
$2$ and $3$.

Failures (random damage) were modeled by the instant removal of a fraction of randomly chosen vertices. The intentional damage (attack) was described by the instant deletion of a fraction of vertices with the highest numbers of connections (degree). The networks were grown and then were instantly damaged. 
In these simulations, the networks were treated as undirected. The following quantities were measured as functions of the fraction $f$ of deleted vertices: 

(i) the average shortest path $\overline{\ell}$ 
between randomly chosen vertices of the network, 
 
(ii) the relative size $S$ of the largest connected component (corresponds to the giant connected component if it exists), and 
 
(iii) the average size $\overline{s}$ of connected components (excluding the giant connected one).

\begin{figure}
\epsfxsize=90mm
\epsffile{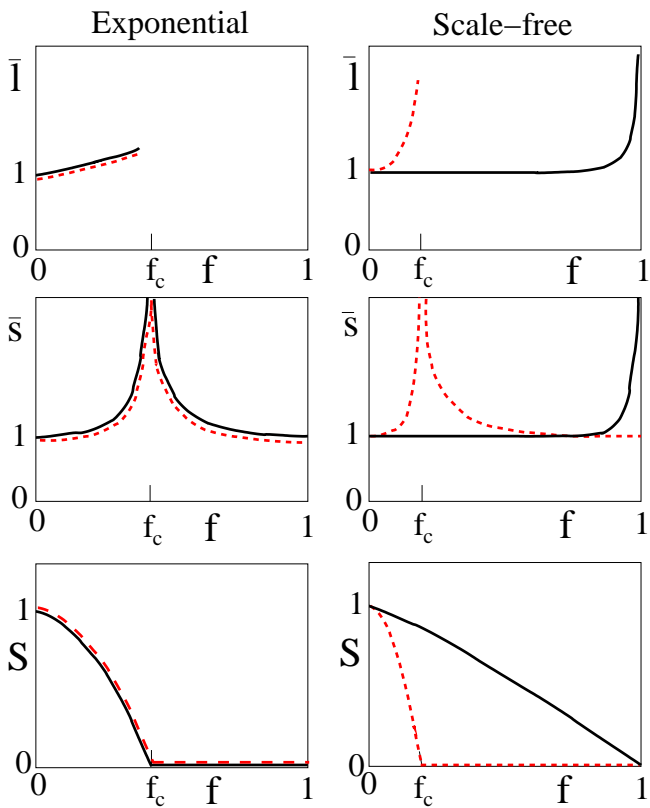}
\caption{
Schematic plots of the effect of intentional and random damage (attack and failures) on the characteristics 
of exponential undirected networks and scale-free undirected ones with exponent $\gamma \leq 3$  
\protect\cite{ba00a}. 
The average shortest path between vertices, $\overline{\ell}$, the size of the largest connected component, $S$, and the average size of isolated clusters, $\overline{s}$  
are plotted vs. the fraction of removed vertices $f \equiv 1-p$. 
The networks are large. The solid lines show the effect of the random damage. The effect of the intentional damage is shown by the dashed lines. For the exponential networks, both kinds of damage produce the same dependences. For the scale-free networks with $\gamma \leq 3$, 
in the event of the random damage, the percolation threshold is at the point $f \to 1$. 
}
\label{f30}
\end{figure}

A striking difference between the scale-free networks and the exponential one was observed. Whereas the exponential network produces the same dependences $\overline{\ell}(f)$, $S(f)$, and $\overline{s}(f)$ for both kinds of damage, for all scale-free nets which are discussed here, these curves are distinct for different types of damage. The qualitative effect of the intentional damage was more or less the same for all four networks (see  Fig. \ref{f30}). 
The average shortest path rapidly grows with growing $f$, the size of the giant connected component turns to be zero at some point $f_c$ indicating the percolation threshold, $S(f_c)=0$. 
Near this point, $S(f) \propto (f_c-f)$ as in the mean-field theory of percolation. 
At $f_c$, $\overline{s}$ diverges. Hence, the behavior is usual for the mean-field (or infinitely dimensional) percolation and for the percolation in the classical random graphs. Nevertheless, one general 
distinct feature of these scale-free nets should be emphasized. The value $f_c$ in them is anomalously low -- several percents, unlike the percolation threshold of the exponential networks, so that such nets are very sensitive to intentional damage. 


The main observation of Ref. \cite{ba00a} is that the random damage has far less pronounced effect on the scale-free nets than the intentional one. The variations of the average shortest distance with $f$ are hardly visible. The size of the giant 
strongly connected component decreases slowly until it disappears in the vicinity of $f=1$. $\overline{s}(f)$ grows smoothly with growing $f$ without visible signs of singularity. This means that these scale-free networks are extremely resilient to random damage. To destroy them acting in such away, that is, to eliminate their giant connected component and to disintegrate them to a set of uncoupled clusters, it is necessary to delete practically all their vertices! 

Similar observations were made for scale-free networks of metabolic reactions \cite{jtaob00}, protein networks \cite{jmbo01a} and food webs \cite{sm00}. 

The effect of the attack on the scale-free networks seems rather natural since vertices of the highest degree determine the structure of these nets, but the vitally important resilience against failures needs detailed explanation. 
Several recent papers have been devoted to the study of this intriguing problem.


\subsection{Resilience against random breakdowns}\label{ss-resilience}

As we saw in Sec. \ref{ss-failure}, the random breakdowns (failures) of networks more or less correspond to the classical site percolation problem, i.e., a vertex of the network is present with probability $p=1-f$, and one has to study how properties of the network vary with changing $p$. Now we have at hand the controlling parameter $p$ to approach the percolation threshold. This parameter can be easily inserted into the general relations of Sec. \ref{ss-theory}. 

The first calculations of the threshold for failures in scale-free networks belong to Cohen, Erez, ben-Avraham, and Havlin \cite{ceah00a}. Here, for logical presentation, we start with the considerations based on the approach of  
Callaway, Newman, Strogatz, and Watts \cite{cnsw00}.   

Let us look, how Eqs. (\ref{29-10}) and (\ref{29-11}) (for undirected equilibrium networks) are modified when each vertex of a network is present with probability $p$ (site percolation). 
Let $P(k)$ be the degree distribution for the {\em original (i.e. undamaged or virgin)} network with $p=1$. 
Again, to find $H_1(y)$, we have to start from a randomly chosen edge, but now we start from a randomly chosen edge of the {\em original network} with $p=1$, so it may be absent when $p<1$.  
Hence we must account for probability $1-p$ that the very first vertex (the vertex at the end of the edges) is absent. This produces the term $(1-p)y^0$ in $H_1(y)$. Then, we can pass through this point with the probability $p$, so the equation for $H_1(y)$ takes the form 
 
\begin{equation} 
H_1(y) = 1-p + p y \Phi_1(H_1(y))
\, .
\label{31-1}
\end{equation} 
Similarly, while calculating $H(y)$, we start from a randomly chosen vertex of the {\em original network} with $p=1$, which is absent in the network under consideration with the probability $1-p$. Hence we obtain the additional term $1-p$ in $H(y)$ and must multiply the remaining contribution (see Eq. (\ref{29-11})) by $p$. Therefore, 
 
\begin{equation} 
H(y) = 1-p + p y \Phi(H_1(y))
\, .
\label{31-2}
\end{equation} 
This pair of equations \cite{cnsw00} actually solves the site percolation problem for networks with random connections. In the event of bond percolation, i.e., when an edge is present with the probability $p$, the form of Eq. (\ref{31-1}) does not change 
(just follow the justification of Eq. (\ref{31-1}) above). 
Nevertheless, in this case, the equation for $H(y)$ differs from Eq. (\ref{31-2}). Indeed, in the bond percolation problem, all vertices are present, so when we start from a randomly chosen vertex, we can repeat the arguments leading to Eq. (\ref{29-11}) of Sec. \ref{ss-theory}. Then, again $H(y) = y \Phi(H_1(y))$. 

For site percolation, proceeding in the way outlined in Sec. \ref{ss-theory}, one gets from Eqs. (\ref{31-1}) and (\ref{31-2}) the following expression for the average size of the connected component above the percolation threshold:
 
\begin{equation} 
\overline{s} = H^\prime(y) = 
p \left(1 + \frac{p\Phi^\prime(1)}{1-p\Phi_1^\prime(1)} \right)
\, .
\label{31-3}
\end{equation}  
For bond percolation, it looks like 
 
\begin{equation} 
\overline{s} =  
\left(1 + \frac{p\Phi^\prime(1)}{1-p\Phi_1^\prime(1)} \right)
\, .
\label{31-4}
\end{equation} 
Therefore, the criterion for the existence of the giant connected component now becomes $\Phi^\prime(1) > 1/p$, i.e., $p\Phi^{\prime\prime}(1) - \Phi^\prime(1) > 0$, for both (!) site and bond percolation. Now, instead of 
the criterion of Molloy and Reed (\ref{29-17}), one has \cite{ceah00a,cnsw00}
  
\begin{equation} 
\sum_k k\left(k - \frac{1+p}{p} \right)P(k) = 
\overline{k^2} - \frac{1+p}{p}\,\overline{k} > 0
\, 
\label{31-5}
\end{equation} 
for both site and bond percolation. 
We again emphasize that, here, $P(k)$ is the degree distribution of the virgin network with $p=1$, and $\overline{k}$ and $\overline{k^2}$ are the average degree and the second moment for the virgin (undamaged) network again. 
If we take the distribution $\tilde P(k)$ of the network with removed vertices or edges, we must use the original relations of Sec. \ref{ss-theory}.  

Criterion (\ref{31-5}) may be rewritten in the form: 
  
\begin{equation} 
p\, z_2 > z_1
\, ,
\label{31-5a}
\end{equation} 
where $z_1$ and $z_2$ are the average numbers of the first and second nearest neighbors in the virgin undamaged network, respectively. Compare Eq. (\ref{31-5a}) with Eq. (\ref{29-17a}).
Hence the percolation thresholds for both site and bond problems are at the same point, 
  
\begin{equation} 
p_c = \frac{1}{(\overline{k^2}/\,\overline{k}) - 1} = \frac{z_1}{z_2}
\, ,
\label{31-6}
\end{equation} 
Notice the beauty of this simple formula! 

Proceeding in a similar way to the derivation in Sec. \ref{ss-theory}, one obtains the relations for the calculation of the relative size $S$ of the giant connected component. In the event of the site percolation problem,  
  
\begin{equation} 
1 - S = H(1) = 1-p + p\Phi(t^\ast)
\, ,
\label{31-7}
\end{equation} 
where $t^\ast$ is the smallest real non-negative solution of the equation 
  
\begin{equation} 
t^\ast = 1-p + p\Phi_1(t^\ast)
\, .
\label{31-8}
\end{equation} 
For the bond-percolation problem, one should apply Eq. (\ref{31-7}) and 
$t^\ast = \Phi_1(t^\ast)$ instead of Eq. (\ref{31-8}).

These relations were used in Ref. \cite{cnsw00} to study the effect of random damage on networks with different degree distributions. The results of the numerical calculations 
support the observations discussed in Sec. \ref{ss-failure}, so that it is really hard to eliminate the giant connected component by this means. 

Basic relations (\ref{31-5}) and the first of (\ref{31-6}) were derived in Ref. \cite{ceah00a} in a different but instructive way. Let us outline it briefly. One starts the derivation applying the original criterion (\ref{29-17}) of Molloy and Reed 
directly to the randomly damaged network. Then, it contains the degree distribution $\tilde P(k)$ of the damaged network. This degree distribution may be expressed in terms of the original distribution in the following way \cite{ceah00a}:
  
\begin{equation} 
\tilde P(k) = \sum_{k^\prime = k}^\infty {k^\prime \choose k} p^k (1-p)^{k^\prime-k} P(k^\prime)
\, .
\label{31-9}
\end{equation} 
Note that this equation is valid both for the deletion of vertices and deletion of edges. 
One may check that the first and second moments of the degree distribution for the damaged network, $\tilde P(k)$, are related to the moments of the degree distribution of the virgin network: 
  
\begin{eqnarray} 
\overline{k}^{\,\prime} & = & p \overline{k}
\, ,
\nonumber
\\[5pt]
\overline{k^2}^{\,\prime} & = & p^2 \overline{k^2} + p(1-p) \overline{k}
\, .
\label{31-10}
\end{eqnarray} 
Substituting these relations into the criterion of Molloy and Reed (\ref{29-17}), one immediately gets Eqs. (\ref{31-5}) and (\ref{31-6}). 


\begin{figure}
\epsfxsize=85mm
\epsffile{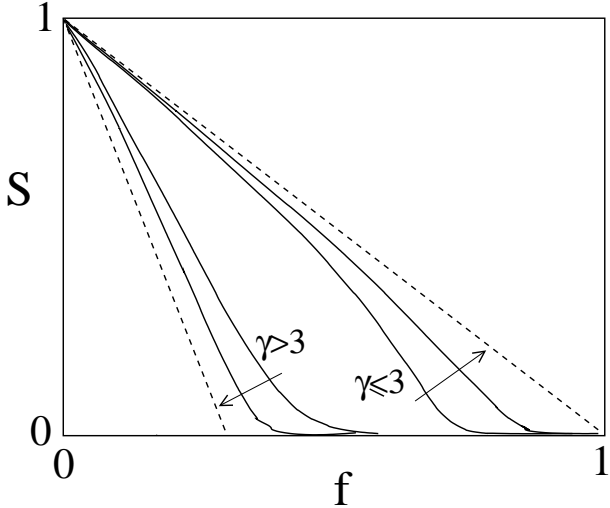}
\caption{
Schematic plot of dependences of the relative size of the largest connected component in the randomly damaged finite size net vs. the fraction of removed vertices $f=1-p$ \protect\cite{ceah00a}. Distinct curves correspond to different network sizes $N$ and two values of the $\gamma$ exponent, 
$2<\gamma\leq 3$ and $\gamma > 3$. Arrows show displacement of the curves with increasing $N$. Dashed lines depict the limits $N \to \infty$. Notice the strong size effects.
}
\label{f31}
\end{figure} 


In Ref. \cite{ceah00a}, for networks with power-law distributions, the variation of the size of the giant connected component with $p$ was studied through simulation (see Fig. \ref{f31}). 
Strong size effects was observed. For $\gamma >3$, the percolation threshold is visible at the point $p_c=1-f_c>0$. For $\gamma \leq 3$, if a network 
is infinite, one sees that $p_c$ approaches zero, so that in this situation, one has to remove (at random) practically all vertices of the network to eliminate the giant connected component! 

This very important observation can be understood if one looks at Eqs. (\ref{31-5}) and (\ref{31-6}). Indeed, from Eq. (\ref{31-6}), it follows that 
{\em the percolation threshold $p_c$ is zero if the second moment of the degree distribution of the virgin (undamaged) network 
diverges}. This occurs in networks with power-law degree distributions when $\gamma \leq 3$, but this is only one particular possibility. For example, the second moment diverges in networks with copying (inheritance) of connections of vertices \cite{dms009,dms01i,vfmv01a} (see Sec. \ref{s-non-scale-free}), so that these network are also super-resilient to failures. 

The condition $\gamma \leq 3$ for the resilience of scale-free networks to random damage and failures makes the values of the exponents of the degree distributions of communications networks quite natural. 
All of them are less than $3$ (see Sec. \ref{ss-communications}). 
Note that many other networks, e.g., biological ones, must be necessarily resilient to failures. Therefore, this condition is of great importance.   

One should note that this result is valid for infinite networks. 
As one can see from Fig. \ref{f31}, for $\gamma <3$, the size effects are very strong, and the curves slowly approach the infinite network 
limit where $p_c(N\to\infty)=0$. Here, $N$ is the size of the network. Let us estimate this size effect 
for $2<\gamma \leq 3$ by introducing the size-dependent threshold $p_c(N)$, 
whose meaning is clear from Fig. \ref{f31}.
When $N \gg 1$, from Eq. (\ref{31-6}) we obtain 
  
\begin{equation} 
p_c(N) = \frac{z_1}{z_2} 
\cong \frac{\overline{k}}{\overline{k^2}} 
\approx 
\frac{\int_{k_0}^{N^{1/(\gamma-1)}}dk\, k k^{-\gamma}}
{\int_{k_0}^{N^{1/(\gamma-1)}}dk\, k^2 k^{-\gamma}}
\, .
\label{31-11}
\end{equation} 
Notice, that, for $2<\gamma\leq3$, the average number of second-nearest neighbors is $z_2\cong\overline{k^2}$, since the second moment diverges as $N \to \infty$, see Eq. (\ref{29-7}). 
The nature of the upper cut-off of the power-law degree distribution, $k_{cut}/k_0 \sim N^{1/(\gamma-1)}$, was explained in Sec. \ref{ss-relations}.  
$k_0>0$ can be estimated as the minimal value of the degree in the network. 
One may expect that 
$k_0\sim1$. 

If $2<\gamma<3$, from Eq. (\ref{31-11}), it readily follows that 
  
\begin{equation} 
p_c(N) = C(k_0,\gamma) N^{-(3-\gamma)/(\gamma-1)}
\, .
\label{31-12}
\end{equation} 
Here, $C(k_0,\gamma)$ 
does not depend on $N$ and is of the order of $1$. 
$C(k_0,\gamma)$ actually depends on the particular form of the degree distribution for small values of degree and is not of great interest here. 
When $\gamma$ is close to $3$, $p_c=0$ can be approached only for a huge network.   
Even if $\gamma=2.5$ and a net is very large, we get noticeable values of the threshold $p_c$, e.g., $p_c(N=10^6) \sim 10^{-2}$ and $p_c(N=10^9) \sim 10^{-3}$. 

This finite size effect is most pronounced when $\gamma=3$. In this case, Eq. (\ref{31-11}) gives 
  
\begin{equation} 
p_c(N) \approx \frac{2}{k_0 \ln N}
\, .
\label{31-13}
\end{equation}  
For example, if $k_0=3$, $p_c(N=10^4) \approx 0.07$, $p_c(N=10^6) \approx 0.05$, and $p_c(N=10^9) \approx 0.03$. 

These estimates demonstrate that in reality, that is, for {\em finite} scale-free networks, the percolation threshold is actually present even if $2<\gamma \leq 3$ (see Fig. \ref{f31}). 
Only if $\gamma \leq 2$, the threshold $p_c(N)$ is of the order of $1/N$ (that is, the value of the natural scale for $p$) and is not observable. From the estimate (\ref{31-12}) one sees, that if $\gamma>2$, it should 
be close enough to $2$ for the extreme resilience of {\em finite} scale-free networks to failures.    

One may find the discussion of the resilience of directed equilibrium networks to random damage in Ref. \cite{dms01a}. 

Strictly speaking, all the results of this section were obtained for equilibrium networks. We do not know any analytical answers for the problem of instant damage in growing networks. However, it seems that the 
results for equilibrium networks describe the observations \cite{ba00a} of {\em instant} damage in growing networks quite well. This is not the case for permanent damage in growing networks \cite{chk01,dms01f} (see Sec. \ref{ss-anomalous}). 



\subsection{Intentional damage}\label{ss-intentional}

The intentional damage (attack), as we have seen in Sec. \ref{ss-failure}, can be defined 
as the instant removal of a fraction of vertices with the largest degrees. The theory 
outlined in Secs. \ref{ss-theory} and \ref{ss-resilience} can be easily generalized to the situation when 
the occupation probability depends on the degree of a vertex of the undamaged network, so that $p=p(k)$ \cite{cnsw00}. Then, for the intentional damage, 
$p(k)=1$ if $k \leq k_{cut}$ and $p(k)=0$ if $k > k_{cut}$.
$k_{cut}$ depends on the value of the fraction $f$ of vertices which are deleted, so that $k_{cut}=k_{cut}(f)$. 

In Ref. \cite{cnsw00}, this approach was used for the computation of the dependence of the size of the giant connected component on $f$. The networks with power-law degree distributions without isolated vertices (i.e. $P(0)=0$) were considered. 
The calculations were performed for the power-law degree distributions with $\gamma = 2.4$, $2.7,$ and $3.0$. 
It was shown that, in accordance with the observations in  
Ref. \cite{ba00a}, the deletion, in such a way, of a rather small fraction of vertices eliminates the giant connected component. The corresponding threshold values of $f$ are really small, 
$f_c(\gamma=2.4)=2.3\times10^{-2}$, $f_c(2.7)=1.0\times10^{-2}$, and 
$f_c(3.0)=0.2\times10^{-2}$. In this respect, the networks are very sensitive to these damage. 

On the other hand, 
the corresponding values of $k_{cut}$, at which the transition takes place, were calculated to be 
$k_{cut}=9$, $10$, and $14$ for $\gamma=2.4$, $2.7$, and $3.0$, respectively. 
A simple estimate  
$\int_{k_{cut}(f)}^\infty P(k) = f$ 
yields $k_{cut}(f) \sim f^{-1/(\gamma-1)}$ and provides the values $k_{cut}(f_c(\gamma))$ 
close to the above ones.
This means that for elimination of the giant connected component in this situation, one has to delete  
even vertices of rather low degree. In this regard, one must produce a really tremendous destruction to disintegrate such networks. Similar observations were made for the WWW \cite{bkm00} (see Sec. \ref{sss-www}). It was found that 
even the deletion of all vertices of {\em in-degree} larger than $2$ does not destroy the giant {\em weakly} connected component of the WWW.

Let us show how the threshold $f_c$ depends on $\gamma$. In Ref. \cite{ceah00b} the dependence $f_c(\gamma)$ was obtained in the framework of the continuum approach. Here we show how one can get the exact results \cite{dm01e}. 

From the relations of Ref. \cite{cnsw00} obtained for the situation when some  
vertices are deleted, and the occupation probability $p(k)$ depends on 
vertex degree, it is easy to derive the following condition for the percolation threshold 
  
\begin{equation} 
\sum_{k=0}^\infty k(k-1)P(k)p(k) = \sum_{k=0}^\infty k P(k)
\, 
\label{33-1}
\end{equation} 
(compare with Eq. (\ref{31-5})).
Here, $P(k)$ is the degree distribution of the undamaged network. The intentional damage cuts off vertices with $k > k_{cut}(f)$, where $k_{cut}(f)$ can be obtained from 
  
\begin{equation} 
f = 1 - \sum_{k=0}^{k_{cut}}P(k)
\, .
\label{33-2}
\end{equation}  
Then, the condition (\ref{33-1}) takes the following form: 
  
\begin{equation} 
\overline{k^2} - 2\overline{k} = \sum_{k=0}^\infty k(k-2)P(k) = 
\sum_{k=k_{cut}(f)+1}^\infty \!\!\!\!\! k(k-1) P(k)
\, .
\label{33-3}
\end{equation}
Here, $\overline{k^2}$ and $\overline{k}$ are the moments of the degree distribution of the undamaged network. 
Equation (\ref{33-3}) may be rewritten in the form \cite{dm01e}
  
\begin{equation} 
\sum_{k=0}^{k_{cut}(f)} \!\!\!  k(k-1)P(k) = 
\sum_{k=0}^\infty k P(k)
\, .
\label{33-3a}
\end{equation} 
This is the generalization of the Molloy-Reed criterion to the case of intentional damage.

Let us derive exact Eqs. (\ref{33-3}) and (\ref{33-3a}) 
in another way using the instructive ideas of Ref. \cite{ceah00b} but avoiding the continuum approximation. 
After the deletion of the most connected vertices, all the edges attached to the deleted vertices must also be removed. Connections 
in the network are random, so the probability $\tilde{f}$ that an edge is attached to one of the deleted vertices equals the ratio of the total number of edges of deleted vertices to the total degree of the network:
  
\begin{equation} 
\tilde{f}(f) = \sum_{k=k_{cut}(f)+1}^\infty P(k) = 
1 - \sum_{k=0}^{k_{cut}(f)} P(k) 
\, .
\label{33-4}
\end{equation}   

Now we can recall that Eq. (\ref{31-6}) for the percolation threshold is also valid for bond percolation. Therefore, it is possible to substitute $\tilde{p_c}=1-\tilde{f_c}$ into it. 
In fact, at first, the vertices of the highest degree were removed (the first step), and only afterwards their connections were deleted (the second step). 
In this event, Eq. (\ref{31-6}) describes only the effect of removal of edges. Then, it seems natural to use in Eq. (\ref{31-6}) the degree distribution with the cut-off $k_{cut}(f)$ arising after the first step. 
Accounting for this, one gets the relation
  
\begin{equation} 
(1 - \tilde{f_c}) \sum_{k=0}^{k_{cut}} k^2 P(k) = 
(2 - \tilde{f_c}) \sum_{k=0}^{k_{cut}} k P(k)
\, 
\label{33-5}
\end{equation} 
from which Eq. (\ref{33-3}) follows immediately. This demonstrates the equivalence of the approaches of Refs. \cite{cnsw00} and \cite{ceah00a,ceah00b}. 

In the particular case of the power-law degree distribution, $P(0)=0$ and 
$P(k \geq 1) = k^{-\gamma}/\zeta(\gamma)$, where 
$\zeta(\gamma) \equiv \sum_{k=1}^\infty k^{-\gamma}$ is the zeta-function, 
Eq. (\ref{33-2}) takes the form 
  
\begin{equation} 
f = 1 - \frac{\sum_{k=1}^{k_{cut}} k^{-\gamma}}{\zeta(\gamma)}
\, ,
\label{33-6}
\end{equation} 
and condition for the percolation threshold (\ref{33-3}) looks like
  
\begin{equation} 
\sum_{k=1}^{k_{cut}} k^{2-\gamma} = 
\zeta(\gamma-1) + \sum_{k=1}^{k_{cut}} k^{1-\gamma}
\, .
\label{33-7}
\end{equation} 

\begin{figure}
\epsfxsize=85mm
\epsffile{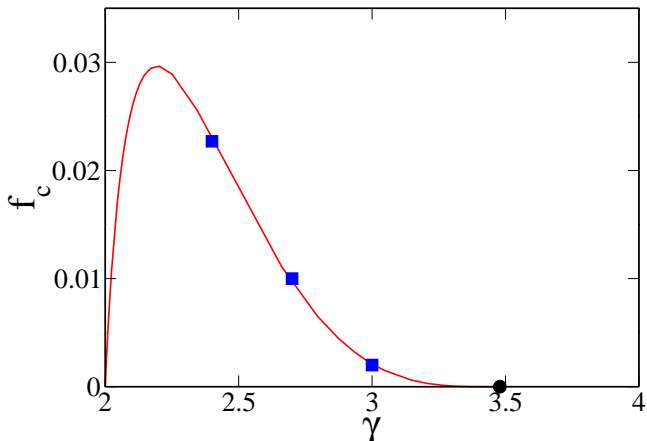}
\caption{
Dependence of the percolation threshold $f_c=1-p_c$ on the value of the $\gamma$ exponent of the large scale-free network for intentional damage 
(attack) \protect\cite{dm01e}. Such a dependence was originally obtained in the framework of a continuum approach \protect\cite{ceah00b}.  
Here we present the exact curve. 
$f$ is the fraction of removed vertices with the largest numbers of connections. 
The degree distribution of the network before the attack is 
$P(k) \propto k^{-\gamma}$ for $k \geq 1$, $P(0)=0$. 
The circle indicates the point $\gamma=3.479\ldots$ above which $f_c=0$. 
The squares represent the results of calculations and simulation in Ref. 
\protect\cite{cnsw00}. 
}
\label{f32}
\end{figure}

From Eqs. (\ref{33-6}) and (\ref{33-7}), one can easily obtain 
$k_{cut}(\gamma)$ and $f_c(\gamma)$ (see Fig. \ref{f32} \cite{dm01e}). 
Note that $f_c>0$ only in the range $2<\gamma<3.479\ldots$. When $\gamma<2$, a finite number of vertices keep a finite fraction of all connections, so their removal should have a striking effect on the network. 
For $\gamma>3.479\ldots$, the giant connected component is absent even before the attack. Indeed, in the undamaged network, the giant connected component exists if 
$\sum_{k=1}^\infty(k^2-2k)k^{-\gamma} = \zeta(\gamma-2)-2\zeta(\gamma-1)>0$ 
[see Eq. (\ref{29-17})]. This corresponds to $\gamma<3.479\ldots$. 

\vspace{35mm}$\phantom{x}$

The dependence $f_c(\gamma)$ has the maximum, $f_c^{max} = 0.030\ldots$. 
This is a really small value, so the network is indeed weak against the attack. 
One should emphasize that this result is very sensitive to the particular form of $P(k)$ in the range of small $k$ and to the number of dead ends in the network. In particular, the range of the values of $\gamma$, where the giant connected component exists, crucially changes when the minimal degree increases. 

In Ref. \cite{ceah00b}, one more interesting observation was made. 
The average shortest-path length between two vertices in random networks is of the order of 
the logarithm of their size (see Sec. \ref{ss-shortest}). The same statement is valid for vertices in their giant connected components. In Ref. \cite{ceah00b}, 
the average shortest-path length between two vertices of the 
giant connected component was studied near the percolation threshold. 
In such a situation, like in ordinary infinite-dimensional percolation, the average shortest-path length was found to be proportional to the square root of the total number of vertices in the giant connected component. 
\vspace{5pt}


\subsection{Disease spread within networks}
\label{ss-spread}

The dynamics of disease spread in undirected networks with exponential and power-law degree distributions was studied in Refs. \cite{pv00,pv01} and then in Refs. \cite{pv01c,mpv01a,db01a} (see popular discussion of these 
problems in Ref. \cite{lm01}). This process is generically related to the percolation properties of the networks. 

For the modeling of the spread of diseases within networks, two standard models were used. In the susceptible-infected-susceptible (SIS) model, 

(i) each healthy (susceptible) vertex is infected with rate $\nu$ when it has at least one infected neighbor, and 

(ii) infected vertices are cured (become susceptible) with rate $\delta$. 

Hence, the main parameter of the model is effective spreading rate $\lambda\equiv\nu/\delta$. 

The susceptible-infected-removed (SIR) model with three states of vertices (susceptible (healthy), infected, and removed (dead or immunized)) is slightly more complex. Nevertheless, the main parameter of interest describing 
the spread of the disease is the same, that is, $\lambda$, and 
main results for the SIR model on networks \cite{pv01c} are similar to those for the SIS model \cite{pv00,pv01,mpv01a}, so here we discuss the simpler case.  

Thus, we speak about the SIS model on equilibrium undirected networks.  
In non-scale-free networks (exponential, Poissonian, the WS network, and others) the situation is very similar to the one for the disease spreading in ordinary homogeneous systems: 
there exists a nonzero epidemic threshold $\lambda_c \sim \overline{k}$, where $\overline{k}$ is the average degree of a network, below which the 
disease dies out exponentially fast. This means that, after a random vertex 
becomes infected, the average density of infected vertices (prevalence) $\rho(t)$ rapidly approaches zero. 

When $\lambda > \lambda_c$, the infection spreads and becomes endemic: 
$\rho(t \to \infty) \equiv \rho  \propto (\lambda - \lambda_c)$. Notice that the dependence is linear in the vicinity of the threshold. 

For equilibrium undirected networks with arbitrary degree distribution $P(k)$, the epidemic threshold is at the point 
  
\begin{equation} 
\lambda_c = \frac{\overline{k}}{\overline{k^2}} 
\, .
\label{33a-1}
\end{equation} 
The relation was obtained in the framework of the dynamical mean-field approach \cite{pv00,pv01}.
Note that, if the ratio $\overline{k}/\overline{k^2}$ is small, the form of Eq. (\ref{33a-1}) naturally coincides with Eq. (\ref{31-6}) for the 
percolation threshold of randomly damaged network, and $\lambda_c \cong p_c$. 
Then the statements of Secs. \ref{ss-failure} and \ref{ss-resilience} about the absence of the percolation threshold in infinite scale-free networks with $\gamma\leq 3$ can be repeated for the epidemic threshold: in infinite networks, if $\gamma\leq 3$, diseases spread and become endemic for any $\lambda>0$. 
In general, {\em in infinite networks, if $\overline{k^2}$ diverges, the epidemic threshold is absent}, so that this result is applicable for a wide class of networks, e.g., see models of networks in Refs. \cite{dms009,dms01i,vfmv01a} (Sec. \ref{s-non-scale-free}).
Thus, although the infinite scale-free networks with $\gamma \leq 3$ are extremely robust against random damage, they are incredibly 
sensitive to the spread of infections. Both these, at first sight, contrasting phenomena have the same origin, the fat tail of a degree distribution. 

One may notice that most of the observed scale-free networks in Nature (see Sec. \ref{s-nature}, Fig. \ref{f20}, and Table \ref{t1}) have $\gamma$ exponent between $2$ and $3$. 
Then, why are we still alive? If the claim about the absence of the epidemic threshold in this networks is perfect, pandemics would never stop.

We emphasize that the strong statement that the threshold is absent can be applied only for {\em infinite} networks. In finite networks, the epidemic threshold is actually present. The estimates (\ref{31-12}) and (\ref{31-13}) also yield the dependence of the effective epidemic threshold $\lambda_c(N) \cong p_c(N)$ on the network size. As we demonstrated in Sec. \ref{ss-resilience}, 
$\lambda_c(N)$ in real networks with $\gamma \leq 3$ may be large enough. Only if $\gamma$ is close to $2$ from above or smaller than $2$, 
the threshold is unobservable. Notice that many real networks have the $\gamma$ exponents in this range. (Another quite plausible answer to 
our question is that the $\gamma$ exponent value of the web of human sexual contacts may be greater than $3$, see Ref. \cite{leasa01}.) 
  
In an infinite scale-free network with $\gamma=3$, 
the prevalence in the endemic state is 
  
\begin{equation} 
\rho \sim \exp(-C/\lambda)
\, ,
\label{33a-2}
\end{equation} 
(see Refs. \cite{pv00,pv01}) where $C$ is a constant. 

When $\gamma>3$ but is close enough to $\gamma=3$, 
  
\begin{equation} 
\rho \sim (\lambda-\lambda_c)^{1/(\gamma-3)}
\, .
\label{33a-3}
\end{equation} 
Finally, according to continuum dynamic mean-field calculations in Refs. \cite{pv00,pv01}, when $\gamma>4$, $\rho \sim (\lambda-\lambda_c)$ (the degree distribution $P(k) \propto k^{-\gamma} \theta(k-m)$ was  used). Note, however, that if $m=1$, for large values of $\gamma$, the giant connected component is absent, the network is a set of 
disconnected clusters, and the disease spread is impossible.  
Here we again repeat that the results of this section was obtained for equilibrium random networks with statistically uncorrelated vertices.  

How can we stop pandemics in widespread scale-free networks with $\gamma\leq3$? In Refs. \cite{pv01c,db01a}, the effects of immunization for these networks were studied. The results obtained \cite{pv01c,db01a} may be 
easily understood if we recall the weak effect of random damage to the integrity of such nets and strong effect of an intentional attack 
(see Secs. \ref{ss-failure}, \ref{ss-resilience}, and 
\ref{ss-intentional}). Analogously, random immunization 
cannot restore the epidemic threshold, but targeted immunization 
programs for highly connected vertices are the most effective way to stop an epidemic.


\subsection{Anomalous percolation on growing networks}
\label{ss-anomalous} 

We have shown (see Secs. \ref{ss-theory} and Fig. \ref{f29a}) that percolation on equilibrium networks displays many features of percolation on infinite-dimension lattices, i.e. of standard mean-field or effective medium percolation. 
One might expect that such a ``mean-field'' behavior is natural for all random networks which have no real metric structure. 
Furthermore, it seems, the abrupt removal of a large fraction of randomly chosen vertices or edges from the growing network presumably leads to rather standard percolation phenomena, see Sec. \ref{ss-failure}, although this issue is still not clear. 
However, as it was found in Ref. \cite{chk01}, this is not the case for networks growing under permanent damage which show quite unusual percolation phenomenon. 

Here we discuss the process of the emergence of a giant connected component in a growing network under the variation of growth conditions. 
In Ref. \cite{chk01}, the simplest model of growing network, in which such a percolation phenomenon is present, was studied (see Sec. \ref{s-notions}): 

(i) At each time step, a new vertex is added to the network. 

(ii) Simultaneously, $b$ new undirected edges are created between $b$ pairs of randomly chosen vertices ($b$ may be non-integer). 

The degree distribution of this simple network is exponential. 
The matter of interest is the probability ${\cal P}(s,t)$ that a randomly chosen vertex belongs to a connected component with $s$ vertices at time $t$. The master equation for this probability has the form \cite{dms01f}

\begin{eqnarray}
& & t\frac{\partial{\cal P}(s,t)}{\partial t} + 
{\cal P}(s,t) = 
\nonumber
\\[5pt]
& & \delta_{s,1} +
b s \sum_{u=1}^{s-1}{\cal P}(u,t){\cal P}(s-u,t) -
2b s{\cal P}(s,t)
\, . 
\label{34-1}
\end{eqnarray} 
This is a basic master equation for the evolution of connected components in growing networks. Note that Eq. (\ref{34-1}) is nonlinear unlike previously discussed master equations for degree distributions (see Eq. (\ref{6-3}) in Sec. \ref{s-exponential}, Eq. (\ref{7-2}) in Sec. \ref{ss-idea}, (\ref{8-4}) in Sec. \ref{ss-masterequation}, (\ref{9-7}) in Sec. \ref{ss-simplestscale-free}, (\ref{27-3}) 
in Sec. \ref{s-non-scale-free}). The first term on the right-hand side of Eq. (\ref{34-1}), i.e., the Kronecker symbol, accounts for the addition of new vertices (single clusters) to the network, the second (gain) term is the contribution from the fusion of pairs of connected components into larger ones, the last (loss) term describes the disappearance of connected components due to the fusion processes. 

Equation (\ref{34-1}) has a stationary solution ${\cal P}(s)$ in the long time limit. From the distribution ${\cal P}(s)$, the relative size of the giant connected component also follows: $W = 1 - \sum_{s=1}^\infty {\cal P}(s)$. 
The following surprising results were found numerically 
and by simulation in Ref. \cite{chk01} and then in the framework of exact analysis in Ref. \cite{dms01f}. 

It was shown that the phase transition of the emergence of giant connected component (percolation transition) in the growing network cannot be described by an effective medium theory. This phase transition is of infinite order, and, near the percolation threshold, the relative size of the giant component behaves as 
  
\begin{equation} 
W(b) = 0.590\ldots 
\exp
\left\{-\frac{\pi}{2\sqrt{2}} \frac{1}{\sqrt{b-b_c}} 
 \right\}
\, ,
\label{34-2}
\end{equation} 
where the constant $0.590\ldots$ may be calculated 
up to 
any desirable precision \cite{dms01f}. 
Here the rate $b$ of the emergence of new edges plays the role of a control parameter. When $b$ is small, the giant component is absent, and, obviously, when $b$ is large, the giant component must be present. The 
phase transition occurs at $b_c=1/8$. All the derivatives of $W(b)$ are zero at this point in sharp contrast to a linear $W(b)$ dependence in the standard mean field theory of percolation. This indicates that this phase transition is of infinite order like the Berezinskii-Kosterlitz-Thouless phase transition \cite{b70,kt73}. 

When $b<1/8$, i.e. in the phase without the giant connected component, the average size of a finite connected component is 
  
\begin{equation} 
\overline{s} = \frac{1 - \sqrt{1-8b}}{4b}
\, .
\label{34-3}
\end{equation}  
When the giant component is present, i.e. when $b>1/8$, 
  
\begin{equation} 
\overline{s} = \frac{1}{2b(1-W)} 
\, .
\label{34-4}
\end{equation} 
This means that the average size of a finite connected component jumps discontinuously at the percolation threshold from $2$ at $b=1/8-0$ (the phase without the giant component) to $4$ at $b=1/8+0$ (the phase, in which the giant component is present) \cite{chk01}. 
This behavior is in contrast to the divergence of $\overline{s}$ 
at the threshold point for standard percolation. 
(In the latter case, the divergence takes place, either one approaches the threshold from above or below.) 
We emphasize that the anomalous percolation transition is not accompanied by any anomaly of the degree distribution\vspace{115pt}. 


\begin{figure}
\epsfxsize=85mm
\epsffile{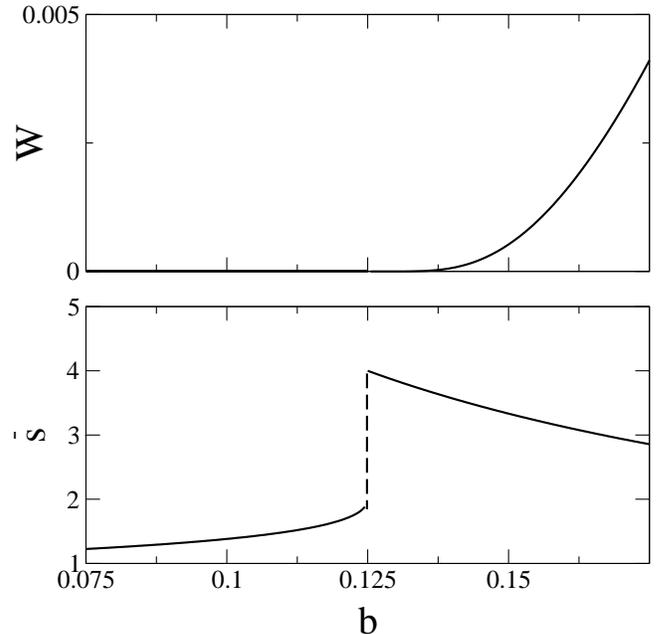}
\caption{
Anomalous percolation in the growing network. 
\ \ 
(a) The giant connected component size $W$ vs. the rate $b$ of the creation of new edges in the growing network (see Eq. (\protect\ref{34-2})). 
\ \ 
(b) The average size $\overline{s}$ of a finite connected component vs. $b$ 
(see Eqs. (\protect\ref{34-3}) and 
(\protect\ref{34-4}))\protect\cite{chk01}. 
\ \ 
The region of anomalous percolation threshold $b_c=1/8$ is shown.   
}
\label{f32a}
\end{figure} 


The probability ${\cal P}(s)$ that a randomly chosen vertex belongs to a finite connected component of size $s$ shows surprising behavior \cite{dms01f}. Recall that in standard percolation and in percolation in equilibrium networks ${\cal P}(s)$ is of a power-law form at the 
percolation threshold and it decreases exponentially both below and above the percolation threshold. More precisely, near the standard percolation threshold, ${\cal P}(s)$ is a power law with an exponential cutoff. In the growing network the situation is quite different:
 
(i) ${\cal P}(s) \propto [s\ln s]^{-2}$ at the percolation threshold, 

(ii) is a power-law function with an exponential cutoff at $s_c \propto 1/W$ in the phase with the giant component and, 

(iii) in contrast to standard percolation, ${\cal P}(s)$ has a power-law tail in the entire phase without the giant component. 

Furthermore, in Ref. \cite{dms01f} this model was generalized, and the percolation in growing scale-free undirected networks was studied (preferential linking mechanism was applied). The results are 
similar to those described above. However, in this case, the percolation threshold $b_c$ and factors in Eq. (\ref{34-2}) depend on the value of the exponent of the degree distribution. 

This anomalous behavior may be interpreted in the following way.      
New edges are being attached to large connected components with higher probability, and large connected components have a better chance to merge and grow. This produces the preferential growth of large connected components even in networks where new edges are attached to randomly chosen vertices, that is, without any preference.  
Such a mechanism of effective preferential attachment of new vertices to large connected components naturally produces power-law distributions of the sizes of connected components and power-law probabilities ${\cal P}(s)$. This ``self-organized critical state'' is realized in the growing network only if the 
giant component is absent. 

As soon as the giant connected component emerges, 
the situation changes radically. A new channel of the evolution of connected components is coming into play, and, with high probability, 
large connected components do not grow up to even larger ones but join to the giant component. Therefore, there are few large connected components if the giant component is present, and then ${\cal P}(s)$ is exponential. 

Thus, in the growing networks, two phases are in contact at the point of the emergence of the giant connected component --- the critical phase without the giant component and the normal phase with the giant component. 
This contact provides the above described effects. There exists another example of a contact of a ``critical phase'' (the line of critical points) with a normal phase, namely, the Berezinskii-Kosterlitz-Thouless phase transition \cite{b70,kt73}. 
Interestingly, critical dependences in both these cases have similar functional forms\vspace{8pt}.

We finish this section with the following remark. 
Other percolation problems for random networks can be considered. 
For instance, in the recent paper \cite{bg01}, ``core'' percolation was introduced. 
Dead-end vertices and their nearest neighbors are removed successively up to the point when no dead ends remain. The remaining giant component (if it exists) is called ``core''. 
For the classical random graph (Erd\"{o}s-R\'{e}nyi network), it was found that the core is present when the average degree $\overline{k}$ is above $e=2.718\ldots$. For comparison, in the same network, the ordinary giant connected component exists if 
$\overline{k}>1$ \cite{bbook85} (see Secs. \ref{s-classical} and \ref{ss-theory}).


\section{Growth of networks and self-organized criticality}\label{s-soc}

We have demonstrated above that the growing networks often {\em self-organize} into scale-free structures. 
The change of parameters controlling their growth removes them from the class of scale-free structures. This is typical for the general self-organized criticality phenomena \cite{btw87,btw88,bbook97}, so the considered processes can be linked with many other problems. 
In the present section, we discuss briefly this linking.


\subsection{Linking with sand-pile problems}\label{ss-sand-pile}

As long as only  a degree distribution, that is, the distribution of a one-vertex characteristic, is studied, the models of networks growing with preferential linking can be reduced to the following general problem \cite{dms001} (see Fig. \ref{f17}). 
At each increment of time, $m$ new particles are distributed between the {\em increasing} number (by one per time step) of boxes according to some rule. Here, the boxes play the role of vertices. The  particles are associated with edges. 
The probability that a new particle gets to a particular box depends on the filling of this box and on the filling numbers of all other boxes.  
In fact, what we made in Secs. \ref{s-exponential}, 
\ref{s-scale-free}, and \ref{s-non-scale-free}, was mainly consideration of 
various versions of this classical model. 

One can enumerate the boxes by age, so that such a system has boundaries, the ``oldest'' box and the new one, and, naturally, the distributions of particles in different boxes are different. 
In fact, the resulting enumerated set of boxes looks like a sand pile with a front (boxes being added) moving with unit rate. The height of the sand pile increases as the box age grows.  

Obvious relations to some other classical problems are also possible. For instance, the arising master equations can be generically related to those for fragmentation phenomena. 
One should also mention the Flory-Stockmayer theory of the polymer growth \cite{f41a,s43a,f71}.


\subsection{Preferential linking and the Simon model}\label{ss-simon}

Reasons for power-law distributions occurring in various systems, 
including systems mentioned in Sec. \ref{ss-sand-pile}, were a matter of interest of numerous empirical and theoretical studies starting from 1897 \cite{pbook897,gbook31,zbook49}.  
An important advance was achieved by    
H.A. Simon (1955), who proposed a simple model producing scale-free distributions \cite{s55,sbook57}. 

The Simon model, can be formulated in the following way \cite{zm00}.  

Individuals are divided to groups. 

(i) At each increment of time, a new individual is added to the system. 

(ii) a) With probability $p$ (Simon used the notation $\alpha$), it establishes a new family;  
b) with the complementary probability $1-p$, it chooses at random some old individual and joins its family. 

The rule (ii) b) simply means that new individuals are distributed among families with probability proportional to their sizes, similar to rules for preferential linking. 
The number of individuals, of course, equals $t$, and, at long times, the number of families is $pt$. Using the master equation approach (see Sec. \ref{ss-masterequation}), and passing to the long-time limit it is possible to get the following stationary equation for the distribution of 
the sizes of families, 
  
\begin{equation} 
P(k) + (1-p) [kP(k) - (k-1)P(k-1)] = \delta_{k,1}
\, .
\label{35-1}
\end{equation} 
Introducing $\gamma=1+1/(1-p)$, we can write the solution of Eq. (\ref{35-1}) in the form
  
\begin{equation} 
P(k) = (\gamma-1) B(k,\gamma) \stackrel{k\gg1}{\propto} k^{-\gamma}
\, ,
\label{35-2}
\end{equation} 
where $B(\ )$ is the beta-function. Therefore, the power-law distribution with exponent $\gamma$ naturally arises. Recently, the non-stationary distribution $P(k,t)$ was also described analytically \cite{kk00} (see also Ref. \cite{kk01b}). 

The Simon model was originally proposed without any relation to networks.
However, it is possible to formulate the Simon model for networks in terms of vertices and, e.g., directed edges \cite{be00}. 
 
(i) At each increment of time, a new edge is added to the network. 
 
(ii) a) Also, with probability $p$ a new vertex is added, and the target end of the new edge is attached to the vertex. b) With the complementary probability $1-p$, the target end of the new edge is attached to the target end of a randomly chosen old edge. 

Here, rule (ii) b) corresponds to the distribution of new edges among vertices with probability proportional to their in-degree. 
One should indicate some difference between the Simon model and the models of growing networks with preferential linking. In the models of Secs. \ref{s-exponential}, 
\ref{s-scale-free}, and \ref{s-non-scale-free}, one vertex was added at each time step. In the Simon model, at each increment of time, one individual (edge) is added, and the number of added families (vertices)
 is not fixed. Of course, this can not change the stationary distributions and the value of $\gamma$ exponent. The behavior of $P(k,t)$ at long times (large network sizes) is also similar. 

In fact, both the original Simon model and the preferential linking concept are based on a quite general principle -- {\em popularity is attractive}. Popular objects (idols) attract more new fans than the unpopular ones. 

Nevertheless, one should note that the matter of interest of Simon was the one-particle distribution, whereas, for networks, this is only a small part of the great problem: what is their topology?




\subsection{Multiplicative stochastic models and the generalized Lotka-Volterra equation}\label{ss-multiplicative}

One may look at the models of the network growth under mechanism of preferential linking from another point of view. 
The variation (increase) of the degree of a vertex is proportional to the degree of this vertex.   
This allows us to relate such models to the wide class of multiplicative stochastic processes. 
Last time, these processes are intensively studied in econophysics and evolutionary biology \cite{sm00z,sl96,sc97,mmz98,bmls98,bpbook00,bm00}. 
In particular, they are used for the description of wealth distribution.  
The most widely known example is the generalized Lotka-Volterra equation \cite{sm00z}, 
  
\begin{equation} 
w_i(t+1) = r_i(t)w_i(t) + A t w(t) - c(w(t),t) w_i(t)
\, .
\label{37-1}
\end{equation}  
Here, $w_i(t)$ may be interpreted as the wealth of agent $i$, $i=1,\ldots,N$, 
$w(t)=\sum_{i=1}^N w_i(t)/N$ is the average wealth at time $t$. 
The distribution of the random noise $ r_i(t)$ is independent of $t$. 
Its average value is $\langle r_i(t)\rangle = B t$ and the standard deviation equals $D t$. 
$A$ is a positive constant, and $c(w,t)$ is proportional to $t$ at long times. Such a dependence on $t$ of the coefficients in Eq. (\ref{37-1}) provides stationary distributions at long times. In the differential form, Eq. (\ref{37-1}) can be written as 
  
\begin{equation} 
\frac{d w_i(t)}{dt} = [r_i(t)-1]w_i(t) + A t w(t) - c\left(w(t),t\right) w_i(t)
\, .
\label{37-2}
\end{equation}  
Interpretation of Eqs. (\ref{37-1}) and (\ref{37-2}) in terms of the wealth distribution is quite obvious. In particular, the last term restricts the growth of wealth. It was shown \cite{bm00,sr00e} that the average wealth $w(t)$ approaches a fixed value $\overline{w}$ at long times, 
and these equations produce the following stationary distribution 
  
\begin{equation} 
P(w_i) \propto \exp\left( -\frac{A}{D}\frac{\overline{w}}{w_i} \right) w_i^{-\gamma}
\, ,
\label{37-3}
\end{equation} 
where the exponent of the power-law dependence is $\gamma = 2+A/D$. Eq. (\ref{37-3}) gives $P(0)=0$.
Note that the resulting distribution is independent of $B$ and $c(w,t)$. 

The main difference of the particular stochastic multiplicative process described by Eqs. (\ref{37-1}) and (\ref{37-2}) from the models considered in Secs. \ref{s-exponential}, \ref{s-scale-free}, \ref{s-non-scale-free}, and \ref{ss-simon} is the fixed number of the involved agents. 
Nevertheless, the outlined general approach can be used for networks (e.g., see Ref. \cite{mh00}). 
On the other hand, the results obtained for the degree distributions of evolving networks (see Secs. 
\ref{s-exponential}, \ref{s-scale-free}, and \ref{s-non-scale-free}) may be interpreted, for example, in terms of the wealth distribution in evolving societies.


\section{Concluding remarks}\label{s-concluding}

The progress in this field is incredibly rapid, and we have 
failed to discuss and even cite a number of recent results. 
In particular, we have missed random, intentional, and 
adaptive walks on networks \cite{plh01,bp01,alph01,t01b,lkk01a} 
which are closely related to problems of the organization of 
effective search in communication networks. We didn't discuss 
the distribution of the number of the shortest paths passing 
through a vertex, which has a power-law form in scale-free 
networks \cite{gkk01,gkk01b}, and many other interesting problems 
(see, e.g., Refs. \cite{mz01a,brst01,cd99,cd00,dc00,spsk01a,mgp01,jgn01,bb01b,wh00b,wh00a,wh00c,adg00,deb01a,khh01a}). 
After this review had been submitted, we learned about 
the review of Albert and Barab\'asi on the statistical 
mechanics of networks under preparation \cite{ab01a}. We call the reader's attention to this paper. 

Most studies which we reviewed focused on structural properties of growing networks. 
Two aspects of the problem can be pointed out. 

(i) Specific mechanisms of the network growth 
produce their structure and, in particular, the degree distributions of their vertices. 
We demonstrated that the preferential linking mechanism (preferential attachment 
of new edges to vertices with a higher number of connections) provides degree distributions with long  tails. 
Such nets are abundant in Nature. In particular, the communications networks have degree distributions just of this kind. 
The preferential linking is the reason of the {\em self-organization} of a number of growing networks into scale-free structures.

(ii) The resulting networks with such long-tailed distributions have quite different properties than classical random graphs with the Poisson degree distribution. 
In particular, they may be extremely resilient to random damage. This substantial property 
partly explains their abundance in Nature. 
From the other hand, diseases may freely spread within these nets. 
The global topology of such networks is described by a theory that actually generalizes the standard percolation theory. 
This theory is based on the assumption of statistical independence of vertices of an {\em equilibrium} network. 
In such an event, the joint in- and out-degree distribution $P(k_{i},k_{o})$ completely determines the structure of the network. 
This is one of the reasons why the knowledge of the degree distribution is so important. 
Despite the evident success of this approach, one can see that its basic assumptions are not valid for {\em non-equilibrium}, growing networks. 

Keeping in mind most intriguing applications to the evolving communications networks, we have to admit that, currently, the most of the discussed models and ideas can be applied to the real networks only on a schematic, qualitative level. These simple models are still far from reality and only address particular phenomena in real networks. 

The title of the seminal paper of Erd\"os and R\'enyi (1960) was ``On the evolution of random graphs'' \cite{er60}. What Erd\"os and R\'enyi called ``random graphs'' were simple equilibrium graphs with the Poisson degree distribution. 
What they called ``evolution'' was actually the construction procedure 
for these graphs.  
Main recent achievements in theory of networks are related with transition to study of the 
{\em evolving, self-organizing networks with non-Poisson degree distributions}. 
The fast progress in this field, in particular, means a very significant step toward understanding the most 
exciting networks of our World, the Internet, the WWW, and basic biological networks.


\section*{Acknowledgements}\label{s-acknowledgements}

SND thanks PRAXIS XXI (Portugal) for a research grant PRAXIS XXI/BCC/16418/98. The authors  
were partially supported by the project POCTI/1999/FIS/33141. We acknowledge G. Tripathy for thorough reading our manuscript. We also thank A.N. Samukhin  
for numerous stimulating discussions and reading the manuscript. Without his help, this review 
would have been impossible.

\vspace{20pt}

\noindent
$^{\ast}$      Electronic address: sdorogov@fc.up.pt\\
$^{\dagger}$   Electronic address: jfmendes@fc.up.pt

\section*{References}\label{s-references}






\end{multicols}

\end{document}